\documentclass[aps,prd,preprint,groupedaddress,nofootinbib,showpacs,eqsecnum]{revtex4}
\usepackage{graphicx,epsf,color,amsmath,hyperref,multirow}
\usepackage[normalem]{ulem}
\usepackage{mathtools}
\newif\ifdraft
\drafttrue
\newif\ifpreprint
\preprinttrue

\def\fig#1{Fig.~{\ref{#1}}}
\def\Fig#1{Fig.~{\ref{#1}}}
\def\figs#1#2{Figs.~{\ref{#1}} and~{\ref{#2}}}

\def\sect#1{Section~{\ref{#1}}}
\def\sects#1#2{Sections~{\ref{#1}} and~{\ref{#2}}}
\def\app#1{Appendix~{\ref{#1}}}

\def\tab#1{Table~{\ref{#1}}}
\def\Tab#1{Table~{\ref{#1}}}

\def\spa#1.#2{\left\langle#1\,#2\right\rangle}
\def\spb#1.#2{\left[#1\,#2\right]}
\def\tree{{\rm tree}}

\def\fiveloop{{\rm 5\hbox{-}loop}}

\def\Tr{\, {\rm Tr}}

\def\eps{\epsilon}

\def\nn{\nonumber}

\def\C{{\cal C}}
\def\E{{\cal E}}

\def\N#1{{N$^{#1}$MC}}
\def\be{\begin{eqnarray}}
\def\ee{\end{eqnarray}}
\def\G{{\rm GR}}

\def\eqn#1{Eq.~(\ref{#1})}
\def\Eqn#1{Equation~(\ref{#1})}
\def\eqns#1#2{Eqs.~(\ref{#1}) and~(\ref{#2})}

\def\NeqOne{{{\cal N}=1}}

\def\NeqFour{{{\cal N}=4}}
\def\NeqFive{{{\cal N} = 5}}
\def\Neqeight{{{\cal N}=8}}
\def\NeqEight{{{\cal N}=8}}

\def\NkMC{{{\rm N}^k{\rm MC}}}

\def\BCJ{{\rm BCJ}}

\def\tn{{\tilde n}}
\def\al{\alpha}
\def\tal{{\tilde \alpha}}
\def\tJ{{\tilde J}}
\def\c{{\cal K}}
\def\E{{\cal E}}
\def\J{{\tilde J}}
\def\tDelta{{\tilde \Delta}}
\def\x#1{{\bullet}}
\def\P{{\rm P}}
\def\NP{{\rm NP}}

\def\bea{\begin{eqnarray}}
\def\eea{\end{eqnarray}}
\def\ba{\begin{eqnarray}}
\def\ea{\end{eqnarray}}

\def\tree{{\rm tree}}

\newbox\charbox
\newbox\slabox
\def\s#1{{      % Feynman slash
\setbox\charbox=\hbox{$#1$}
\setbox\slabox=\hbox{$/$}
\dimen\charbox=\ht\slabox
\advance\dimen\charbox by -\dp\slabox
\advance\dimen\charbox by -\ht\charbox
\advance\dimen\charbox by \dp\charbox
\divide\dimen\charbox by 2
\raise-\dimen\charbox\hbox to \wd\charbox{\hss/\hss}
\llap{$#1$} }}

\begin{document}
\hfuzz 15 pt

\ifpreprint
\hbox{\hskip0.3cm UCLA/17/TEP/105 \hskip 2.8 cm  NORDITA-2017-071  \hskip3.5 cm
UUITP-29/17}
\fi

\title{The Five-Loop Four-Point Integrand of $\NeqEight$ Supergravity\\ as a Generalized Double Copy}

\author{Zvi~Bern${}^{a}$, John Joseph Carrasco${}^{b}$, Wei-Ming Chen${}^a$, 
Henrik Johansson${}^{c,d}$, Radu Roiban${}^{e}$ and Mao Zeng${}^a$}
\affiliation{
$\null$\\
${}^a$Mani L. Bhaumik Institute for Theoretical Physics\\
UCLA Department of Physics and Astronomy
Los Angeles, CA 90095, USA \\
\vskip -4 mm
${}^b$Institute of Theoretical Physics (IPhT), \\
CEA-Saclay and University of Paris-Saclay\\ 
F-91191 Gif-sur-Yvette cedex, France\\
\vskip -4 mm
${}^c$Department of Physics and Astronomy, Uppsala University, 75108 Uppsala, Sweden \\
\vskip -4 mm
${}^d$Nordita, Stockholm University and KTH Royal Institute of Technology,
Roslagstullsbacken 23, 10691 Stockholm, Sweden \\
\vskip -4 mm
${}^e$Institute for Gravitation and the Cosmos, Pennsylvania State University, 
    University Park, PA 16802, USA \\
}

%\date{\today}

\begin{abstract}
We use the recently developed generalized double-copy procedure to
construct an integrand for the five-loop four-point amplitude of
$\NeqEight$ supergravity.  This construction starts from a naive
double copy of the previously computed corresponding amplitude of
$\NeqFour$ super-Yang-Mills theory.  This is then systematically
modified by adding contact terms generated in the context of the
method of maximal unitarity cuts.  For the simpler generalized
cuts, whose corresponding contact terms tend to be the most complicated,
we derive a set of formulas relating the contact contributions 
to the violations of the dual Jacobi identities in the relevant
gauge-theory amplitudes.
For more complex generalized unitarity
cuts, which tend to have simpler contact terms associated with them, we
use the method of maximal cuts more directly.  The five-loop
four-point integrand is a crucial ingredient towards future
studies of ultraviolet properties of $\NeqEight$ supergravity
at five loops and beyond.  We also present a nontrivial check of the
consistency of the integrand, based on modern approaches for 
integrating over the loop momenta in the ultraviolet region.
\end{abstract}

\pacs{04.65.+e, 11.15.Bt, 11.25.Db, 12.60.Jv \hspace{1cm}}

\maketitle
%\tableofcontents

\section{Introduction}

In recent years there has been enormous progress in our ability to
construct supergravity scattering amplitudes at high loop orders. This
progress flows primarily from three classes of conceptual and technical
advances.  The first is the development of the unitarity
method~\cite{GeneralizedUnitarity,MaximalCutMethod}, which offer a
straightforward algorithmic approach to constructing and verifying
multiloop integrands using only on-shell tree amplitudes.  The second
is the discovery of the Bern-Carrasco-Johansson (BCJ)
color-kinematics duality and associated double-copy
procedure~\cite{BCJ,BCJLoop}.  The third is the progress in loop
integration methods, specifically integration-by-parts
(IBP) reduction~\cite{SmirnovBook, IBPRefs, IBPAdvances,
  Larsen2015ped, Zhang2016kfo}, which has been critical to extracting
ultraviolet information, as in Refs.~\cite{ThreeFourloopN8,
  SimplifyingBCJ, N4GravFourLoop,
  N4GravThreeLoops,N4SugraMatter,N5GravFourLoops}.

In this paper we will describe in more detail the generalized
double-copy procedure recently introduced in
Ref.~\cite{GeneralizedDoubleCopy}, which combines elements of
generalized unitarity and color-kinematics duality to convert generic
gauge-theory loop integrands into gravity ones.  We use the method to
construct the five-loop four-point integrand of $\NeqEight$
supergravity~\cite{NeqEightSugra}, which is an important stepping
stone towards unraveling the ultraviolet properties of this theory.
The organization of the resulting amplitude is provided by the method
of maximal cuts~\cite{MaximalCutMethod}.  A number of other related
on-shell methods have also been developed for constructing multiloop
integrands, especially for supersymmetric theories in four
dimensions~\cite{OtherUnitarity}.  There are also promising methods
for directly constructing integrated expressions for amplitudes,
especially for $\NeqFour$ super-Yang-Mills theory in four
dimensions~(e.g. see Ref. \cite{Bootstrap}).

The duality between color and kinematics plays a central role in our
construction.  Whenever representations of gauge-theory integrands are
constructed which manifest the duality between color and kinematics,
corresponding gravity integrands follow directly via the double-copy
procedure~\cite{BCJLoop}, which replaces color factors with kinematic
factors.  The duality applies to wide classes of gauge and gravity
theories~\cite{BCJ,BCJLoop,DoubleCopyTheories,FundMatter,
  DoubleCopyTheoriesFund, Conformal}, where, in many cases, the
duality has been proven at tree
level~\cite{Square,KiermaierTalk,BjerrumMomKernel,BCJTreeProof,BCJRelationProof,
  SelfDualYM, NLSMaction}. At loop level the duality has conjectural
status, supported by case-by-case explicit calculations.  The duality
has been crucial in the construction of numerous gravity multiloop
amplitudes~\cite{BCJLoop, SimplifyingBCJ, OtherExamples, TwoLoopSQCD},
where it has been used to identify new nontrivial ultraviolet
cancellations in $\NeqFour$ and $\NeqFive$
supergravity~\cite{N4GravThreeLoops,N5GravFourLoops}, known as
`enhanced cancellations'.  Apart from offering a simple means for
obtaining loop-level scattering amplitudes in a multitude of
(super)gravity theories, the duality also addresses the construction
of black-hole and other classical solutions~\cite{ClassicalSolutions}
including those potentially relevant to gravitational-wave
observations~\cite{RadiationSolutions}, corrections to gravitational
potentials~\cite{Donoghue}, the relation between supergravity  symmetries and
gauge-theory ones~\cite{SugraSyms, DoubleCopyTheories,
  DoubleCopyTheoriesFund}, and the construction of multiloop form
factors~\cite{MultiLoopFormFactor}.  The duality has also been identified
in a wider class of quantum field and string
theories~\cite{BLGBCJ,NLSMBCJ,abelianZ,NLSMaction,
  DiskandHeteroticStringBCJ}.  For recent reviews, see
Ref.~\cite{Review}.

However, experience shows that it can sometimes be difficult to find
multiloop integrands where the duality is
manifest~\cite{BCJDifficulty}.  The best known example is the
five-loop four-point integrand of $\NeqFour$ super-Yang-Mills
theory~\cite{FiveLoopN4}, which has so far resisted all attempts to
construct a BCJ representation where the duality between color and
kinematics manifestly holds. This amplitude is crucial for unraveling
ultraviolet cancellations that are known to exist in supergravity
theories but for which no symmetry explanation has been
given~\cite{N5GravFourLoops}.  Because of the complexity of
gravity amplitudes at high loop orders, alternative methods have offered no
path forward; the only currently-known practical means for constructing
the five-loop amplitude is to use a double-copy procedure that
recycles the corresponding gauge-theory
amplitude~\cite{FiveLoopN4}. More generally, we  would like to have a
technique that converts any form of a gauge-theory integrand into the
corresponding gravity ones.

A solution to this technical obstruction has been recently proposed in
Ref.~\cite{GeneralizedDoubleCopy}, which introduced a generalized
double-copy procedure which makes use of general representations of
the gauge-theory integrand. This new approach builds on the central
premise of double-copy construction, but relies only on the proven
existence of BCJ duality at tree level. Generic representations of
gauge-theory integrands that use $f^{abc}$ color factors are
double copied, giving a `naive double copy'.  If algebraic relations
obeyed by the color factors are not mirrored by the kinematic factors,
however, this alone does not result in a correct gravity integrand.
Violations of the kinematic algebra (dual to the color Lie algebra)
must be compensated.  These violations, or `BCJ discrepancy
functions', are the building blocks for new formulas that give
corrections to the naive double copy.  The correction formulas merge
seamlessly with the method of maximal cuts to constructively build
gravity predictions from generic gauge-theory integrands.  These
correction formulas also have a double-copy structure, being bilinear
in the discrepancy functions of each gauge theory.

The starting point of our construction of the five-loop four-point
amplitude of $\NeqEight$ supergravity is the representation of the
$\NeqFour$ super-Yang-Mills amplitude given in Ref.~\cite{FiveLoopN4},
with a slight rearrangement of a few terms. The supergravity amplitude
is constructed via the generalized double-copy procedure.  In
principle, there could have been up to
70,690 diagrammatic contributions with up to millions of terms each.
Fortunately the vast majority of these diagrams either vanish or are
much simpler than naive power counting suggests. Still the expressions
are lengthy, and the final result is collected in a {\em
  Mathematica}-readable attachment~\cite{AttachedFile}.

To confirm our integrand, we carried out a number of
nontrivial checks.  Besides the generalized cuts used in the
construction, we also check consistency of large numbers of
additional generalized unitary cuts.  We numerically confirmed in all
cases where the new formulas are used that a less efficient evaluation
of the gravity unitarity cuts, based on Kawai-Lewellen-Tye tree-level
relations, gives identical results.
We also present nontrivial checks based on integrating the expressions
in spacetime dimension $D=22/5$, where we expect it to be finite, yet individual
terms in our expression diverge.  To carry out these checks we develop
techniques based on modern developments in
integration~\cite{IBPAdvances, Zhang2016kfo, Harley2017, Bosma2017}.
We carry out the check using both unitarity-compatible IBP
methods as well as a new method of direct integration
described in \app{sec:IntegrationAppendix}. 

We leave for the future the much more interesting---and much more
difficult---case of integrating in dimension $D=24/5$, where symmetry
arguments suggest that a divergence could be present~\cite{BjornssonAndGreen,SevenLoopGravity}.  The discovery
of enhanced ultraviolet cancellations in closely-related supergravity
theories~\cite{N4GravThreeLoops,N5GravFourLoops} suggests, however,
that the five-loop amplitude might nonetheless be finite in $D=24/5$.
A direct integration of our integrand would settle the issue.

This paper is organized as follows.  First, in \sect{Review}, we present
a brief review of the method of maximal cuts and the double-copy
construction.  Then, in \sect{BCJFormulaOverview}, we give an overview
of the derivation of the new formulas for obtaining correction to the
naive double copy in terms of BCJ discrepancy functions.  In
\sect{N2LevelJFormulasSection}, we derive the explicit formulas giving
the contact term corrections, involving two four-point contact
interactions or one five-point interaction.  This is generalized to
infinite classes of contact interactions in
\sect{HigherLevelJFormulasSeaction}.  The results for the five-loop
four-point integrand of $\NeqEight$ supergravity are described in
\sect{FiveLoopResultsSection}.  In \sect{sec:vacuumExpansion}, we
series expand the integrand in large loop momenta and perform
nontrivial integration checks demonstrating its consistency.
Our conclusions and outlook are given in
\sect{ConclusionSection}. Two appendices are included; the first gives
correction formulas useful for contact diagrams with four canceled
propagators and the second describes a unitarity-compatible direct
integration of vacuum diagrams generated by series expanding the
integrand.

%%%%%%%%%%%%%%%%%%%%%%%%%%%%%%%%%
\section{Review}
\label{Review}

In the mid 1980s string-theory investigations by Kawai, Lewellen
and Tye (KLT)~\cite{KLT}  exposed remarkable relations between closed- and
open-string tree-level scattering amplitudes.  Since string-theory
tree-level amplitudes have smooth low-energy limits to gauge and
gravity field theory amplitudes, this had a number of implications for
field-theory predictions~\cite{BGK,OneloopN8}. With the advent of
unitarity methods~\cite{GeneralizedUnitarity}, these tree-level
insights have direct impact on our ability to calculate at loops as
well as on our basic understanding of the structure of gravity loop
amplitudes~\cite{BDDPR,OneloopN8}.  With the understanding of the
duality between color and kinematics, much simpler and powerful means
for generating gravity loop amplitudes from gauge theory became
available~\cite{BCJ,BCJLoop}.  We begin with a lightning review of
double-copy structure of gravity amplitudes, before discussing
application of the method of maximal cuts relevant for our
construction~\cite{GeneralizedDoubleCopy}.

\subsection{Tree-level gravity amplitudes from gauge theory}
\label{subsec:TreeLevelReview}

\subsubsection{BCJ duality and double-copy amplitudes}

All tree-level amplitudes in any $D$-dimensional gauge theory coupled 
to fields in the adjoint representation, may be written as
\begin{equation}
\mathcal{A}^{\tree }_{m} = g^{m-2} \sum_{j \in \Gamma_{3,m}}
\frac{c_{j} n_{j}}{D_j} \,,
\label{CubicRepresentation}
\end{equation}
where sum is over the set of $(2m-5)!!$ distinct, 
$m$-point graphs with only cubic (trivalent) vertices,
which we denote by $\Gamma_{3,m}$.
These graphs are sufficient because the contribution of any diagram with
quartic or higher vertices can be assigned to a graph with only
cubic vertices by multiplying and dividing by appropriate
propagators.  The nontrivial kinematic information is contained in the
numerators $n_{j}$ and  generically depends on momenta, polarizations, and spinors.
The color factor $c_j$ is obtained by dressing every vertex in graph
$j$ with the relevant gauge-group structure constant,
$\tilde{f}^{abc}=i\sqrt{2}f^{abc} = \mathrm{Tr}([T^{a},T^{b}] T^{c})$,
where the Hermitian generators of the gauge group are normalized via
$\mathrm{Tr}(T^{a}T^{b})=\delta^{a b}.$ The denominator $1/D_j$
contains the Feynman propagators of the graph $j$
\begin{equation}
\frac{1}{D_j} \equiv \frac{1}{\prod_{i_{j}} d_{i_{j}}}\,,
\end{equation}
where $i_j$ runs over the propagators for diagram $j$, each of which we denote
by $1/d_{i_{j}}$.  The gauge-theory coupling constant is $g$.  If an
on-shell superspace is used the numerators will also depend on
anticommuting parameters.

%%%%%%%%%%%%%% FIGURE %%%%%%%%%%%%% 
\begin{figure}
\includegraphics[scale=.38]{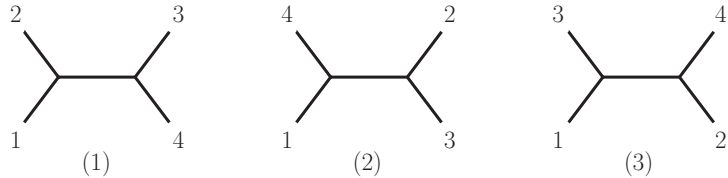}
\vskip -.3 cm 
\caption{The three diagrams with only cubic vertices
contributing to a four-point tree amplitude.
   }
\label{FourBlobFigure}
\end{figure}
%%%%%%%%%%%%%%%%%%%%%%%%%%%    

In a BCJ representation, kinematic numerators obey the same generic
algebraic relations as the color
factors~\cite{BCJ,BCJLoop,SimplifyingBCJ, Review}.  For theories with
only fields in the adjoint representation there are two
properties.  The first property is {\em antisymmetry} under graph
vertex flips: 
\begin{equation}
 c_{\overline{i}} = - c_{i}\ \Rightarrow\ n_{\overline{i}}=- n_{i}\, ,
\label{BCJFlipSymmetry}
\end{equation}
where the graph $\overline{i}$ has same graph connectivity as
graph $i$, except an odd number of vertices have been cyclically
reversed.  The second property is the requirement that all {\em Jacobi
  identities} are satisfied,
\begin{equation}
c_{i}+c_{j}+c_{k} = 0\ \Rightarrow\ n_{i}^\BCJ+n_{j}^\BCJ+n_{k}^\BCJ = 0 \,,
\label{BCJDuality}
\end{equation}
where $i$, $j$, and $k$ refer to three graphs which are identical except for one internal edge.
For example at four points the color factors of
the three diagrams listed in \fig{FourBlobFigure} obey the Jacobi identity.

Once corresponding gauge-theory loop integrands have been arranged
into a form where the duality is manifest~\cite{BCJ,BCJLoop}, it is
then easy to obtain gravity loop integrands: one simply replaces the
color factors of a gauge-theory integrand with the kinematic
numerators of another gauge-theory integrand,
\begin{equation}
c_{i}\ \rightarrow\ \tilde{n}_{i}\,.
\label{ColorSubstitution}
\end{equation}
This immediately gives the double-copy form of a gravity tree amplitude,
\begin{equation}
\mathcal{M}^{\tree}_{m} = i \left(\frac{\kappa}{2}\right)^{m-2}
\sum_{j \in \Gamma_{3,m}}\frac{\tilde{n}_{j}n_{j}}{D_j} \,,
\label{DoubleCopy}
\end{equation}
where $\kappa$ is the gravitational coupling and 
 $\tilde{n}_j$ and $n_j$ are the kinematic numerator factors of the two gauge theories.
Only one of the two sets of numerators needs to manifestly satisfy the duality
(\ref{BCJDuality})~\cite{BCJLoop,Square} in order for the
double-copy form (\ref{DoubleCopy}) to  be valid.

\subsubsection{Ordered partial amplitudes}

The color factors, $c_{i}$ in \eqn{CubicRepresentation}, can be
expressed in a color-trace basis. Collecting associated kinematic
factors yields,
\begin{equation}
\mathcal{A}^{\tree }_{m}=
g^{m-2} \sum_{\rho \in {\cal S}_{m-1}} \mathrm{Tr}\left(T^{\rho_1} T^{\rho_2} \ldots T^{\rho_m}\right) 
A^{\tree}_{m}(\rho_1,\rho_2, \ldots, \rho_m)\,.
\label{PartialPresentation}
\end{equation}
where the sum runs over the set ${\cal S}_{m-1}$ of non-cyclic permutations.
The $A^{\tree}_m(\rho)$ are called color-ordered partial amplitudes.,
The terminology {\em ordered} refers to 
the fact that all graphs contributing to any given
$A^{\tree}_{m}(\rho)$ have the same ordering or external legs
as the cyclic ordering of the color trace ${\rm Tr}(\rho)$.
We can write the color-ordered amplitudes in terms of graphs via
\begin{equation}
A^{\tree}_{m}(1,\rho_2, \ldots, \rho_m) = 
\sum_{i \in \Gamma_{\rho}} \frac{n_i}{D_i}\,,
\label{orderedAmplitudes}
\end{equation}
where $\Gamma_{\rho}$ refers to the graphs with cubic vertices where
the legs are ordered following the color ordering.

%%%%%%%%%%% FIGURE %%%%%%%%%%
\begin{figure}
  \centering
  \includegraphics[width=0.45\textwidth]{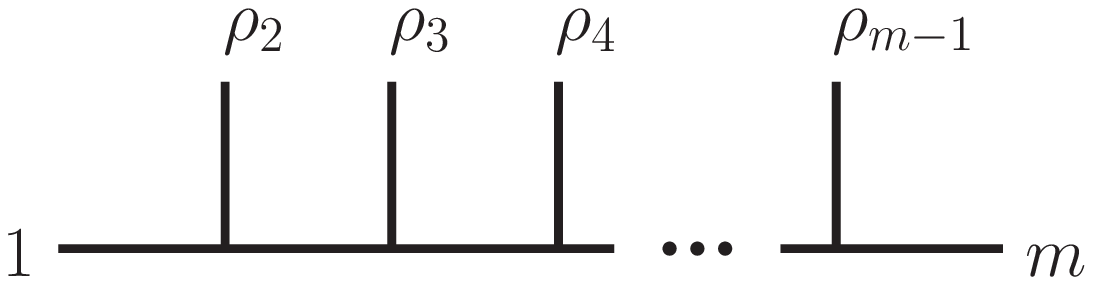}
  \vskip -.3 cm 
  \caption{A half-ladder tree graph, used to define the color factor in \eqn{PartialRepresentation}.}
  \label{fig:HalfLadder}
\end{figure}
%%%%%%%%%%%%%%%%%%%%%%%%%    

The partial-ordered amplitudes in \eqn{PartialPresentation} are not
independent and can be reduced to a sum over $(m-2)!$ partial amplitudes
using color-Jacobi identities~\cite{DDM}
\begin{equation}
\mathcal{A}^{\tree }_{m}= g^{m-2} \sum_{\rho \in {\cal S}_{m-2}}
c(1|\rho_2, \ldots, \rho_{m-1} |m) A^{\tree}_{m}(1,\rho_2, \ldots,
\rho_{m-1}, m)\,,
\label{PartialRepresentation}
\end{equation}
where $c(1|\rho_2, \ldots, \rho_{m-1}|m)$ is the color factor of the half-ladder diagram in \fig{fig:HalfLadder}.
Replacing $c(1|\rho |m)$ by a color-dual kinematic numerator of the
same half-ladder graph $n^{\BCJ}(1|\rho| m)$, and taking into
account the appropriate ratio of coupling constants indeed yields
another representation of gravity tree amplitudes~\cite{Square}:
\begin{equation}
\mathcal{M}^{\tree }_{m}=
 i \left(\frac{\kappa}{2}\right)^{m-2} \sum_{\rho \in {\cal S}_{m-2}} 
\tilde n^{\BCJ}(1 | \rho_2, \ldots, \rho_{m-1} | m) A^{\tree}_{m}(1,\rho_2, \ldots, \rho_{m-1}, m)\,.
\label{PartialRepresentationG}
\end{equation}

\subsubsection{KLT relations}

The KLT relations~\cite{KLT} give direct relations between gravity and
gauge-theory tree amplitude.  The KLT formulas can be obtained from
BCJ duality, by using Jacobi identities to express all kinematic
numerators in \eqn{orderedAmplitudes} in terms of a basis of $(m-2)!$
numerators, called master numerators.  One can then (pseudo-)invert
the relationship between a minimal basis of $(m-3)!$ independent
ordered amplitudes to solve for the master numerators in terms of
partial amplitudes.  Indeed, the availability of a color-dual form for
kinematic numerators is responsible for the reduction to a basis of
$(m-3)!$~\cite{BCJ}.  As the propagator matrix is singular, such
pseudo-inversions are not unique, so there are many possibilities.

The first such formula valid for an arbitrary number of legs was given
in Appendix A of Ref.~\cite{OneloopN8}. It remains as a particularly
sparse and efficient form, so we use it for directly 
constructing gravity unitarity cuts.  The tree-level relation is 
\begin{eqnarray}
{\cal M}^\tree_m(1,2,\ldots,m)&=&i(-1)^{m+1}  \Bigl( {\kappa \over 2} \Bigr)^{m-2} 
\Bigl[A^\tree_m(1,2,\ldots,m)\displaystyle
\sum_{\rm perms}f(i_1,\ldots,i_j)\overline{f}(l_1,\ldots,l_{j'}) \nn\\
&&\hskip 3 cm \null\times
\tilde{A}^\tree_m(i_1,\ldots,i_j,1,m-1,l_1,\ldots,l_{j'},m)\Bigr]\nn\\
&& \null \hskip 1 cm
 +{\rm Perm}(2,\ldots,m-2) \,,
\label{KLTmpoint}
\end{eqnarray}
where $A_m$ and $\tilde A_m$ are two $m$-point gauge-theory amplitudes 
from each of the two copies. The sum is over all permutations $\{i_1,\ldots,i_j\} \in 
{\rm Perm}\{2,\ldots,\lfloor m/2\rfloor\} $ and $\{l_1,\ldots,l_{j'}\}\in
{\rm Perm}\{\lfloor m/2\rfloor+1,\ldots,m-2\} $ with $j=\lfloor m/2  
\rfloor-1$ and $j'= \lfloor m/2 \rfloor-2$, which gives a total of
$(\lfloor m/2 \rfloor-1)! \times (\lfloor m/2 \rfloor-2)!$ terms inside the
square brackets. The notation ``$+\,{\rm Perm}(2,\ldots,m-2)$'' signifies
a sum over the expression for all permutations of legs
$2,\ldots,m-2$. The functions $f$ and $\overline{f}$ are given by,
\begin{eqnarray}
f(i_1,\ldots,i_j)&=&s_{1,i_j}\prod_{m=1}^{j-1}
\left(s_{1,i_m}+\sum_{k=m+1}^{j} g(i_m,i_k) \right),\nn\\
\overline{f}(l_1,\ldots,l_{j'})&=&s_{l_1,m-1}\prod_{m=2}^{j'}
\left(s_{l_m,m-1}+\sum_{k=1}^{m-1} g(l_k,l_m) \right) ,
\end{eqnarray}
and 
\begin{equation}
g(i,j)=\left\{
\begin{array}{ll}
        s_{i,j}  & \mbox{if $i> j$,}\\
        0& \mbox{otherwise.}
\end{array} \right.
\end{equation}

By applying BCJ amplitude relations~\cite{BCJ}, many different versions
of KLT relations can be constructed~\cite{BjerrumMomKernel},
including a tidy recursive definition~\cite{abelianZ}.  The general form
of the KLT relations in terms of a basis of gauge-theory amplitudes
may be written as,
\begin{align}
\mathcal{M}^{\tree }_{m}= & \null
 i \left(\frac{\kappa}{2}\right)^{m-2}   
\smashoperator{\sum_{\tau,\rho \in {\cal S}_{m-3}}}K(\tau|\rho) 
\tilde A^{\tree}_{m}\left(1,\rho_2, \ldots, \rho_{m-2}, m, (m-1)\right)\nonumber \\
& \hskip 2 cm \times
A^{\tree}_{m}\left(1,\tau_2, \ldots, \tau_{m-2}, (m-1), m\right) \, ,
\label{KLT}
\end{align}
where the sum runs over $(m-3)!$ permutations of external legs.  The
KLT matrix $K(\tau|\rho)$, indexed by the elements of the  
two permutation orderings of
the relevant partial amplitudes, also called a momentum kernel,
depends only on momentum invariants arising from inverse propagators.

Not only do the various versions of KLT kernels follow from
color-kinematics duality, but a comparison of \eqn{PartialRepresentationG} and
\eqn{KLT} gives a useful non-local representation of color-dual BCJ
numerators from color-ordered partial amplitude~\cite{KiermaierTalk,
  BjerrumMomKernel}.  This gives a set of explicit nonlocal BCJ
numerators,
\begin{align} 
n^\BCJ(1 | \tau_2, \ldots,& \tau_{m-2} , \tau_{m-1} | m)\nonumber  \\
&  = \left\{
 \begin{array}{lc}
   \displaystyle    \sum_{\rho \in {\cal S}_{m-3}} K(\tau|\rho) A^{\tree}_{m}(1, \rho_2, \ldots
\rho_{m-2}, m, (m-1))\,,  & \hskip .6cm \hbox{if } \tau_{m-1} = m-1  \,, \\
       0\,, & \hskip .6 cm \hbox{if } \tau_{m-1} \not = m-1 \,.
  \end{array}  \right. 
 \label{ExplicitBCJFromKLT}
\end{align} 
In this formula the permutations of $(m-2)$ legs of the half ladder is
effectively reduced to a permutation sum over $(m-3)$ legs, because some
of the numerators vanish.  Numerators of diagrams which are not of the
half-ladder form in \fig{fig:HalfLadder} follow from the dual Jacobi
relations~\eqref{BCJDuality}.  \Eqn{ExplicitBCJFromKLT} is useful
below to derive KLT forms of unitarity cuts from BCJ forms.

\subsection{Method of maximal cuts}
\label{ConstructionSection}

We now review the method of maximal cuts~\cite{MaximalCutMethod}
applied to building a double-copy gravity integrand.  The method of
maximal cuts is a refinement of the generalized unitarity
method~\cite{GeneralizedUnitarity}.  We organize the maximal cut
method in a constructive way, assigning new contributions to new
contact diagrams as one proceeds. (For recent examples, see
Refs.~\cite{Nonplanar5PtN4, GeneralizedDoubleCopy}.)  In subsequent
sections we will describe how to make this procedure efficient for
gravity theories at high loop orders, by recycling gauge-theory 
results.

%%%%%%%%%%%%%% FIGURE  %%%%%%%%%%%%%%
\begin{figure}[t]
  \includegraphics[clip,scale=0.54]{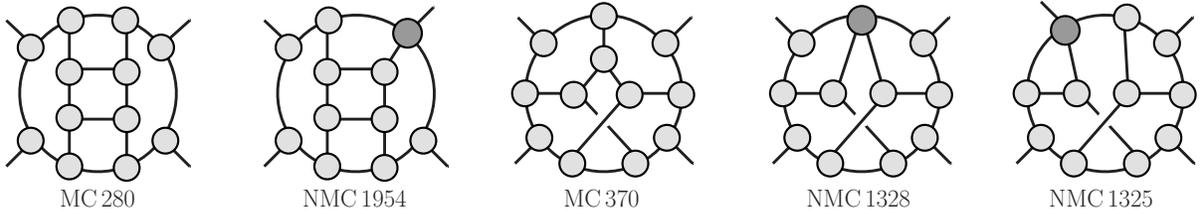}
\vskip -.2 cm 
\caption[a]{Sample maximal and next-to-maximal cuts that are
  determined by the naive double copy.  The exposed lines connecting
  the blobs are on shell.  The labels refer to those used in the {\em
    Mathematica} attachment~\cite{AttachedFile}.  }
\label{MCandNMCFigure}
\end{figure}
%%%%%%%%%%%%%%%%%%%%%%%%%%%%%%%%%%%%%

%%%%%%%%%%%%%% FIGURE  %%%%%%%%%%%%%%
\begin{figure}[t]
  \includegraphics[clip,scale=0.63]{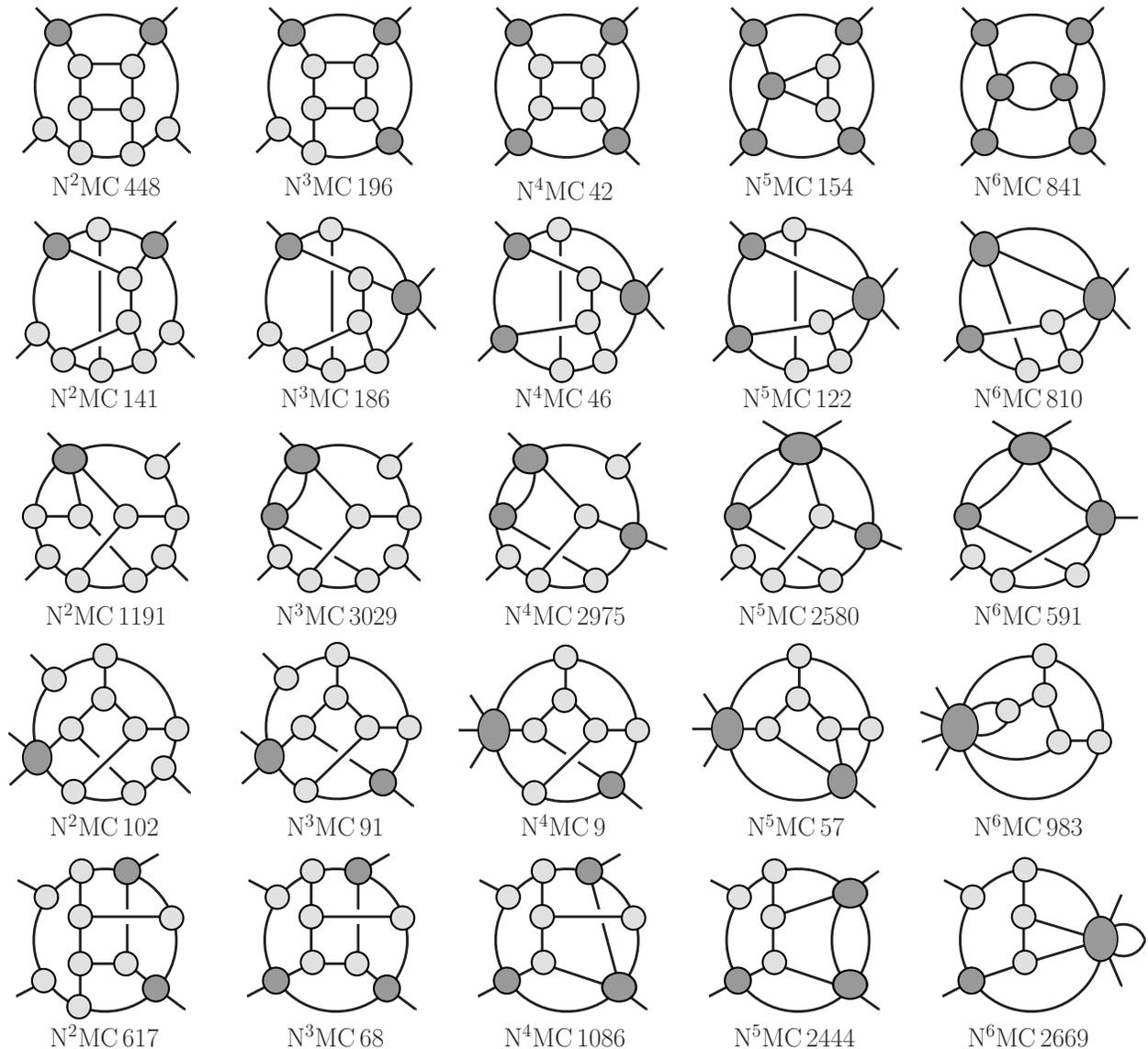}
\vskip -.2 cm 
\caption[a]{Sample \N{k}s for a five-loop four-point amplitude.  The
  exposed lines connecting the blobs are on shell.  The labels refer
  to those used in the {\em Mathematica} attachment~\cite{AttachedFile}.
}
\label{NkMaxCutsFigure}
\end{figure}
%%%%%%%%%%%%%%%%%%%%%%%%%%%%%%%%%%%%%

The method of maximal cuts~\cite{MaximalCutMethod} constructs
multiloop integrands from generalized unitarity cuts.  These cuts
cluster in levels according to the number of internal propagators $k$ allowed to remain off shell,
\begin{align}
{\cal C}^\NkMC &= \sum_{\rm states} {\cal A}_{m(1)}^{\rm tree} \cdots {\cal A}_{m(p)}^{\rm tree}\,,
&    k \equiv \sum^p_{i=1}m(i)-3p \,,
\label{GeneralizedCut}
\end{align}
where the ${\cal A}_{m(i)}^{\rm tree}$ are tree-level $m(i)$-multiplicity amplitudes corresponding to
the blobs,  illustrated in \figs{MCandNMCFigure}{NkMaxCutsFigure}.  
This is valid for either gauge or gravity amplitudes.  
In the gauge-theory case, the state sum also includes sums over internal color.
As illustrated in the first diagram in \fig{MCandNMCFigure}, at the maximal cut (MC)
level the maximum number of propagators are replaced by on shell
conditions and all tree amplitudes appearing in \eqn{GeneralizedCut}
are three-point amplitudes.  At the next-to-maximal-cut (NMC) level a
single propagator is placed off shell and so forth.  We will categorize 
different cuts at level $k$ by the contained tree amplitudes with four
or more legs: an $m_1\times m_2 \times \cdots \times m_q$ cut
contains one tree amplitude with $m_1$ legs, one with $m_2$ legs and so forth.

In the method of maximal cuts, the integrands for $L$-loop
amplitudes are obtained by first establishing an integrand whose maximal
cuts are correct, then adding to it terms so that NMCs
are all correct and systematically proceeding through the next$^k$ maximal cuts (\N{k}s),
until no further contributions can be found.  Where this happens is dictated
by the power counting of the theory and by choices made at each level.
For example, if a minimal power counting is assigned to each contribution, for
$\NeqFour$ super-Yang--Mills four-point amplitudes, cuts through NMCs, \N2s and \N3s are
sufficient at three~\cite{BCJLoop},
four~\cite{SimplifyingBCJ} and five loops~\cite{FiveLoopN4}, respectively.  

%%%%%%%%%%%%%% FIGURE  %%%%%%%%%%%%%%%%
\begin{figure}[t]
\includegraphics[clip,scale=0.44]{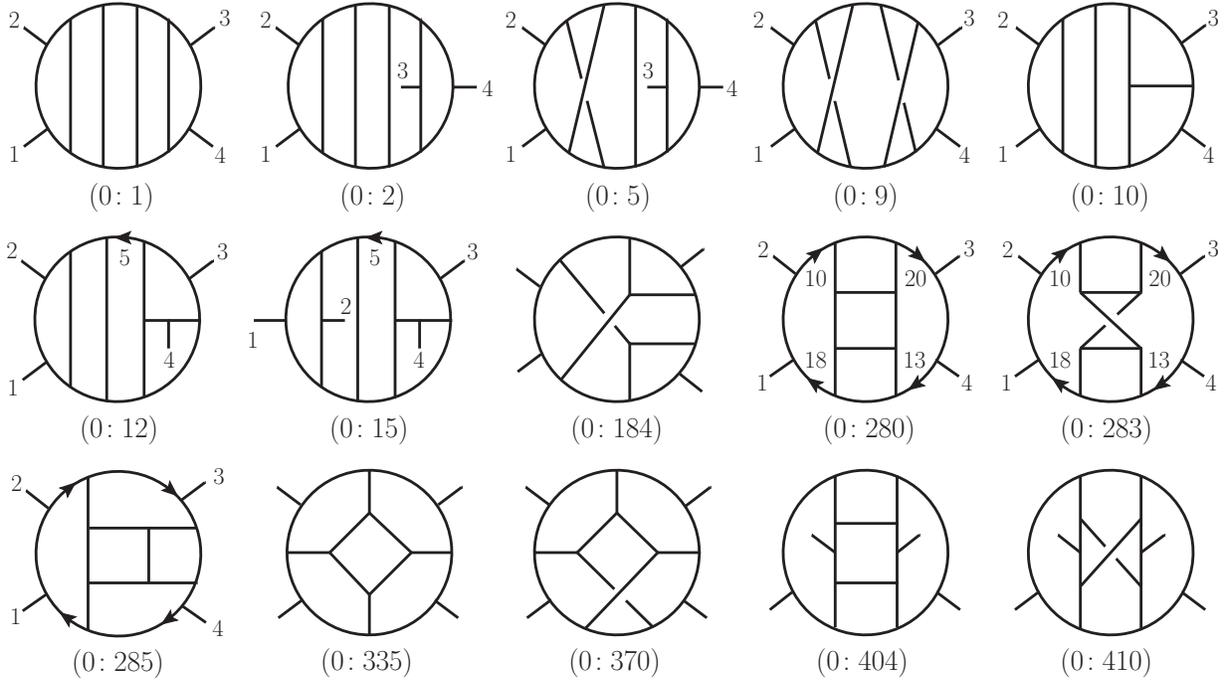}
\caption[a]{Examples of parent diagrams used in the naive double copy.
  These are diagrams with only cubic vertices and 16 propagators
  carrying loop momentum. In our construction of the five-loop
  four-point amplitude of $\NeqEight$ supergravity there are a total
  of 410 such nonvanishing diagrams. The labeling $(0\! : j)$ indicates
  that it is a level 0 diagram with no collapsed propagators and $j$
  is the diagram number, following the labels in the {\em
    Mathematica} attachment.  }
\label{FiveLoopExampleGraphsFigure}
\end{figure}
%%%%%%%%%%%%%%%%%%%%%%%%%%%%%%%%%%%%%%%%%%

Most previous calculations~ (see e.g., Refs.~\cite{ThreeFourloopN8,
  SimplifyingBCJ, N4GravFourLoop, N4GravThreeLoops, N4SugraMatter,
  N5GravFourLoops,BCJDifficulty,MultiLoopFormFactor}) found it
convenient to organize the results in terms of purely cubic diagrams,
assigning all higher-order missing \N{k} data to the parent graphs
with only cubic vertices, such as the five-loop ones illustrated in
\fig{FiveLoopExampleGraphsFigure}.  
Representations with only cubic diagrams have useful
advantages: they are useful for establishing minimal 
power counting in each diagram, and the number of graphs used to describe the result
proliferate minimally with loop order and
multiplicity\protect\footnote{Though still factorially.}.  A disadvantage is
that Ans\"atze are required to impose the higher-order data on each
graph while respecting power counting, symmetry, and the multiple
unitarity cuts to which a given diagram contributes.  As the loop order
increases, it becomes cumbersome to solve the requisite system of equations
that imposes these constraints.

\subsection{Naive double copy and contact diagram corrections}

For our purposes of constructing the five-loop four-point integrand it is
better to directly assign new cut data to contact graphs in
one-to-one correspondence to the \N{k}, as in the original
method of maximal cuts construction~\cite{MaximalCutMethod}, avoiding
Ans\"atze for the amplitudes.  We now describe this organizational
principle in the context of obtaining high-loop-order gravity integrands.

The starting point in our gravity construction is a gauge-theory
integrand, whose terms are assigned to only graphs with cubic
vertices.  The actual gauge-theory amplitude would be given:
\begin{equation}
{\cal A}_m^{L\text{-loop}} = i^L g^{2L+m-2} 
\sum_{{\cal S}_m} \sum_{i \in \Gamma_{3,m,L}} 
\int \prod_{j}^L \frac{d^D l_j}{(2 \pi)^D} \frac{1}{S_i}
 \frac{c_i n_i} {D_i}  \,,
\label{genericGaugeLoopIntegral}
\end{equation}
where the first sum runs over the set ${\cal S}_m$ of external leg
permutations. The second sum runs over the set of diagrams $
\Gamma_{3,m,L}$ with only three vertices, $m$ external points and $L$
loops.  The symmetry factors $S_i$ for each diagram $i$ remove
overcounts, including those arising from internal automorphism
symmetries with external legs fixed.  As in
\sect{subsec:TreeLevelReview}, the color factors $c_i$ of all graphs
are obtained by dressing every three-vertex in the graph with a factor
of $\tilde f^{abc} = \Tr([T^a, T^b] T^c)$, where the gauge group
generators $T^a$ are normalized via $\Tr(T^a T^b) = \delta^{ab}$.  As
before, the gauge coupling is $g$.  The kinematic numerators, $n_i$,
are functions of momenta, spinors, and polarization vectors. As usual,
the $1/D_i$ signify the product of Feynman propagators of diagram $i$.

Our construction starts with a naive double copy, which 
we call the `level 0' or 'top-level' contribution,
\begin{equation}
{\cal M}_m^{L\text{-loop}} = i^{L+1} \left( \frac{\kappa}{2}\right)^{2L+m-2} 
\sum_{{\cal S}_m} \sum_{i \in \Gamma_{3,m,L}}
\int \prod_{j}^L \frac{d^D l_j}{(2 \pi)^D} 
\frac{1}{S^{(0)}_i} \frac{N^{(0)}_i} {D_i}  \,,
\label{NDCLoopIntegral}
\end{equation}
where the level 0 numerators are just double copies of gauge-theory
numerators,
\begin{equation}
N^{(0)}_i = n_i \tilde n_i \,.
\label{NumeratorSquareGeneric}
\end{equation}
If the gauge-theory $n_i$ satisfy the BCJ relations
(\ref{BCJDuality}), then we have the complete gravity integrand and we
would be done~\cite{BCJLoop}.  However, when the gauge-theory
integrand (\ref{genericGaugeLoopIntegral}) does not manifest BCJ
duality, our naive double copy requires corrections to become a
gravity integrand, as we can systematically determine by evaluating
generalized cuts.

First we should note that all maximal cuts (MCs) and all next to
maximal cuts (NMCs) will be automatically satisfied by our naive
double copy. The reason is that on-shell ($D$-dimensional)
supergravity three-point amplitude is just the square of the
$\NeqFour$ super-Yang-Mills ones,
\begin{equation}
{\cal M}_3^{\NeqEight\,\tree}(1,2,3) = i \frac{\kappa}{2} 
\Bigl[A_3^{\NeqFour\,\tree}(1,2,3) \Bigr]^2\,,
\end{equation}
for all states of the theory.
All NMCs are also automatically satisfied because color-kinematics
duality automatically holds for the four-point tree
amplitudes~\cite{BCJ}.  Examples of MCs and NMCs are  given in
\fig{MCandNMCFigure}.

\begin{figure}[t]
\includegraphics[clip,scale=0.55]{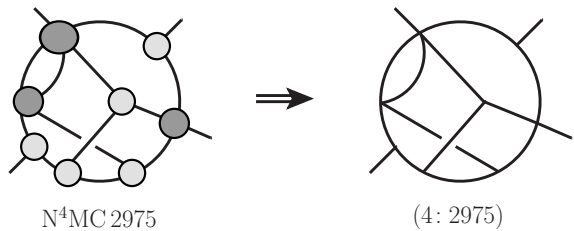}
\caption[a]{After subtracting contributions from lower cut
  levels, as in \eqn{ContactEquation}, only a local contact term remains. }
\label{CutToContactsFigure}
\end{figure}

\begin{figure}[t]
\includegraphics[clip,scale=0.63]{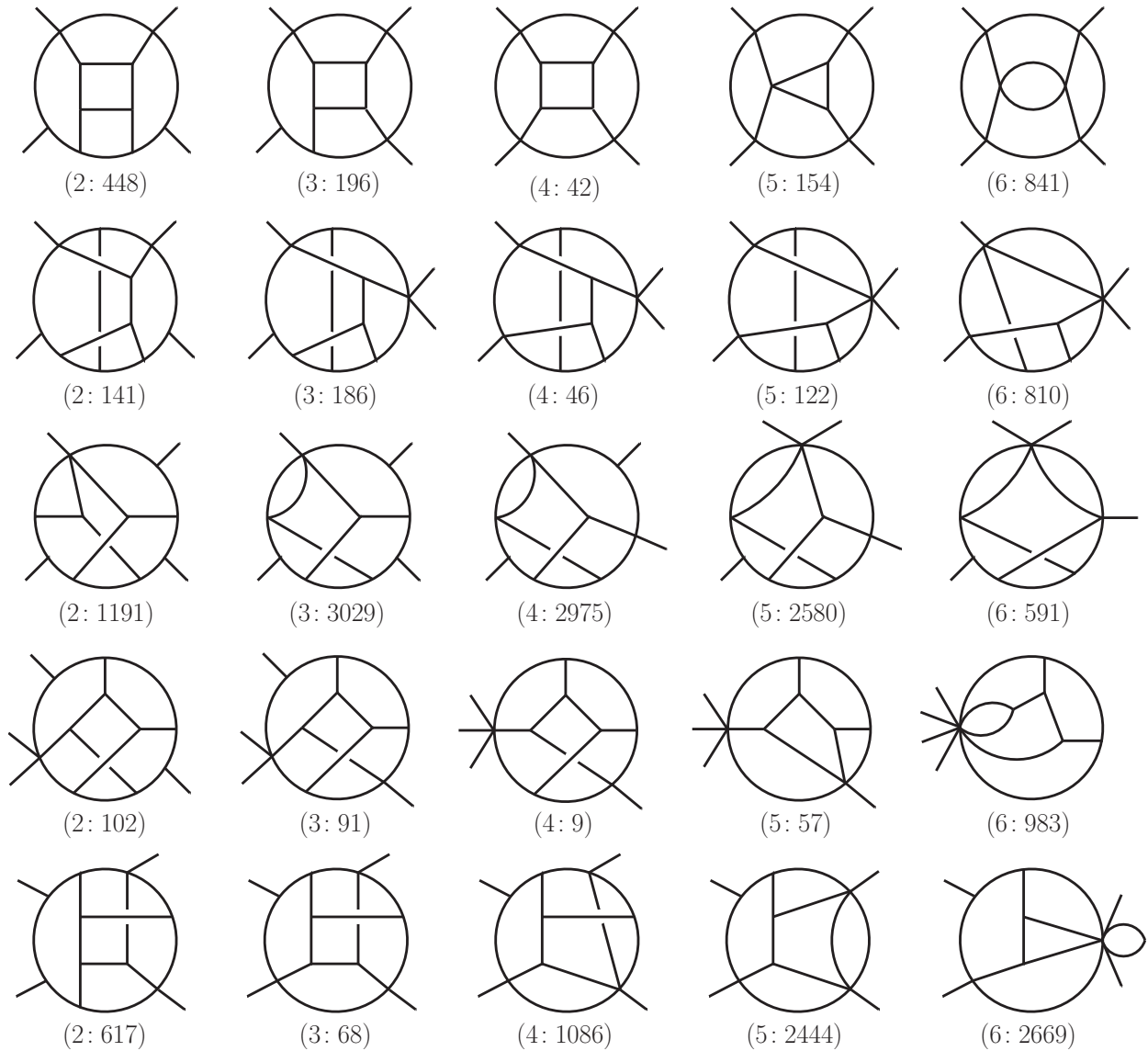}
\caption[a]{Contact diagrams corresponding to each N$^k$-maximal cut in
  \fig{NkMaxCutsFigure} cuts for $k=2,\ldots,6$.  The exposed lines are 
   off shell in this figure. 
   }
\label{NkMaxContactsFigure}
\end{figure}

Starting with the N$^2$MCs, the cuts of the naive-double copy
no longer generically match the actual cuts of the double-copy
gravity theory.  Because the naive double copy automatically gives the
correct MCs and NMCs, the correction terms are necessarily contact
terms involving two or more collapsed propagators.  The cut conditions
are then solved starting from the \N2s and proceeding towards the higher $k$
\N{k}s. At each new cut level the only new information is
captured by contact terms as illustrated in \fig{CutToContactsFigure}. 
\Fig{NkMaxContactsFigure} displays the contact diagrams
representing the new information contained in the generalized
cuts of \fig{NkMaxCutsFigure}.

%%%%%%%%%%%%%% FIGURE  %%%%%%%%%%%%%%%%
%%%%%%%%%%%%%%%%%%%%%%%%%%%%%%%%%%%%%%%

%%%%%%%%%%%%%% FIGURE  %%%%%%%%%%%%%%%%
%%%%%%%%%%%%%%%%%%%%%%%%%%%%%%%%%%%%%%%%%

The contact terms are defined as differences between a cut of the
complete gravity amplitude and the cut of our partially-constructed
gravity amplitude.  The gravity generalized cuts can in principle all
be obtained by plugging gravity tree amplitudes obtained from the KLT
tree relations into \eqn{KLTmpoint} into the generalize cut
\eqref{GeneralizedCut}, although this is rather inefficient.  We also
define an incomplete integrand ${\cal I}^k$ given by starting from the
naive double copy and including all contact terms through level
$(k-1)$. At any level $k$ we define the incomplete integrand to be the
sum over all diagrams from level zero to level $(k-1)$,
\begin{equation}
{\cal I}^{(k)} = \sum_{\ell =0}^{k-1} \sum_{{\cal S}_m} \sum_{i_\ell} 
   \frac{1}{S^{(\ell)}_{i_\ell}} \frac{N^{(\ell)}_{i_\ell}} {D^{(\ell)}_{i_{\ell}}} \,,
\label{GravityLoopIncompleteIntegrand}
\end{equation}
where the sum over $\ell$ is over the contact term levels up to level
$k-1$ and the sum over $i_\ell$ is over diagrams at level $\ell$.
The $N^{(\ell)}_i$, $S^{(\ell)}_i$ and $D^{(\ell)}_i$ are respectively the
numerators, symmetry factors and kinematic denominators for diagram
$i$ at level $k$.  As usual the sum over ${{\cal S}_m}$ represents the
sum over the $m!$ permutations of external legs. The kinematic
denominators are composed of products of Feynman propagators for each
diagram.

Starting from the gravity cut, ${\cal C}^\NkMC$, and subtracting from
it the cut of the incomplete integrand
\eqref{GravityLoopIncompleteIntegrand}, gives us the missing piece in
the cut,
\begin{equation}
\c^\NkMC = {\cal C}^\NkMC - {\cal I}^{k} \Bigr|_{\NkMC} \,,
\label{ContactEquation}
\end{equation}
where a \N{k} is taken.  
This difference can be assigned to a contact diagram because all the
nonlocal contributions are accounted for at earlier levels.  In this way
for each cut for $k\ge 2$ there is a contact term diagram, as
illustrated in \fig{CutToContactsFigure}.  See also
\Fig{NkMaxContactsFigure} for examples of contact diagrams that are in
one-to-one correspondence to the generalized unitarity cuts in
\fig{NkMaxCutsFigure}. 

We promote these contact terms to off-shell expressions simply by
removing all on-shell constraints,
\begin{equation}
\c^\NkMC \rightarrow \c^\NkMC \Bigr|_{\rm  off\hbox{-}shell} \,.
\label{OnShellToOffshell}
\end{equation}
This then defines a level-$k$ contact term assigned to a given graph,
as illustrated in \fig{CutToContactsFigure}.  We take the final
contact diagram to be one where no cut conditions are imposed.  Each
non-vanishing contact graph generated this way is then incorporated
into the partially constructed integrand.  The generated contact
diagram is not unique because one can add or subtract terms that
vanish prior to releasing the cut conditions.  An important constraint
is that the constructed contact terms should always respect diagram
symmetry, even after on-shell constraints are removed.  A simple way
to impose the symmetry on an arbitrary off-shell continuation is to
explicitly average over all diagram symmetries.  Different choices of
off-shell continuations can alter higher-level contact terms.  An
important feature of this construction is that each contact term
depends only on choices made at previous lower-$k$ levels.

The construction proceeds level by level in the cuts until no further
contact terms are found.  Where this happens is dictated by the power
counting of the gravity theory.\protect\footnote{For the five-loop
four-point amplitude of $\NeqEight$ supergravity no new 
contact terms are found beyond level $k=6$.}

In general, \eqn{ContactEquation} can be quite complicated to
simplify, especially when the gravity cut is obtained from the KLT
version of generalized cuts.  It however is an efficient means to
generate expressions for numerical evaluation.  Far more efficient ways
to analytically generate these contributions will be described in
Sections~\ref{BCJFormulaOverview}-\ref{HigherLevelJFormulasSeaction}.

The final amplitude is obtained at the end of this process, when we
reach a level $k_{\rm max}$ in the incomplete integrand
(\ref{GravityLoopIncompleteIntegrand}) beyond which there are no
further nonvanishing contributions. After assembling the naive double
copy and contact diagrams the resulting gravity amplitude is obtained
by summing over all nonvanishing levels and integrating,
\begin{equation}
{\cal M}_m^{L\text{-loop}} = i^{L+1} \left( \frac{\kappa}{2}\right)^{2L+m-2} 
 \sum_{\ell}^{k_{\rm max}}  \sum_{{\cal S}_m} \sum_{i} 
\int \prod_{j}^L \frac{d^D l_j}{(2 \pi)^D} \frac{1}{S^{(\ell)}_i} \frac{N^{(\ell)}_i}
   {D^{(\ell)}_i}  \,,
\label{GravityLoopIntegralLevels}
\end{equation}
where $k_{\rm max}$ is the highest level containing nonvanishing
diagrams.

%%%%%%%%%%%%%%%%%%%%%%%%%%%%
\subsection{Double copy and gravity unitarity cuts}

In order to use \eqn{ContactEquation} to obtain the missing contact
diagram, we need efficient means to obtain the gravity cuts.  In this
subsection we explain how gauge-theory generalized cuts can be
converted directly to gravity cuts, without having to go back to
gravity tree amplitudes via the KLT relations \eqref{KLTmpoint}.  Once
these steps have been carried out in the corresponding gauge-theory
amplitudes we simply recycle them into gravity. This bypasses the
nontrivial steps of having to perform state sums~\cite{SuperSums}
ensuring that results are valid in $D$-dimensions~\cite{DDimensions}.

Consider a generalized unitarity cut in \eqn{GeneralizedCut} and
\fig{NkMaxCutsFigure} for gauge theory.  We can express each tree
amplitude in terms of diagrams with only cubic vertices as in
\eqn{CubicRepresentation},
\begin{align}
{\cal C}_{\rm YM} \equiv \sum_{\text{states}} \prod^p_{j=1} {\cal A}^{\tree}_{m(j)}
    & = {\sum_{\text{states}}} 
\prod^p_{j=1}  \smashoperator[r]{\sum_{g(j) \in \Gamma_{3,m (j)}}}
\, \frac{c_{g(j)} \, {n}_{g(j)}}{D_{g(j)}}\,,
\label{YMDiagramsCut}
\end{align}
where $j$ specifies the tree amplitude, $g(j)$ represents a graph of
the $j$th tree amplitude from the set of graphs $\Gamma_{3,m(j)}$,
including the trivial three-vertex for the three-point amplitude.
For simplicity we have suppressed the coupling constants here and in 
all subsequent formulas for generalized cuts.
The denominators $1/D_{g(j)}$ are composed of the Feynman propagators
of the graph $g(j)$. 

By applying the color decomposition in \eqn{PartialRepresentation} to each tree amplitude 
we obtain a color-decomposed form of the unitarity cut,
\begin{align}
{\cal C}_{\rm YM}  & =  {\sum_{\text{states}}}\, 
\prod^p_{j=1} \, \sum_{\rho^{(j)} \in {\cal S}_{m(j)-2}} c(\rho^{(j)}) A_{m(j)}^\tree (\rho^{(j)})\,,
\end{align}
where $\rho^{(j)}$ refers to the arguments in
\eqn{PartialRepresentation}, but for the $j$th tree. The
permutation ${\cal S}_{m(j)-2}$ act on $(m(j)-2)$ of the 
legs of the $j$th tree amplitude. For three-point
trees the permutation sum is trivial.  As before, the internal color
indices are included in the state sum.

Now consider generalized gravity cuts.  A crucial property is
that the states of double-copy theories factorize into the outer
product of states of their constituent single-copy theories.  In
particular, for $\NeqEight$ supergravity in four dimensions, every
gravity state is indexed by `left' and `right' $\NeqFour$ super-Yang--Mills states:
\begin{equation}
(\NeqEight {\text{ SG state}}) = (\NeqFour {\text{ sYM state}})_{\rm L} 
 \otimes  (\NeqFour{\text{ sYM state}})_{\rm R} \,.
\label{StateSumFactorization}
\end{equation}
In fact, the state sum over the entire supergravity multiplet is a
double-sum over the entire super-Yang-Mills multiplet,
\begin{equation}
\sum_{{\NeqEight\ {\text{SG}} \atop \text{states}}} = 
\sum_{{\NeqFour\ {\text{sYM}} \atop {\text{L-states}}}}
  \times \sum_{{\NeqFour\ {\text{sYM}} \atop {\text{R-states}}}} \,.
\end{equation}
This holds in $D\le 10$ dimensions where $\NeqFour$ super-Yang--Mills
theory is defined as a dimensional reduction of the $D=10$, $\NeqOne$
theory.

Each gravity tree amplitude in the cut, such as those in
\fig{NkMaxCutsFigure}, can be arranged into a BCJ double-copy form
\begin{align}
{\cal C}_{\G}  \equiv \sum_{\text{states}} \prod^p_{j=1} M^{\tree}_{m(j)}
 = i^p \sum_{\text{states}_\text{L}} \, \sum_{\text{states}_\text{R}} \,
\prod^p_{j=1}  \smashoperator[r]{\sum_{g(j) \in \Gamma_{3,m(j)}}} \;
\frac{n^{\text{\BCJ}}_{g(j)} \, \tilde{n}^{\text{\BCJ}}_{g(j)}}{D_{g(j)}}\,,
\end{align}
where we have suppressed the gravitational coupling and 
 $n^{\text{\BCJ}}$ and $\tilde{n}^{\text{\BCJ}}$ are kinematic
numerators of the left and right gauge theories.  For each tree
amplitude one can always find BCJ forms for the numerators.  For
example, the explicit BCJ numerators in \eqn{ExplicitBCJFromKLT} for
each tree amplitude immediately give the gravity amplitude starting
from a gauge-theory amplitude.

We can then rearrange the cut into a KLT form, using the tree-level
results from the previous subsection. Given the that BCJ form of the 
numerators have exactly the same algebraic properties as color factors,
we write the cut in precisely the same form as the color decomposed
gauge-theory cut \eqref{ExplicitBCJFromKLT}
\begin{align}
{\cal C}_{\G}  & =  i^p {\sum_{\text{states}}}\, 
\prod^p_{j=1} \, \sum_{\rho^{(j)} \in {\cal S}_{m(j)-2}} \tn^\BCJ(\rho^{(j)}) A_{m(j)}^\tree (\rho^{(j)}) \,,
\end{align}
where the numerator is that of the half-ladder diagram specified in
\fig{fig:HalfLadder}.  In this formula the numerators
$n^\BCJ(\rho^{(j)})$ correspond to the half-ladder diagrams with an
ordering of legs specified by the permutation $\rho^{(j)}$.  Here the tree
subscripts $m(j)$ encode the multiplicity of the $j$-th tree, and $m$ 
and $L$ are the overall multiplicity and loop order of the amplitude.
Plugging in the specific BCJ numerators in \eqn{ExplicitBCJFromKLT}
reduces each permutation sum from acting on $(m(j)-2)$ legs to
$(m(j)-3)$ legs, given the numerator vanishings in \eqn{ExplicitBCJFromKLT}.

Substituting in the explicit expression for BCJ numerators in  \eqn{ExplicitBCJFromKLT}
immediately gives the KLT form of the gravity generalized cut,
\begin{align}
{\cal C}_{\G} &= i^p \sum_{ \text{states} } 
\prod^p_{j=1} \,  \smashoperator[r]{\sum_{\rho^{(j)},\tau^{(j)} \in {\cal S}_{m(j)-3}}} \,  K(\rho^{(j)} | \tau^{(j)} ) \, 
A^{\tree}_{m(j)}(\rho^{(j)})\, \tilde A^{\tree}_{m(j)}(\tau^{(j)})\nonumber\\
&=i^p {\sum_{\vec{\rho},\vec{\tau} }} 
 K(\vec{\rho}\,  |\vec{ \tau})
 \Biggl( \sum_{ \text{states}_\text{L}} A^{\tree}_{m(1)}(\rho^{(1)})\cdots A^{\tree}_{m(p)}(\rho^{(p)}) \Biggr)
 \Biggl( \sum_{ \text{states}_\text{R}} \tilde A^{\tree}_{m(1)}(\tau^{(1)})\cdots 
    \tilde A^{\tree}_{m(p)}(\tau^{(p)}) \Biggr) \,,
\label{doubleCopyCutKLT}
\end{align}
where we have suppressed overall factors of the $(\kappa/2)$ gravitational coupling
and 
\begin{equation}
 K(\vec{\rho} \, |\vec{ \tau}) \equiv K(\rho^{(1)}|\tau^{(1)} ) \cdots  K(\rho^{(p)} | \tau^{(p)} ) \,,
\end{equation}
and we used the factorization of the state sums as in
\eqn{StateSumFactorization}.  For each gauge-theory tree amplitude,
the permutation sum follows that in \eqn{KLT}.  For the three- and
four-point cases the permutation sum is a single term.

\Eqn{doubleCopyCutKLT} allows us construct gravity generalized
unitarity cuts from corresponding gauge-theory tree amplitudes.
However, it is much more efficient to apply \eqn{doubleCopyCutKLT}
directly to cuts of previously constructed gauge-theory loop
amplitudes, rather than using tree amplitudes. That is, we take
\eqn{doubleCopyCutKLT} as a recipe for assembling color-ordered
gauge-theory cuts into gravity cuts.  In this way the states sums, and
other simplifications are automatically inherited from the
gauge-theory loop integrands.  Another enormous technical advantage is
that we need the cuts and the constructed loop integrand to be valid
in $D$ dimensions, not just in four dimensions.  In particular,
explicit checks confirm the validity of the five-loop four-point
amplitude of $\NeqFour$ super-Yang--Mills~\cite{FiveLoopN4} for $D\le
6$~\cite{DDimensions}.  This is then automatically imported into the
corresponding $\NeqEight$ supergravity amplitude.  It is course
crucial to guarantee the validity of the expressions outside of $D=4$
dimensions, given we are interested primarily in its ultraviolet
behavior in higher dimensions. 

Unfortunately, even after applying \eqn{doubleCopyCutKLT} to convert
cuts of gauge-theory loop amplitudes, the analytic expressions
inherited from the KLT construction are rather complicated.  This
makes it difficult to analytically simplify the contact terms in
\eqn{ContactEquation} at high loop orders. However, it does provide a
rather efficient means for numerically evaluating any cut, by first
numerically evaluating the gauge-theory unitarity cuts and then
carrying out the matrix multiplication in \eqn{doubleCopyCutKLT}
numerically.  This will prove very useful in
\sect{FiveLoopResultsSection}, where the five-loop four-point
amplitude on $\NeqEight$ supergravity is constructed. While the
numerical analysis is quite helpful, especially for confirming the
correctness of expressions, the required Ans\"atse are impractical.
Much more efficient means for analytically constructing gravity
contact terms are given in the next sections.

%%%%%%%%%%%%%%%%%%%%%%%%%%%%%%%%%%%%%%%%

\section{Contact terms from BCJ duality}
\label{BCJFormulaOverview}

In the previous section we reviewed a constructive method for building
up a supergravity amplitude starting from a naive double copy of a
corresponding gauge-theory amplitude.  However, it is still nontrivial
to extract the contact terms at high loop orders, given the
analytic complexity of generalized cuts obtained as obtained from
\eqn{doubleCopyCutKLT}.  To deal with this,
Ref.~\cite{GeneralizedDoubleCopy} outlined a method for obtaining
correction terms to the naive double copy directly from corresponding
gauge-theory expressions, without having to construct gravity
unitarity cuts.  This enormously simplifies the task.  Here we elaborate
on the details of this method.

%%%%%%%%%%%%%%%%%%%%%%
\subsection{Overview of gravity cuts from BCJ discrepancy functions}

As noted in the previous section, at high loop orders it can be difficult to find
representations of the amplitudes that manifest BCJ duality.   
Instead, we start from the ``naive double copy'' in \eqn{NDCLoopIntegral},  
obtained by replacing the color factors with numerators that do not satisfy the duality, and correct it 
until it reproduces all the generalized cuts of the gravity amplitude.
The properties of three and four-point gauge-theory amplitudes guarantee that the naive double copy 
has the correct maximal and next-to-maximal cuts. The method of maximal cuts provides a means
to systematically construct the contact terms corresponding to the \N{k} with $k\ge2$.

The building blocks for the corrections terms are BCJ discrepancy
functions, which are defined in terms of the violation of BCJ
duality by a given representation of the gauge-theory amplitude,
\begin{equation}
J = n_i + n_j + n_k\,,
\end{equation}
where graphs $i$, $j$, and $k$ are a Jacobi triplet of graphs, 
as in \eqn{BCJDuality}. As already noted in Ref.~\cite{GeneralizedDoubleCopy}, we
find that the corrections are quadratic in the discrepancy functions
\begin{equation}
\E_\G \sim \sum_{a,b} g_{ab} J_a \J_b \,,
\label{ExtraQuadraticJ}
\end{equation}
where $J_a$ and $\J_a$ are discrepancy functions from the two
gauge-theory copies and $g_{ab}$ are appropriate rational functions of
kinematic invariants.

The bilinear structure of the correction terms in
\eqn{ExtraQuadraticJ} is suggested by the fact that the corrections
should all vanish if BCJ duality were manifest in {\em either} the
first or second copy.  A further heuristic argument for the
bilinearity of ${\cal E}_\G$ in discrepancy functions relies on an
understanding of the structure of the terms that need to be added to
the naive double copy in order to restore linearized diffeomorphism
invariance. Since diffeomorphism invariance of the double-copy theory
is related to the gauge invariance of the two single copies \cite{BCJ,
  Square, FundMatter, BCJGaugeSym, CGJR}, we first explore the latter.
At loop level gauge invariance may require
nontrivial changes of variables;  we avoid this difficulty by restricting the
integrand to its generalized cuts, which are given in terms of tree-level
amplitudes.
To mimic the properties of the naive double copy we suspend enforcing
the color-Jacobi identities. Then, under a gauge transformation of the first gluon,
\begin{equation}
\varepsilon_1^\mu \mapsto \varepsilon_1^\mu +  k_1^\mu\,,
\label{GaugeTransformation}
\end{equation}
the color-dressed cut of a gauge-theory amplitude shifts by, 
\begin{equation}
\delta {\cal A}\big|_{\rm cut} = \sum_{\{i,j,k\}} g_{ijk}(
\widehat \varepsilon_1, \varepsilon_2,\dots, p_1,\dots) (c_i+c_j+c_k) \big|_{\rm cut} \ ,
\label{ColorGauge}
\end{equation}
where $\big|_{\rm cut}$ denotes that cut conditions are imposed
and the hat means that $\varepsilon_1$ is absent (having been replaced by
$p_1$, per \eqn{GaugeTransformation}).  The sum runs over the
triplets of graphs $i, j, k$ such that, under 
Jacobi relations, 
\begin{equation}
c_i+c_j+c_k = 0 \, .
\end{equation}
The $g_{ijk}$ are rational functions of all momenta and polarization
vectors except that of the first gluon.

In gravity the scattering amplitudes also enjoy an on-shell gauge invariance.  They must be invariant under
\begin{equation}
\varepsilon^{\mu\nu}_1 \mapsto \varepsilon^{\mu\nu}_1  +  k_1^\mu \varepsilon^\nu_1 \,, 
\quad
\text{where}
\quad
\varepsilon_1\cdot k_1 = 0\,, 
\label{diffTransform1}
\end{equation}
and
\begin{equation}
 \varepsilon^{\mu\nu}_1 \mapsto  \varepsilon^{\mu\nu}_1 + k_1^\nu \tilde \varepsilon^\mu_1\,, 
\quad
\text{where}
\quad
\tilde \varepsilon_1 \cdot k_1 = 0\,,
\label{diffTransform2}
\end{equation}
which capture both linearized diffeomorphism and the gauge symmetry
of the antisymmetric tensor field.
If we start from the BCJ double-copy construction, and as for the gauge-theory 
case suspend enforcing the Jacobi relations, 
the variation of the double-copy cut under the gauge transformation is then,
\be
\delta {\cal M}^\text{naive} \big|_{\rm cut}&=& 
\sum_{\{i,j,k\}} {g_{ijk}(\,\widehat {\varepsilon}_1, \varepsilon_2,\dots, p_1,\dots)}
({\tilde n}_i+{\tilde n}_j+{\tilde n}_k) \big|_{\rm cut}
\cr
&+&
\sum_{\{i,j,k\}} {{\tilde g}_{ijk}(\, \widehat {\tilde \varepsilon}_1,\tilde\varepsilon_2,\dots, p_1,\dots)}
(n_i+n_j+n_k) \big|_{\rm cut}\ ,
\label{Unwanted}
\ee where cut conditions are imposed as in the gauge-theory case.
Thus, to restore the linearized diffeomorphism invariance we must add
terms whose gauge transformation cancels $\delta {\cal
  M}^\text{naive}\big|_{\rm cut}$ to the naive double copy.  The
variation of a contribution quadratic in the discrepancies $J$, as in
\eqn{ExtraQuadraticJ}, would be of the right form to cancel the
unwanted contributions \eqref{Unwanted}.

%%%%%%%%%%%%%% FIGURE %%%%%%%%%%%
\begin{figure}
\includegraphics[scale=.45]{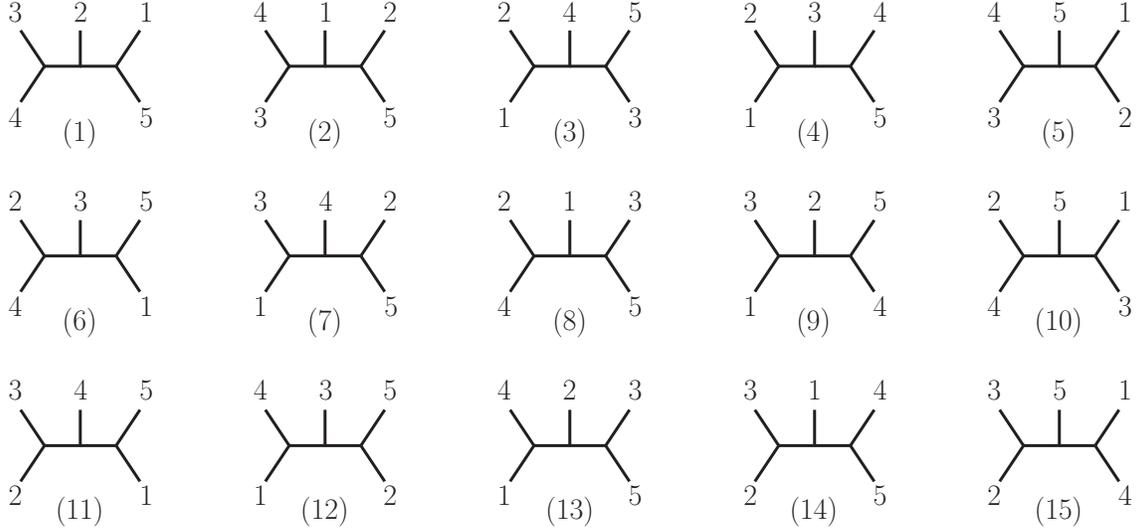}
\caption{The 15 diagrams with cubic vertices for the five-point tree amplitude.}
\label{Tree5Figure}
\end{figure}
%%%%%%%%%%%%%%%%%%%%%%%%%%%%%%%%%

\subsection{Defining BCJ discrepancy functions}

Following Ref.~\cite{GeneralizedDoubleCopy}, we introduce some
notation for tracking different contributions and for tracking
kinematic Jacobi relations.  Consider a cut~\eqref{YMDiagramsCut} of a
gauge-theory amplitude.  We can expand each tree amplitude that
composes the cut in terms of diagrams with only cubic vertices and
then use the labels of each tree diagram to label our numerators,
\begin{equation}
{\cal C}_{\rm YM} = \sum_{i_1,\dots,i_q}
\frac{c_{i_1, i_2, \ldots, i_q} \, n_{i_1, i_2, \ldots, i_q}}  {D_{i_1}\dots D_{i_q}}\,,
\label{CutWithLabels}
\end{equation}
where as usual we drop factors of the coupling and where the $c_{i_1,
  i_2, ...i_q}$ and $n_{i_1, i_2, ...i_q} $ and are the color factors
and kinematic numerators associated with each cut diagram.  Each index
corresponds to a diagram of a tree amplitude contained in the cut with
four or more legs.  Labels for the three-point tree amplitude are not
included since there is only a single fixed vertex for each in a given
cut.  (The three-point amplitudes in the cut also do not play a direct
role in the describing BCJ discrepancy functions.)  The indices follow
an ordering, $1,\dots,q$, of these amplitude factors, and an ordering
of the graphs contributing to each such factor.  For an $m_1 \times
m_2 \cdots \times m_q$ cut, the index $i_v$ runs over the $(2 m_v -
5)!!$ diagrams in the $v$th tree amplitude.  That is, for
four-point tree amplitudes the index $i_m$ runs from 1 to 3, for
five-point tree amplitudes from 1 to 15, for six-point tree amplitudes
from 1 to 105 and so forth.  The $1/D_{i_v}$ are products of Feynman
propagators for graph $i_v$ of the $v$th tree amplitude in the cut.

Generic representations of cut amplitudes do not satisfy the Jacobi relations.
To track the violations of a 
kinematic Jacobi relation on the $\lambda_A$-th propagator of graph
$A$ of $v$-th amplitude factor, we employ a notation similar to that
in \eqn{CutWithLabels}:
\begin{align}
J_{i_1,\dots, i_{v-1}, \{A, \lambda_A\} , i_{v+1},\dots, i_q} = &
s_A \, n_{i_1,\dots, i_{v-1}, A, i_{v+1},\dots, i_q}
+ s_B \, n_{i_1,\dots, i_{v-1}, B, i_{v+1},\dots, i_q} \nn \\
& \hskip 1 cm \null
 + s_C \, n_{i_1,\dots, i_{v-1}, C, i_{v+1},\dots, i_q} \,, 
\label{Jdef}
\end{align}
where graphs $B$ and $C$ are connected to graph $A$ by the color
Jacobi relation on the $\lambda_A$-th propagator of graph $A$.  The
relative signs $s_A$, $s_B$ and $s_C$ between terms are taken to be
those of the corresponding color-Jacobi relation.  As for the
numerators, the indices refer to the diagram number in each amplitude
contributing to the cut.

To simplify the notation whenever the $v$-th amplitude factor is a four-point tree amplitude,
so that graph $A$ has only a single propagator, we simplify the
notation by suppressing the index completely,
because for a four-point
tree amplitude each graph has a single propagator,
 we can always choose
the signs to be all positive, and the Jacobi identity is the same one 
independent of whether we choose diagram $A$, $B$, or $C$:
\begin{equation}
J_{i_1,\dots, i_{v-1}, \x{v} , i_{v+1},\dots, i_q} \equiv
J_{i_1,\dots, i_{v-1}, \{A, \lambda_A\} , i_{v+1},\dots, i_q}\,,
\hskip 1 cm  \hbox{tree $v$ is four point}.
\label{JdefSimp}
\end{equation}

To make the notation systematic, including also relative signs in the
Jacobi relations, we define functions that 
organize the graphs in Jacobi triplets $A, B, C$,  connected by
Jacobi transformations around propagator $\lambda_A$  of diagram $A$:
\begin{equation}
t(A, \lambda_A)=\{A,B,C\} ~~\text{and}~~ 
s(A, \lambda_A) = \{s_A, s_B, s_C\}\,,
\label{tripletsandsigns}
\end{equation}
such that 
\be
s_A\, c_A+ s_B \,c_B+ s_C \, c_C = 0 \, ,
\label{colorJ}
\ee
 where $c_A, c_B$, and $c_C$ are the color factors of diagrams
$A,B$ and $C$.  The triple $\{s_A, s_B$, $s_C\}$ simply gives the
signs in the Jacobi relation. 
 Of course, the overall sign of the function $s$ is arbitrary,
and we will always choose $s_A= 1$.

The BCJ discrepancy functions associated to a (connected) tree-level
graph or to a connected component of a cut are then defined as
\begin{equation}
J_{\{A, \lambda_A\}} = s(A, \lambda_A)_1\, n_A
  + s(A, \lambda_A)_2\, n_B
  + s(A, \lambda_A)_3\, n_C \,,
\end{equation}
where the $s(A, \lambda_A)_1$, $s(A, \lambda_A)_2$ and $s(A,
\lambda_A)_3$ are the three components of the triplet of signs in the
Jacobi relation~\eqref{tripletsandsigns}.  As usual, the momenta in
the numerators are expressed in terms of the momenta common to the
three graphs.  More formally, the discrepancy functions are defined
as
\begin{equation}
{\vec J} = \sigma\cdot {\vec n} \, ,
\label{Jvsn}
\end{equation}
where ${\vec n}$ is the vector of kinematic numerators and the matrix $\sigma$ is defined as
\be
\sigma_{\{j, \lambda_j\}}{}^i = 
\left\{
\begin{array}{ll}
  s_i & \text{if } i =  t(j, \lambda_j)_1  \text{ or } i =  t(j, \lambda_j)_2 
             \text{ or } i =  t(j, \lambda_j)_3 \,,
 \\
0 &\text{otherwise.} 
\\
\end{array}
\right.
\label{sigma}
\ee 
This matrix has $(m_p-3)(2m_p-5)!!$ rows since every
$m_p$-point tree amplitude has $(2m_p-5)!!$ diagrams with only cubic
vertices and each diagram has $(m_i-3)$ propagators.  The number of
columns in the matrix is just the number of diagrams in 

For a cut composed of several tree amplitudes, the analogous matrix is defined as
\begin{equation}
\sigma^{j_1,\dots, j_{p-1}, j_p , j_{p+1},\dots, j_q}
      _{i_1,\dots, i_{p-1}, \{i_p, l_{i_p}\} , i_{p+1},\dots, i_q}
= \delta^{j_1}_{i_1}\dots \delta^{j_{p-1}}_{i_{p-1}} \sigma_{\{i_p, l_{i_p}\}} {}^{j_p} 
  \delta^{j_{p+1}}_{i_{p+1}} \delta^{j_{q}}_{i_{q}} \,,
\end{equation}
where the index $p$ runs from 1 to $q$, i.e. over all tree amplitudes in the cut.

\subsection{Contact terms and properties of generalized gauge transformations}

For any field theory, like the maximally supersymmetric gauge theory,
for which BCJ representations are known to exist for all tree
amplitudes, any generalized cut that decomposes a loop integrand into
a sum of products of tree amplitudes can be written as
\begin{equation}
{\cal C}_\G = \sum_{i_1,\dots,i_q} 
\frac{n_{i_1, i_2, ...i_q}^\text{BCJ} \tn_{i_1, i_2, ...i_q}^\text{BCJ}}  {D_{i_1}\dots D_{i_q}}\,,
\label{BCJCut}
\end{equation}
where the $n^\BCJ$ and $\tn^\BCJ$ are the BCJ numerators associated
with each of the two copies. The notation for the indices is the same
as in \eqn{CutWithLabels}.  These numerators are related to those of
an arbitrary representation, such as that in \eqn{CutWithLabels}, by a
generalized gauge transformation,
\begin{eqnarray}
n_{i_1, i_2, ...i_q} &=& n_{i_1, i_2, ...i_q}^\text{BCJ} 
  + \Delta_{i_1, i_2, ...i_q} \,.
\label{GeneralizedGaugeTrans}
\end{eqnarray}
The only constraint on the shifts $\Delta$ is that the corresponding
cut of the gauge-theory amplitude is unchanged, that is
\begin{equation}
\sum_{i_1,\dots,i_q} \frac{\Delta_{i_1, i_2, ...i_q} c_{i_1, i_2, ...i_q} }
{D_{i_1}\dots D_{i_q}} = 0 \, .
\label{gt}
\end{equation}
Using this constraint and the properties of the BCJ numerators, it is
not difficult to see that the cut ${\cal C}_\G$ of the
gravity amplitude can be written as
\begin{equation}
 {\cal C}_\G = 
  \sum_{i_1,\dots,i_q} \frac{n_{i_1, i_2, ...i_q} \tn_{i_1, i_2, ...i_q} }
  {D_{i_1}\dots D_{i_q}} +{\cal E}_\G\, .
\label{CutNDC}
\end{equation}
Indeed, the first term is, clearly, the corresponding cut of the naive
double copy while the extra contribution ${\cal E}_\G$ is
\begin{equation}
{\E}_\G =  - \sum_{i_1,\dots,i_q} \frac{\Delta_{i_1, i_2, ...i_q}
  \tDelta_{i_1, i_2, ...i_q}}
  {D_{i_1}\dots D_{i_q}} \,,
\label{Extra}
\end{equation}
where the $\Delta$ and $\tDelta$ are the shifts associated with each
of the two copies.  The cross terms $(n^\text{BCJ} \tDelta)$ and
$(\tn^\text{BCJ} \Delta)$ which appear when plugging
Eq.~\eqref{GeneralizedGaugeTrans} in Eq.~\eqref{BCJCut} cancel because
$n^\text{BCJ}$ and $\tn^\text{BCJ}$ have the same algebraic properties
as the corresponding color factors.

%%%%%%%%%%%%%%%%%%

While \eqn{Extra} gives the extra contribution which transforms the
cut of the naive double copy into the cut of a gravity amplitude, it
is not in a particularly practical form because of the nontriviality
of determining the generalized-gauge-transformation parameters. The
essential step for efficiently determining these missing pieces is
expressing \eqn{Extra} in terms of the BCJ discrepancy functions $J$
and $\tJ$, as suggested in \eqn{ExtraQuadraticJ}.

The relation between $\vec J$ and $\vec \Delta$ follows by multiplying
Eq.~\eqref{GeneralizedGaugeTrans} by the matrix $\sigma$ defined in
Eq.~\eqref{sigma},
\begin{equation}
{\vec J} = \sigma\cdot {\vec \Delta} \, ,
\label{JofDelta}
\end{equation}
where ${\vec \Delta}$ is the vector of shifts (analogous to the vector
of kinematic numerators). We also use the defining property of BCJ
numerators, $\sigma\cdot {\vec n}^\text{BCJ}=0$.
What makes inverting this equation difficult is that both the
$\Delta$s and $J$s satisfy nontrivial constraints.  While the solution
to the constraint equation for $\Delta$s, Eq.~\eqref{gt}, is generally
unenlightening, we can derive relatively simple formulas for the extra
pieces in terms of an over-complete set of
$J$s~\cite{GeneralizedDoubleCopy}.  When expressed in terms of the
independent discrepancy functions ${\cal E}_\G$ can appear without a
clear pattern simply because, by applying the constraint equations, we
can easily take an expression with a simple structure and complicate
it.  In this and the next subsections we describe the general
structure; in the next section we give specific case by case solutions
that reveal simple patterns.  Since the constraints on the $J$'s
follow, in part, from the constraints on
generalized-gauge-transformation parameters, we begin by discussing
the latter and postpone the former for the next subsection.

For a cut with a single tree-level amplitude with four or more
external legs (i.e. for $q=1$) a solution to Eq.~\eqref{gt} is that
\be
\Delta_{A} = \sum_{\lambda_A \in {\cal D}(A)} d_A^{(\lambda_A)} \alpha_{\{A, \lambda_A \}} \, ,
\label{sol1}
\ee
where $\lambda_A$ is an element in the set of labels ${\cal D}(A)$
for the propagators of diagram $A$.  The factor $d_A^{(\lambda_A)}$ is
the inverse propagator corresponding to this label.  The parameters
$\alpha_{\{A, \lambda_A \}}$ satisfy further constraints,
\begin{equation}
s(A, \lambda_A)_1 \, \alpha_{\{A, \lambda_A\}} = s(A, \lambda_A)_2 \, \alpha_{\{B, \lambda_B\}} 
=  s(A, \lambda_A)_3 \, \alpha_{\{C, \lambda_C\}} \, ,
\label{remainingGT}
\end{equation}
where graphs $\{A,B,C\}$ and graph propagators $\{\lambda_A,\lambda_B,
\lambda_C\}$ form the Jacobi triplet.  While other solutions may
exist, the one described above has the advantage of being natural for
maintaining the locality of kinematic numerator
factors and making easier to solve \eqn{gt}.\protect\footnote{It is worth mentioning that the developments
  described here and elaborated on in later sections do not rely on
  manifest locality of numerator factors.}  Eqs.~\eqref{sol1} and
\eqref{remainingGT} ensure that, when $\Delta$ is plugged into
Eq.~\eqref{gt} for $q=1$, its vanishing is an immediate consequence of
the color-Jacobi relations \eqref{colorJ}.\protect\footnote{We note
  that for a four-point amplitude the index on $ \alpha$ on the
  right-hand side is superfluous; in this case all signs can be chosen
  to be positive and Eq.~\eqref{remainingGT} implies that the three
  functions are all equal.  }

The solution to Eq.~\eqref{gt} for the case of multiple tree
amplitudes each with four or more legs ($q>1$) is similar: one simply
repeats the construction above for each of the tree-level amplitude
factors.  
\begin{equation} \Delta_{i_1, i_2, ...i_q} = \sum_{v=1}^q \sum_{\lambda_v \in  D(i_v)} 
   d_v^{(i_v, \lambda_v)}  \alpha_{i_1,\dots, i_{v-1}, \{{i_v}, \lambda_v\} ,i_{v+1}, \dots   i_q} \, ,
 \label{solGT}
\end{equation}
where $ d_v^{(i_v,\lambda_v)} $ is the $\lambda_v$th inverse propagator of the $i_v$th diagram of 
the $v$th amplitude.
The remaining generalized gauge invariance constraints relate, as before, the parameters corresponding to triplets of graphs 
connected by Jacobi relations. If graphs $A,B,C$ belong to the $v$-th blob then
\begin{align}
  s(A, \lambda_A)_1 \, \alpha_{i_1,\dots, i_{v-1}, \{A, \lambda_a\}, i_{v+1}, \dots i_k} 
= & s(A, \lambda_A)_2 \, \alpha_{i_1,\dots, i_{v-1}, \{B, \lambda_B\},  i_{v+1}, \dots i_k} \nn \\
= s(A, \lambda_A)_3 \, \alpha_{i_1,\dots, i_{v-1}, \{C, \lambda_C\}, i_{v+1}, \dots i_k}  \,.
\label{remainingGTfull}
\end{align}

For later convenience it is useful to rewrite \eqn{solGT} evaluated on the solution to Eqs.~\eqref{remainingGTfull} in matrix form,
\be
\vec \Delta = \zeta\cdot {\vec \alpha}^\text{independent} \, ,
\label{independentalpha}
\ee
where ${\vec \alpha}^\text{independent}$ is the vector of independent
functions parametrizing the solution to Eqs.~\eqref{remainingGTfull}
and $\zeta$ is a (rectangular) matrix whose nonzero entries are (sums
of) inverse propagators.

\subsection{Constraints and properties of BCJ discrepancy functions}

As already mentioned in the previous subsection, the BCJ discrepancy
functions possess certain properties stemming from their presentation
in terms of kinematic numerators as well as from their relation to the
parameters of the generalized gauge transformations relating the
initial (generic) numerators to the BCJ numerators. We describe them
here in some detail and outline the steps for inverting
Eq.~\eqref{JofDelta} and constructing the extra contributions ${\cal
  E}_\G$ in \eqn{CutNDC} in terms of BCJ discrepancy functions.

Relations between the discrepancy functions arise from the following
sources:
\medskip

 The first source is just a simple overcount arising
from our way of defining the discrepancy functions.
For convenience and symmetry, we define one discrepancy
function for each propagator of each graph. Since 
Jacobi relations group graphs in triplets, the BCJ
discrepancy functions are equal (up to overall irrelevant
signs) in sets of three---corresponding to circular permutations of each
such triplet as in \eqn{tripletsandsigns}.

\medskip

A second source of relations between the $J$ functions is
that they are expressed in terms of kinematic numerators. To see this,
let us consider a cut involving a single $m$-point amplitude (and all
other being three-point amplitudes). There are $(2m-5)!!$ kinematic
numerators that are used to construct $(m-3)(2m-5)!!/3$ BCJ
discrepancy functions\protect\footnote{That is, for each graph and
  each propagator we construct a $J$ and we remove the overcount by
 a factor of 3 described in the previous paragraph. }. For $m>6$ the latter is larger than the former and thus, in this case there
must exist relations between $J$s coming from them being linear
combinations of kinematic numerators. These are analogous in spirit to
the Kleiss--Kuijf relations for tree-level amplitudes~\cite{KK}; in that case the
$(n-1)!/2$ color-ordered partial amplitudes are expressed in terms of
the kinematic dependence of the $(2n-5)!!$ color-dressed graph.  The
generalization to cuts with two or more amplitude factors is
straightforward.

These relations can be formalized in terms of the matrix $\sigma$
introduced above. As stated in Eq.~\eqref{Jvsn}, the vector of BCJ
discrepancy functions are given by
\begin{equation}
{\vec J} = \sigma\cdot {\vec n} \, .
\end{equation}
The matrix $\sigma$,  necessarily has left zero-eigenvectors,
\be
v_0^{(k)}\cdot \sigma = 0 \, ,
\label{ZeroEigenVector}
\ee
where the $v_0^{(k)}$ have numerical entries.
All linear relations with constant coefficients between discrepancy functions are therefore given 
by these eigenvectors,
\begin{equation}
v_0\cdot {\vec J} = 0 \, .
\end{equation}
Among them are, of course, those corresponding to the triple overcount
described above. They correspond to particularly simple zero
eigenvectors, with only two nonvanishing entries.

\medskip

A third source of relations between BCJ discrepancy functions is their
expression in terms of the independent parameters of the generalized
gauge transformation connecting the generic and
color-kinematics-satisfying numerators.  The relations $J(\alpha)$ are
obtained by acting on Eq.~\eqref{GeneralizedGaugeTrans} with the
matrix $\sigma$; since $\sigma\cdot {\vec n}^\text{BCJ} = \vec 0$ and
further using Eq.~\eqref{independentalpha} we are left with
\be
{\vec J} = \sigma\cdot \zeta\cdot {\vec \alpha}^\text{independent} \, ,
\label{JofAlIndep}
\ee
where as before ${\vec \alpha}^\text{independent}$ is the vector of independent functions specifying the generalized gauge parameters. 
This is closely related to the discussion in Ref.~\cite{HenryConstraints} for the case of tree amplitudes.

Apart from the left zero-eigenvectors of the matrix $\sigma$, the fact
that there are fewer ${\vec \alpha}^\text{independent} $ than
kinematic numerators implies that the matrix $\zeta$ has further left
zero-eigenvectors; for an $n$-point amplitude factor, the entries of
the relevant vectors involve $(n-1)$ propagators. In later sections we
shall see examples of such relations.

To summarize, the strategy to solve Eq.~\eqref{JofDelta} and to
construct ${\cal E}_\G$ is: we first express the generalized gauge
parameters $\Delta$ in terms of the independent ones by solving
\eqref{GeneralizedGaugeTrans}; this leads us to
Eq.~\eqref{JofAlIndep}. We then choose as many independent equations
from Eq.~\eqref{JofAlIndep} as the number of components of
${\vec\alpha}^\text{independent}$, solve them, and apply the solutions
to the remaining equations. If the chosen equations are independent,
the remaining equations are the constraints obeyed
by the BCJ discrepancy functions.
Finally, plugging gauge parameters in Eq.~\eqref{Extra} casts the
extra terms in the form~\eqref{ExtraQuadraticJ} with $g_{ab}$ being
rational functions of momentum invariants.  We may further use the
constraint equations (or their solution) to reorganize the entries of
$g_{ab}$ so that kinematic denominators are in one-to-one
correspondence to the graphs with only cubic vertices that appear in
the cut.

%%%%%%%%%%%%%%%%%%%%%%%%%%%%%%%%%%%%%%%%%%%%

\section{Formulas for level 2 contact terms}
\label{N2LevelJFormulasSection}

In this section we derive formulas for the corrections to the naive
double copy on a case-by-case basis, putting the results into 
symmetric forms.
We  organize the cuts not only by the level but also by number of legs in each tree
amplitude with more than three legs in the cut.  As discussed already in the previous 
section, a cut which is composed of $m_1, m_2, \ldots m_q$-point tree amplitudes
with $m_j \ge 4$ will be referred to as an $m_1\times m_2\times\cdots\times
m_q$ cut.  

As discussed in the previous section, the naive double copy reproduces the maximal and 
next-to-maximal cuts of the corresponding gravity amplitude. Thus, the first correction 
term ${\cal E}_\G$ \eqref{Extra} is at the \N2 level. Moreover, since all double (maximal) 
and single (next-to-maximal) propagator contributions to such cuts are already accounted for 
by the naive double copy, ${\cal E}_\G$ for all \N2s are local and
gives directly a contact term, without further subtractions.

We now discuss separately the two classes of \N2s---those containing two four-point 
tree amplitudes and those containing a single five-point tree amplitude.

\subsection{Two four-point tree amplitudes in cut}

%%%%%%%%%%%%%% FIGURE %%%%%%%
\begin{figure}
\includegraphics[scale=.54]{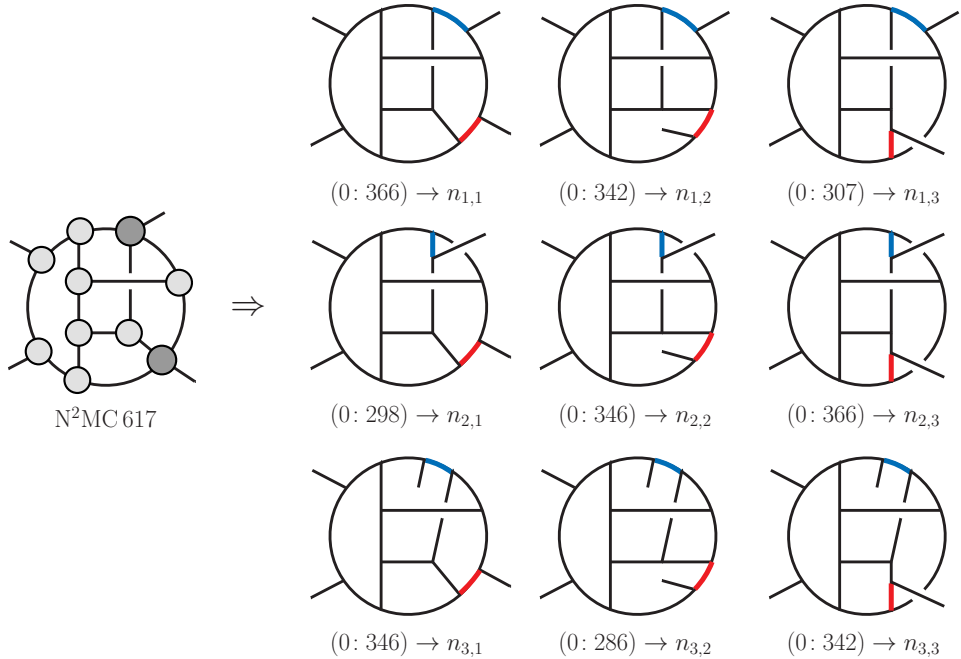}
\caption{Expanding each of the two four-point blob gives a total of
  nine diagrams.  The labels refer to the level and diagram numbers,
  and the $n_{i,j}$ correspond to the cut labels. The shaded thick
  (blue and red) lines are the propagators around which BCJ
  discrepancy functions are defined.  }
\label{FourxFourBlobFigure}
\end{figure}
%%%%%%%%%%%%%%%%%%%%%%%%%%%  

Consider a $4\times4$ cut, for which an example is illustrated in the first cut
on the first line of \fig{NkMaxCutsFigure}.  Each four-point tree
amplitude can be expanded in terms of three four-point diagrams with
only cubic vertices, as illustrated in \fig{FourBlobFigure}.
Expanding both tree amplitudes into such diagrams gives a total of nine
diagrams, as illustrated in \fig{FourxFourBlobFigure} (some of whose
numerators may vanish).  We label the contributing graphs -- and hence
their color and kinematic numerator factors -- by the label of the
off-shell propagators they contain, $c_{i_1, i_2}$ and $n_{i_1, i_2}$.
The first index refers to the 
 diagram in the (arbitrarily-chosen) first
tree amplitude and the second index refers to the (remaining) second
tree amplitude.  Thus, this cut of the gauge-theory amplitude is
written as
\begin{equation}
{\cal C}^{4\times 4}_\text{YM} = \sum_{i_1, i_2}^3 \frac{n_{i_1, i_2}
  c_{i_1, i_2}}{d_{i_1}^{(1)}d_{i_2}^{(2)}}\,,
\label{C4x4}
\end{equation} 
where $1/d_{i_1}^{(1)}$ is the propagator of diagram $i_1$ of the
first four-point tree amplitude factor and $1/d_{i_2}^{(2)}$ is the
propagator of diagram $i_2$ of the second four-point tree amplitude factor.

As discussed in the previous section, the construction of the
correction ${\cal E}_\G$ to the naive double copy relies on using the
generalized gauge transformation -- i.e. shifts of numerator factors
which preserve tree amplitudes and generalized cuts. For the cut
\eqref{C4x4}, the solution \eqref{solGT} and \eqref{remainingGTfull} to
the constraints on these shifts is
\begin{equation}
\Delta_{i_1,i_2} = n_{i_1, i_2} -  n^\BCJ_{i_1,i_2} 
= d^{(1)}_{i_1} \al^{(1)}_{i_2} + d^{(2)}_{i_2} \al^{(2)}_{i_1}\, .
\label{GeneralizedGT4x4}
\end{equation}
The form of the generalized gauge transformation in
\eqn{GeneralizedGT4x4}, is chosen so as to maintain locality of the 
two numerators $n_{i_1 i_2}$ and $n^\BCJ_{i_1,i_2}$.  With this,
 the color-Jacobi identities
\begin{equation}
\sum_{i_1=1}^3 c_{i_1 i_2} = 0\,,  \hskip 2 cm  
\sum_{i_2=1}^3 c_{i_1 i_2} = 0\,,
\end{equation}
ensure that the cut \eqref{C4x4} is invariant: 
\be
 \sum_{i_1, i_2=1}^3 \frac{\Delta_{i_1, i_2} c_{i_1, i_2}}{d_{i_1}^{(1)}d_{i_2}^{(2)}} = 
    \sum_{i_2=1}^3 \frac{\alpha^{(1)}_{i_2} }{d_{i_2}^{(2)}}  \sum_{i_1=1}^3 c_{i_1, i_2} 
 + \sum_{i_1=1}^3 \frac{\alpha^{(2)}_{i_1} }{d_{i_1}^{(1)}}  \sum_{i_2=1}^3 c_{i_1, i_2} = 0 \, .
\ee

Thus, the term \eqref{Extra} that corrects the \N2 cut of
the naive double copy \eqref{CutNDC} to a gravity cut is
\begin{equation}
\E^{4\times4}_\G  =  - \sum_{i_1,i_2 =1}^3 \frac{\Delta_{i_1, i_2}
  \tDelta_{i_1, i_2}}  {d^{(1)}_{i_1} d^{(2)}_{i_2} }
 = - \sum_{i_1,i_2=1}^3 \frac{d^{(1)}_{i_1} d^{(2)}_{i_2} 
    (\al^{(1)}_{i_2}\,\tal{}^{(2)}_{i_1} 
  + \al^{(2)}_{i_1}\,\tal^{(1)}_{i_2})} {d^{(1)}_{i_1} d^{(2)}_{i_2} } \,,
\label{Extra4x4Start}
\end{equation}
where we also used that the sum of the inverse propagators in each four-point amplitude vanishes.
The propagators cancel leaving
\begin{align}
{\E}^{4\times 4}_\G 
& = - \sum_{i_2=1}^3 \al^{(1)}_{i_2}\, \sum_{i_1=1}^3 \tal^{(2)}_{i_1}
   -  \sum_{i_1=1}^3 \al^{(2)}_{i_1}\, \,\sum_{i_2=1}^3 \tal^{(1)}_{i_2}\, .
\end{align}
To rewrite ${\E}^{4\times 4}_\G $ in terms of BCJ discrepancy functions we must solve the Eqs.~\eqref{JofDelta} for this cut.
They read
\begin{equation}
J_{\x1, i_2} \equiv \sum_{i_1=1}^3 n_{i_1 i_2} =
        d^{(2)}_{i_2}  \sum_{i_1} \al^{(2)}_{i_1}\,,
\hskip 1.5 cm 
J_{i_1, \x2} \equiv \sum_{i_2=1}^3 n_{i_1 i_2} =
        d^{(1)}_{i_1}  \sum_{i_2} \al^{(1)}_{i_2} \,.
\label{LocalJs}
\end{equation}
Similar formulas hold for the $\tJ$. 
We notice here a manifestation of the constraints described in the
previous section: on the one hand the right-hand side depends on only
particular combinations of gauge parameters and on the other hand
existence of solutions to these equations requires that the BCJ
discrepancy functions be related to each other,
\begin{equation}
 \sum_{i_1=1}^3 \al^{(2)}_{i_1} = \frac{J_{\x1, 1}}{d^{(2)}_{1}}= \frac{J_{\x1, 2}}{d^{(2)}_{2}}= \frac{J_{\x1, 3}}{d^{(2)}_{3}}\,,
\hskip 1.5 cm 
\sum_{i_2=1}^3 \al^{(1)}_{i_2} = \frac{J_{1, \x2}} {d^{(1)}_{1}}= \frac{J_{2, \x2}} {d^{(1)}_{2}}= \frac{J_{3, \x2}} {d^{(1)}_{3}}\,.
\label{sol_and_rels}
\end{equation}
We therefore find a simple expression of the extra contribution in terms of discrepancy functions,
\begin{equation}
{\E}^{4\times 4}_\G = 
- \frac{1}{d^{(1)}_{1} d^{(2)}_{1}} 
     \Bigr(J_{\x1, 1}\tJ_{1, \x2}
        + J_{1,\x2} \tJ_{\x1,1} \Bigr) \,.
\label{Extra4x4}
\end{equation}
The relations \eqref{sol_and_rels} between the discrepancy functions allow us to write 
a more symmetric version of the extra contribution by averaging over the all three choices
for each of the two sums of gauge parameters:
\begin{equation}
{\E}^{4\times 4}_{\G} =
- \frac{1}{9} \sum_{i_1,i_2 =1}^3 \frac{1}{d^{(1)}_{i_1} d^{(2)}_{i_2}} 
  \Bigl(J_{\x1, i_2} \tJ_{i_1, \x2} 
     +J_{i_1, \x2} \tJ_{\x1, i_2} \Bigr) \,.
\label{Extra4x4_symmetric}     
\end{equation}
These expressions for the extra contributions are actually local 
because $J$ and $\tJ$ are proportional to inverse propagators, 
as indicated in \eqn{LocalJs}, canceling the propagators.

%%%%%%%%%%%%%%%%%%%%
\subsection{One five-point tree amplitude in cut}
\label{sec:5ptcontact}

The second class of \N2s contains one five-point tree amplitude
\begin{equation}
{\cal C}^{5}_\text{YM} = 
 \sum_{i=1}^{15} \frac{n_{i} c_{i}}{d_{i}^{(1)}d_{i}^{(2)}} \,.
\end{equation}
The sum runs over the 15 five-point tree-level graphs with only cubic
vertices, illustrated in \fig{Tree5Figure}, that build the five-point
tree-level amplitude.  Here $d_{i}^{(j)}$ signifies the $j$th inverse
propagator of the $i$th graph.  (More generally we will include an
extra upper index on the inverse propagators to specify which tree
amplitude it belongs to, but here we suppress it because there is only
a one five-point tree amplitude in the cut.)  Unlike the case of the two
four-point tree insertions, the two propagators are now
correlated. We use the labeling of diagrams in \fig{Tree5Figure},
corresponding to the pairs of inverse propagators,
\begin{align}
&
\{s_{34}, s_{15}\},
\{s_{34}, s_{25}\},
\{s_{12}, s_{35}\},
\{s_{12}, s_{45}\},
\{s_{34}, s_{12}\},
\{s_{24}, s_{15}\},
\{s_{13}, s_{25}\},
\{s_{24}, s_{35}\},
\nonumber \\
&
\{s_{13}, s_{45}\},
\{s_{24}, s_{13}\},
\{s_{23}, s_{15}\},
\{s_{14}, s_{25}\},
\{s_{14}, s_{35}\},
\{s_{23}, s_{45}\},
\{s_{23}, s_{14}\}\, ,
\label{props_5pt}
\end{align}
where $s_{ij} \equiv (k_i + k_j)^2$.  In each pair, we refer to the first entry as the `first propagator' 
and the second entry as the `second propagator'; that is, $d_{i}^{(1)}$ is the first entry of
the $i$th pair and $d_{i}^{(2)}$ is the second entry of the $i$th pair.

The gauge transformation \eqref{sol1} connecting the color-kinematics-satisfying
numerators to some arbitrary ones is
\begin{equation}
\Delta_i = n_{i} - n^\BCJ_{i} = d^{(1)}_{i} \al^{(1)}_{i}
  +  d^{(2)}_{i} \al^{(2)}_{i} \, .
\end{equation}
As in the general case discussed in the previous section, the inverse propagators 
allow the gauge-theory amplitude to be invariant under generalized gauge transformations through 
the appearance of the color-Jacobi relations while also maintaining the  locality of numerator factors.
The functions $\al^{(1)}_{i}$ and $\al^{(2)}_{i}$ are not independent;
rather, they are linearly related to each other by \eqn{remainingGT}
so that the amplitude is invariant under the generalized gauge
transformations once the color-Jacobi relations are solved.

%%%%%%%%%%%%%%%  TABLE %%%%%%%%%%%
\begin{table}[tb]
\begin{tabular}{c||c||c}
diagram & \text{ 1st propagator Jacobi triplet }  & \text{ 2nd propagator Jacobi triplet } \cr
 \hline
 \hline
 1 & $\{1, 6, 11\}$, $\{1, 1, 1\}$  & $\{1, 5, 2\}$, $\{1, 1, 1\}$ \cr
 2 & $\{2, 7, 12\}$, $\{1, 1, 1\}$  & $\{2, 1, 5\}$, $\{1, 1, 1\}$\cr
 3 & $\{3, 8, 13\}$, $\{1, 1, 1\}$  & $\{3, 5,    4\}$, $\{1, -1, 1\}$\cr
 4 & $\{4, 9, 14\}$, $\{1, 1, 1\}$  & $\{4, 3, 5\}$, $\{1,    1, -1\}$\cr
 5 & $\{5, 3, 4\}$, $\{1, -1, -1\}$ & $\{5, 2, 1\}$, $\{1, 1, 1\}$\cr
 6 & $\{6, 11, 1\}$, $\{1, 1, 1\}$  & $\{6, 10, 8\}$, $\{1, 1, 1\}$ \cr
 7 & $\{7, 12, 2\}$, $\{1, 1, 1\}$  & $\{7, 9, 10\}$, $\{1, 1, -1\}$ \cr
 8 & $\{8, 13, 3\}$, $\{1, 1, 1\}$  & $\{8, 6,    10\}$, $\{1, 1, 1\}$ \cr
 9 & $\{9, 14, 4\}$, $\{1, 1, 1\}$  & $\{9, 10, 7\}$, $\{1, -1,    1\}$ \cr
10 & $\{10, 9, 7\}$, $\{1, -1, -1\}$& $\{10, 8, 6\}$, $\{1, 1, 1\}$ \cr
11 & $\{11, 1, 6\}$, $\{1, 1, 1\}$  & $\{11, 15, 14\}$, $\{1, 1, 1\}$ \cr
12 & $\{12, 2, 7\}$, $\{1, 1, 1\}$  & $\{12, 15, 13\}$, $\{1, -1, 1\}$ \cr
13 & $\{13, 3, 8\}$, $\{1, 1, 1\}$  & $\{13,    12, 15\}$, $\{1, 1, -1\}$ \cr
14 & $\{14, 4, 9\}$, $\{1, 1, 1\}$  & $\{14, 11, 15\}$, $\{1,    1, 1\}$ \cr
15 & $\{15, 12, 13\}$, $\{1, -1, -1\}$& $\{15, 14, 11\}$, $\{1, 1, 1\}$ \cr
\hline
 \end{tabular}
\caption{Five-point diagrams and associated Jacobi triplets. 
   For each of the two propagators in each diagram, the
  triplet of diagrams participating in the Jacobi identity
   is specified by the first triplet of numbers in each entry.
  The second triplet gives the relative signs in the Jacobi
  relations.}
\label{TripletsAndSignsTable}
\end{table}
%%%%%%%%%%%%%%%%%%%%%%%%%%%%%%%%%%%%

Each graph has two associated Jacobi relations, corresponding to its
two propagators.  \Tab{TripletsAndSignsTable} gives these pairs and
the triplet of signs with which the color or numerator factor enters
the Jacobi relation.  For example, for the graph 15, defined by the
pair of propagators $\{s_{23}, s_{14}\}$ (cf. Eq.~\eqref{props_5pt}),
the two color-Jacobi relations are
\begin{equation}
c_{15}-c_{12}-c_{13}=0
\qquad
c_{15}+c_{14}+c_{11}=0 \, .
\end{equation}
Of the 30 functions $\al^{(1)}_{i}$ and $\al^{(2)}_{i}$, 6 are
determined by the requirement \eqref{gt} that the gauge-theory
amplitude is invariant; thus, there are superficially 24 remaining
generalized gauge functions.

The extra terms \eqref{Extra} completing the cut of the naive double copy to the gravity cut \eqref{CutNDC} are given by 
\begin{equation}
{\E}^{5}_{\G} =
- \sum_{i} \frac{(d^{(1)}_{i} \al^{(1)}_{i}  + d^{(2)}_{i} \al^{(2)}_{i})
      (d^{(1)}_{i} \tal^{(1)}_{i}+d^{(2)}_{i} \tal^{(2)}_{i})}
      {d^{(1)}_i d^{(2)}_i} \,.
\label{Extra5Start}
\end{equation}

As for the previous case, the task is to convert \eqn{Extra5Start} so
that instead of being given in terms of gauge parameters it is
expressed in terms of the simpler discrepancy functions.
The same triplets of graphs and signs above define the violations of
the kinematic Jacobi relations. 
For example,
\be
J_{\{i,1\}} &=& s(i, 1)_1 \,n_{t(i, 1)_1} + s(i, 1)_2 \,n_{t(i, 1)_2}
  + s(i, 1)_3 \, n_{t(i, 1)_3} \,,
\\
J_{\{i,2\}} &=& s(i, 2)_1\, n_{t(i, 2)_1} + s(i, 2)_2 \, n_{t(i, 2)_2} 
  + s(i, 2)_3\, n_{t(i, 2)_3}\,,
\ee
are the discrepancy functions corresponding to propagators $1$ and $2$ of the $i$-th graph.
The three terms correspond to the three numerators participating in the Jacobi relation. 
More explicitly, from \tab{TripletsAndSignsTable} for the first three diagrams we have
\begin{align}
J_{\{1,1\}} & = n_{1} + n_6 + n_{11}\,,
\hskip 1.5 cm
J_{\{1,2\}} = n_{1} + n_{5} + n_{2} \,, \nonumber\\
J_{\{2,1\}} & = n_{2} + n_7 + n_{12}\,,
\hskip 1.5 cm
J_{\{2,2\}} = n_{2} + n_{1} + n_{5} \,, \nonumber\\
J_{\{3,1\}} & = n_{3} + n_8 + n_{13}\,,
\hskip 1.5 cm
J_{\{3,2\}} = n_{3} - n_{5} + n_{4} \,.
\end{align}
The remaining 24 discrepancy functions, including the associated signs,
 can be read off from \tab{TripletsAndSignsTable}.

As described in detail in the previous section and illustrated in the case of the $4\times 4$ cut, the  discrepancy functions are 
not independent.  First there are simple relations coming from simple overcount such as,
\begin{equation}
J_{\{1,1\}}  =  J_{\{6, 1\}} \,, \hskip 1 cm 
J_{\{2,1\}}  =  J_{\{12,1\}} \,, \hskip 1 cm 
J_{\{1,2\}}  =  J_{\{2,2\}}  \,, \hskip 1 cm 
J_{\{10,1\}} = - J_{\{9,2\}} \, .
\label{J5SimpleRelations}
\end{equation}
The remaining such relations are easily read off from
\tab{TripletsAndSignsTable}. All told there are 20 such relations.  
In addition to these, there are 
five momentum-dependent nontrivial constraints, corresponding to zero eigenvectors 
of the $\zeta$ matrix defined in Eq.~\eqref{independentalpha}. A simple and symmetric choice is 
\begin{align}
0 & =   \frac{J_{\{1, 2\}}}{d^{(1)}_1} + \frac{J_{\{3, 1\}}}{d_{3}^{(2)}}
     - \frac{J_{\{6, 2\}}}{d_{6}^{(1)}} - \frac{J_{\{2, 1\}}}{d^{(2)}_2}
    =  \frac{J_{\{1, 2\}}}{s_{3 4}} + \frac{J_{\{3, 1\}}}{s_{35}}
     - \frac{J_{\{6, 2\}}}{s_{2 4}} - \frac{J_{\{2, 1\}}}{s_{2 5}} 
    \,,
    \nonumber \\
0 & =   \frac{J_{\{2, 1\}}}{d^{(2)}_2} + \frac{J_{\{3, 2\}}}{d_{3}^{(1)}}
      - \frac{J_{\{7, 2\}}}{d_{7}^{(1)}} - \frac{J_{\{3, 1\}}}{d_{3}^{(2)}}
    =   \frac{J_{\{2, 1\}}}{s_{2 5}} + \frac{J_{\{3, 2\}}}{s_{1 2}}
      - \frac{J_{\{7, 2\}}}{s_{1 3}} - \frac{J_{\{3, 1\}}}{s_{3 5}} 
 \,, 
    \nonumber \\
0 & =   \frac{J_{\{1, 1\}}}{d_{1}^{(2)}} + \frac{J_{\{3, 2\}}}{d_{3}^{(1)}}
      - \frac{J_{\{11, 2\}}}{d_{11}^{(1)}} - \frac{J_{\{3, 1\}}}{d_3^{(2)}} 
    =   \frac{J_{\{1, 1\}}}{s_{1 5}} + \frac{J_{\{3, 2\}}}{s_{1 2}}
      - \frac{J_{\{11, 2\}}}{s_{2 3}} - \frac{J_{\{3, 1\}}}{s_{3 5}} \,,
    \nonumber  \\
0 & =  \frac{J_{\{1,  2\}}}{d^{(1)}_1} + \frac{J_{\{3, 1\}}}{d_{3}^2}
       -\frac{J_{\{12, 2\}}}{d_{12}^{(1)}} - \frac{J_{\{1, 1\}}}{d_{1}^{(2)}}
    =  \frac{J_{\{1,  2\}}}{s_{3 4}} + \frac{J_{\{3, 1\}}}{s_{3 5}}
       -\frac{J_{\{12, 2\}}}{s_{1 4}} - \frac{J_{\{1, 1\}}}{s_{15}}  \,,
         \nonumber \\
0 & = \frac{J_{\{1, 1\}}}{d_{1}^{(2)}} - \frac{J_{\{6, 2\}}}{d_{6}^{(1)}} + 
      \frac{J_{\{3, 2\}}}{d^{(3)}_1} - \frac{J_{\{4, 1\}}}{d_{4}^{(2)}}
    = \frac{J_{\{1, 1\}}}{s_{15}} - \frac{J_{\{6, 2\}}}{s_{24}} + 
      \frac{J_{\{3, 2\}}}{s_{12}} - \frac{J_{\{4, 1\}}}{s_{45}} \, .
\label{J5Relations}
\end{align}
Each denominator corresponds to the other propagator in the diagram
around which the BCJ identity is being performed.
Similar equations for the five-point tree amplitude were constructed in
Refs.~\cite{HenryConstraints,PierreConstraints} from the requirement
that BCJ amplitude relations hold.

After imposing all the constraints on the discrepancy functions only 5
of the initial 30 are independent and thus 5 combinations of the generalized
gauge-transformation parameters $\al^{(1)}_{i}$ and
$\al^{(2)}_{i}$ are determined. The rest simply drop out of ${\cal
  E}_\G^5$.  This pattern is similar to the one of solving for
kinematic numerators in terms of amplitudes~\cite{BCJ}: some
numerators are determined in terms of amplitudes while others drop out
of any expression for other amplitudes.

Plugging the solution for the gauge parameters into the expression \eqref{Extra5Start} for the extra term 
correcting the naive double copy we find that ${\E}^{5}_\G$ is given by
\begin{align}
{\E}^{5}_\G = & 
\biggl( \frac{J_{\{3, 2\}}}{d_{3}^{(1)}} + \frac{J_{\{3, 1\}}}{d_{3}^{(2)}}\biggr) 
  \biggl(\frac{\tJ_{\{1, 2\}}}{d_{1}^{(1)}} -\frac{\tJ_{\{1, 1\}}}{d_{1}^{(2)}} 
    - \frac{\tJ_{\{2, 1\}}}{d^{(2)}_2} +  \frac{\tJ_{\{3, 1\}}}{d_{3}^{(2)}} 
  - \frac{\tJ_{\{3, 2\}}}{d_{3}^{(1)}} \biggr)
\nonumber \\
& \null  +  \frac{J_{\{1, 1\}}}{d_{1}^{(2)}} \biggl( \frac{\tJ_{\{1, 1\}}}{d_{1}^{(2)}} - \frac{\tJ_{\{3, 1\}}}{d_{3}^{(2)}} 
    + \frac{\tJ_{\{3, 2\}}}{d_{3}^{(1)}}   \biggr)
    + \frac{J_{\{2, 1\}}}{d^{(2)}_2} \biggl(\frac{\tJ_{\{2, 1\}}}{d^{(2)}_2} 
     - \frac{\tJ_{\{3, 1\}}}{d_{3}^{(2)}} + \frac{\tJ_{\{3, 2\}}}{d_{3}^{(1)}} \biggr)
\nonumber \\
& \null
 +  \frac{J_{\{1, 2\}}}{d^{(1)}_1} \biggl(\frac{\tJ_{\{3, 1\}}}{d_{3}^{(2)} } 
                                     - \frac{\tJ_{\{3, 2\}}}{d_{3}^{(1)}} \biggr) 
+ \frac{J_{\{3, 1\}} \tJ_{\{3, 1\}}} {( d_{3}^{(2)})^2 } \,.
\label{Asymmetric5}
\end{align}
Because of the relations that the $J$s satisfy there are many equivalent forms of ${\E}^{5}_\G$.
The most symmetric one gives the full gravity cut as
\begin{align}
{\cal C}^{5}_\G =\sum_{i=1}^{15} \frac{n_{i}  \tn_{i}}{d^{(1)}_{i} d^{(2)}_{i} } 
 + {\E}^{5}_\G
 \qquad
 \text{with}
 \qquad
 {\E}^{5}_\G & = - \frac{1}{6}\sum_{i=1}^{15}
 \frac{J_{\{i,1\}} \tJ_{\{i,2\}} + J_{\{i,2\}} \tJ_{\{i,1\}} }
  { d^{(1)}_{i} d^{(2)}_{i}} \,.
\label{Extra5}
\end{align}
This symmetric solution is found by using an ansatz with the desired
symmetry and matching it to the solution \eqref{Asymmetric5} for
${\E}^{5}_\G$ in a basis of $J$s.  This symmetric form has the added
advantage that the organization of the terms follows individual
diagrams.  While it is desirable to have symmetric formulas such as
\eqn{Extra5}, this is not essential for it to be useful for
constructing cuts of high-loop order gravity
amplitudes. Eq.~\eqref{Asymmetric5} is perfectly usable 
in the construction of the five-loop four-point 
$\NeqEight$ supergravity amplitude.

Although \eqns{Asymmetric5}{Extra5} have explicit propagators, these
expressions are actually local and correspond directly to the desired
contact term corrections.  In fact, in this relatively simple case,
each term is individually local because each diagram has only two
propagators.  Indeed, the violation of manifest BCJ duality must be
proportional to the off-shell invariant of the propagator which {\em
  does not} participate in the Jacobi relation, {\it i.e.}
\begin{equation}
  J_{\{i,1\}} \propto d^{(2)}_i \,, \hskip 1.5 cm  J_{\{i,2\}} \propto d^{(1)}_i\,.
\end{equation}
Thus, in both \eqns{Asymmetric5}{Extra5}, the propagators cancel term by term 
against the numerators.

%%%%%%%%%%%%%%%%%%%%%%%%%

\section{Formulas for N$^k$MCs with $k\ge 3$}
\label{HigherLevelJFormulasSeaction}

In this section we discuss certain classes of \N{k} cuts with $k\ge 3$. These have a much more intricate 
structure than the \N2 cuts analyzed in the previous section. They also have the important feature 
that, unlike   \N2 cuts, ${\cal E}_\G$ is no longer local so the extraction of the contact term 
is somewhat more intricate.

\subsection{Three four-point tree amplitudes}

Consider an $4\times4\times 4$ \N3.
Following the labeling discussed in previous sections, in terms of the 27 parent diagrams, this cut is
\be
{\cal C}^{4\times 4 \times 4 }_\text{YM} = \sum_{i_1, i_2, i_3 = 1}^3 
\frac{n_{i_1, i_2, i_3} c_{i_1, i_2, i_3}}
     {d_{i_1}^{(1)} d_{i_2}^{(2)} d_{i_3}^{(3)}} \, ,
\label{C4x4x4}
\ee
where each index in the sum takes three values corresponding to the three diagrams
of each four-point tree amplitude in the cut.  The upper index in the propagator $1/d^{(j)}_i$ refers the $j$th tree amplitude.
The gauge transformation \eqref{solGT} connecting the 
color-kinematics-satisfying numerators to some arbitrary 
ones is
\be
\Delta_{i_1,i_2,i_3} = n_{i_1,i_2,i_3} - n^\BCJ_{i_1,i_2,i_3} 
= d^{(1)}_{i_1} \al^{(1)}_{i_2,i_3} 
+ d^{(2)}_{i_2} \al^{(2)}_{i_1, i_3}
+ d^{(3)}_{i_3} \al^{(3)}_{i_1,i_2} \,,
\label{Delta4x4x4}
\ee
where $\al^{(x)}_{i_y, i_z}$ obey the $4\times 4\times 4$ version of the relations \eqref{remainingGTfull}. 
Their solution together with the color-Jacobi relations
\be
\sum_{i_1=1}^3 c_{i_1, i_2, i_3} = \sum_{i_2=1}^3 c_{i_1, i_2, i_3} =
  \sum_{i_3=1}^3 c_{i_1, i_2, i_3} = 0\,,
\ee 
and momentum conservation, 
\begin{equation}
\sum_{i_x =1}^3 d^{(x)}_{i_x} = 0 \,,
\label{stueq0}
\end{equation}
guarantee that the gauge-theory $4\times 4\times 4$ cut is invariant
under \eqn{GeneralizedGaugeTrans} with parameters \eqref{Delta4x4x4}.
From \eqn{Extra} we then have the extra contribution that corrects the
naive double copy,
\begin{align}
{\E}^{4\times 4\times 4}_\G & = 
- \sum_{i_1, i_2, i_3 =1}^3
\frac{1}{d_{i_1}^{(1)}d_{i_2}^{(2)}d_{i_3}^{(3)}}
  \Bigl(d^{(1)}_{i_1} \al^{(1)}_{i_2,i_3} 
   + d^{(2)}_{i_2} \al^{(2)}_{i_1, i_3} 
   + d^{(3)}_{i_3} \al^{(3)}_{i_1,i_2} \Bigr)
\nonumber \\
&  \hskip 3 cm  \times \Bigl( d^{(1)}_{i_1} \tal^{(1)}_{i_2,i_3} 
 + d^{(2)}_{i_2} \tal^{(2)}_{i_1, i_3} 
 + d^{(3)}_{i_3} \tal^{(3)}_{i_1,i_2} \Bigr) \, .
\label{Extra4x4x4Start}
\end{align}
Numerator terms proportional to $(d_{i_x}^{(x)})^2$ cancel out because
of the momentum conservation identity \eqref{stueq0}.

For the $4\times 4 \times 4$ cut, the equations \eqref{JofDelta}
relating the discrepancy functions and the gauge parameters we have
\begin{align}
J_{\x1, i_2, i_3}\equiv \sum_{i_1=1}^3 n_{i_1, i_2, i_3} =&
    d^{(2)}_{i_2} \sum_{i_1=1}^3 \al^{(2)}_{i_1, i_3}
  + d^{(3)}_{i_3} \sum_{i_1=1}^3 \al^{(3)}_{i_1, i_2}\,, \nonumber
\\
J_{i_1, \x2, i_3} \equiv \sum_{i_2=1}^3 n_{i_1, i_2, i_3} =& 
    d^{(1)}_{i_1} \sum_{i_2=1}^3 \al^{(1)}_{i_2, i_3} 
  + d^{(3)}_{i_3} \sum_{i_2=1}^3 \al^{(3)}_{i_1, i_2} \,, \nonumber
\\
J_{i_1, i_2, \x3} \equiv \sum_{i_3=1} n_{i_1, i_2, i_3} =&
    d^{(1)}_{i_1} \sum_{i_2} \al^{(1)}_{i_2,i_3} 
  + d^{(3)}_{i_2} \sum_{i_3} \al^{(3)}_{i_1,i_3} \, .
\label{Jacobi4x4x4}
\end{align}
As in the simpler case of the $4\times 4$ cut, these relations capture the fact that 
the discrepancy functions are not independent but rather obey certain relations with 
momentum-dependent coefficients. They also capture the fact that only certain linear 
combinations of gauge parameters can be determined in terms of $J$.
More precisely, there are 27 $\al$-functions and 27 $J$s, but only 15
different combinations of $\al$s appear on the right-hand side of
Eqs.~\eqref{Jacobi4x4x4}. Moreover, only 12 combinations of $\al$s are
determined in terms of 12 $J$s and remaining 15 $J$s are also
determined in terms of these 12.  The undetermined $\al$ functions
drop out of ${\cal E}^{4 \times 4 \time 4}_\G$.

Here and for subsequent cases it is useful to also define ``double discrepancy functions'':
\begin{align}
J_{\x1, \x2, i_3} & \equiv \sum_{i_2 = 1}^3 J_{\x1, i_2, i_3}
   =\sum_{i_1=1}^3 J_{i_1, \x2, i_3} 
   = d^{(3)}_{i_3} \sum_{i_1, i_2=1}^3 \al^{(3)}_{i_1,i_2} \nonumber \,,
\\
J_{\x1, i_2, \x3} & \equiv \sum_{i_3=1}^3 J_{\x1, i_2, i_3} 
  = \sum_{i_1=1}^3 J_{i_1, i_2, \x3 } 
  = d^{(2)}_{i_2} \sum_{i_1, i_3=1}^3 \al^{(2)}_{i_1, i_3} \,, \nonumber
\\
J_{i_1, \x2, \x3} & \equiv \sum_{i_2=1}^3 J_{i_1, i_2, \x3} 
   =\sum_{i_3=1}^3 J_{i_1, \x2, i_3} 
   = d^{(1)}_{i_1} \sum_{i_2, i_3=1}^3 \al^{(1)}_{i_2, i_3} \, ;
\label{DoubleJacobi4x4x4} 
\end{align}
they are particular linear combinations of discrepancy functions. In this case, their main property is that 
they are proportional to a specific inverse propagator. They are also the common value of different 
combinations of $J$s corresponding to different zero eigenvectors in \eqn{ZeroEigenVector} 
of the matrix $\sigma$ defined in Eq.~\eqref{sigma}.
By inspecting these equations it is straightforward to see that 
\begin{align}
&   \sum_{i_2, i_3=1}^3 \al^{(3)}_{i_2,i_3}=\frac{J_{1, \x2, \x3}} {d^{(1)}_1}
   =\frac{J_{2, \x2, \x3}} {d^{(1)}_2}
   =\frac{J_{3, \x2, \x3}} {d^{(1)}_3} \,, \nonumber
\\
&    \sum_{i_1, i_3=1}^3 \al^{(2)}_{i_1, i_3}=\frac{J_{\x1, 1, \x3}} {d^{(2)}_1}
   = \frac{J_{\x1, 2, \x3}} {d^{(2)}_2}
   = \frac{J_{\x1, 3, \x3}} {d^{(2)}_3} \,, \nonumber
\\
&\sum_{i_1, i_2=1}^3 \al^{(1)}_{i_1, i_2}=\frac{J_{\x1, \x2, 1}}{d^{(3)}_1}
   = \frac{J_{\x1, \x2, 2}}{d^{(3)}_2}=\frac{J_{\x1, \x2, 3}}{d^{(3)}_3} \,.
\label{relations_3}
\end{align}

To write the extra contributions $\E_\G^{4\times 4\times 4}$ to the
naive double copy in terms of the discrepancy functions we first solve
Eqs.~(\ref{Jacobi4x4x4}) and (\ref{DoubleJacobi4x4x4}) for the 12
independent gauge parameters which thus become functions of $J$s and
substitute the result in Eq.~\eqref{Extra4x4x4Start}. Upon using momentum
conservation identities, the undetermined gauge parameters drop out and
the terms correcting the naive double copy become
\begin{equation}
{\E}^{4\times 4\times 4}_\G = T_1 + T_2\,,
\label{Extra4x4x4}
\end{equation}
where
\begin{align}
T_1 &= - \sum_{i_3=1}^3 \frac{J_{\x1, 1, i_3} \tJ_{1, \x2, i_3}} 
                {d^{(1)}_1 d^{(2)}_1 d^{(3)}_{i_3}} 
       - \sum_{i_2=1}^3 \frac{J_{\x1, i_2, 1} \tJ_{1, i_2, \x3}}
                  {d^{(1)}_1 d^{(2)}_{i_2} d^{(3)}_1} 
        -\sum_{i_1=1}^3 \frac{J_{i_1, \x2, 1} \tJ_{i_1, 1,  \x3}}
                 {d^{(1)}_{i_1} d^{(2)}_1 d^{(3)}_1}  + \{J \leftrightarrow \tJ\} \,, 
\nonumber \\
T_2 &=   \frac{J_{\x1, 1, 1}\tJ_{1, \x2, \x3}}{d^{(1)}_1 d^{(2)}_1 d^{(3)}_{1}}
         + \frac{J_{1, \x2, 1}\tJ_{\x1, 1, \x3}}{d^{(1)}_1 d^{(2)}_{1} d^{(3)}_{1}}
         + \frac{J_{1,1, \x3}\tJ_{\x1, \x2, 1}} 
            {d^{(1)}_1 d^{(2)}_1 d^{(3)}_{1}}  + \{J \leftrightarrow \tJ\} \, ,
\end{align}
and  we used \eqn{relations_3} to simplify $T_2$.

Unlike the extra terms for the $4\times 4$ and $5$ \N2 cuts, this
expression is no longer local so to extract the corresponding $4\times
4\times 4$ contact term we need to subtract the contribution of the
\N2 contact terms to this cut.  Subtraction terms are easily
constructed from nonlocal terms corresponding to $4\times 4$ contact
terms.  This needs to be done consistently across all higher-level
cuts where a given $4\times 4$ cut enters when putting on-shell
propagators of the $4\times 4\times 4$ cut.  The issue is
that the $4\times 4$ contact terms are not unique, but depend on the
off-shell continuation (\ref{OnShellToOffshell}).
With this understanding, we can formally write the subtraction terms as
\begin{equation}
\E_\G^{4\times 4\times 4}\Big|^\text{subtraction} = 
\sum_{i_1=1}^3 \frac{1}{d_{i_1}^{(1)}} ( \E_\G^{(i_1) 4\times 4} )_{2,3} +
\sum_{i_2=1}^3 \frac{1}{d_{i_2}^{(2)}} ( \E_\G^{(i_2) 4\times 4} )_{1,3} +
\sum_{i_3=1}^3 \frac{1}{d_{i_3}^{(3)}} ( \E_\G^{(i_3) 4\times 4} )_{1,2}  \,, 
\end{equation}
where $(\E_\G^{(i_a)4\times 4})_{j,k}$ is the extra contact
contributions derived from the $4\times 4$ cut built from two tree
amplitudes $j$ and $k$ in the $4\times 4\times 4$ cut. The superscript
$(i_a)$ takes into account the differing residues on each pole.  By
construction, this subtracts the nonlocality in the extra terms.  We
stress that $(\C_\G^{4\times 4})_{i,j}$ is best obtained by relabeling
already chosen \N2 contact terms rather than re-applying the formula
\eqref{Extra4x4_symmetric}.  Otherwise, care is needed to ensure
that a uniform off-shell continuation is used every time the
contribution of a previously-determined contact term is subtracted.
The same principles, of course, hold in general whenever
non-localities are subtracted by lower-level contact terms (which 
are part of the cut of the incomplete integrand, cf. \eqn{ContactEquation}).

\subsection{One five-point and one four-point tree amplitude}

A much more interesting and intricate case is that of a cut with one
five-point and one four-point tree amplitude.  For the five-point tree
we follow the same labeling as for the case of a cut with a single
five-point tree amplitude discussed in sec.~\ref{sec:5ptcontact}.  The
four-point amplitude factor will be labeled as before. Thus, the
gauge-theory cut is
\begin{eqnarray}
{\cal C}^{5\times 4}_\text{YM} &=& 
       \sum_{i=1}^{15} \sum_{j=1}^3 
\frac{n_{i j} c_{i j}}{d_{i}^{(1,1)}d_{i}^{(1,2)} d^{(2)}_{j}} \, .
\end{eqnarray}
The indices $i$ and $j$ run over the 15 and 3 diagrams of the
five-point and four-point amplitude factors, respectively; we shall
refer to the five-point amplitude as the first factor and the
four-point amplitude as the second factor.  The first upper index on
the propagators labels whether the propagator belongs to the first or
the second tree amplitude factor; the second upper index locates the
propagator in the ordered list of propagators of each graph. For the
five-point amplitude factor this list is in \eqn{props_5pt}; as before, for the
four-point amplitude factor we suppress this index since this amplitude has a
single propagator per graph.

The color-Jacobi identities are
\be
&& s(i, 1)_1 c_{t(i, 1)_1 , j} + s(i, 1)_2 c_{t(i, 1)_2 , j} + s(i, 1)_3 c_{t(i, 1)_3 , j} = 0 \,, \nonumber
\\
&& s(i, 2)_1 c_{t(i, 2)_1 j}  + s(i, 2)_2 c_{t(i, 2)_2 , j} + s(i, 2)_3 c_{t(i, 2)_3 , j} = 0 \nonumber\,,
\\
&& c_{i , 1}+c_{i , 2}+c_{i , 3}=0\, \hskip 1cm  i=1,\dots,15    \qquad j=1,2,3 \,,
\label{colorJac5x4}
\ee
where we used the triplet and sign functions in \eqn{tripletsandsigns}.  The values of these
functions are found in \tab{TripletsAndSignsTable}. 

The generalized gauge transformation relating $n_{i,j} $ to 
color-kinematics duality-satisfying ones is: 
\begin{equation}
\Delta_{i,j} \equiv n_{i, j} - n^\BCJ_{i, j} = d^{(1,1)}_{i} \al^{(1,1)}_{i,j}
         + d^{(1,2)}_{i} \al^{(1,2)}_{i,j} +  d^{(2)}_{j} \al^{(2)}_i \,.
\end{equation}
There are $2\times 15\times3 + 15 = 105$ functions; of these 12 are
determined by the requirement \eqref{GeneralizedGaugeTrans} (or
alternatively, \eqref{remainingGTfull}) that the cuts are invariant
under such shifts.  This leaves 93 functions, some of which will be
determined in terms of BCJ discrepancy functions.

From \eqn{Extra} the extra contribution besides the naive double copy
in terms of $\alpha$ functions~is
\begin{align}
{\E}^{5\times 4}_\G & =
- \sum_{i=1}^{15} \sum _{j=1}^3
\frac{1}{d^{(1,1)}_{i} d^{(1,2)}_{i}  d^{(2)}_{j} } \, 
\Bigl(  d^{(1,1)}_{i} \al^{(1,1)}_{i,j}
      + d^{(1,2)}_{i} \al^{(1,2)}_{i,j} +  d^{(2)}_{j} \al^{(2)}_i \Bigr) \nonumber \\
& \hskip 4 cm \times 
\Bigl(  d^{(1,1)}_{i} \tal^{(1,1)}_{i,j}
      + d^{(1,2)}_{i} \tal^{(1,2)}_{i,j} +  d^{(2)}_{j} \tal^{(2)}_i \Bigr) \, ,
\label{Extra5x4Start}
\end{align}
where, to keep the equation short, we did not substitute the 12
$\alpha$ functions determined by the requirement of invariance under
generalized gauge transformations.  Our task is to re-express
Eq.~\eqref{Extra5x4Start} in terms of easy-to-obtain BCJ discrepancy
functions, defined by substituting kinematic numerators in place of
color factors in Eq.~\eqref{colorJac5x4}:
\begin{align}
J_{\{i, 1\}, j} & =s(i, 1)_1 n_{t(i, 1)_1 , j} + s(i, 1)_2 n_{t(i, 1)_2 , j} 
    + s(i, 1)_3 n_{t(i, 1)_3 , j}\,,
\cr
J_{\{i, 2\}, j}&=s(i, 2)_1 n_{t(i, 2)_1 j} +s(i, 2)_2 n_{t(i, 2)_2 , j} 
   + s(i, 2)_3 n_{t(i, 2)_3 , j}  \,,
\cr
J_{i, \x{j}}&=n_{i , 1}+n_{i , 2}+n_{i , 3}\,, \hskip 1 cm  i=1,\dots,15\,,    \qquad j=1,2,3 \, .
\label{SingleDiscrep5x4}
\end{align}
As in the case of the color-Jacobi identities, the triplet and sign
functions $t$ and $s$ are taken from \tab{TripletsAndSignsTable}. Once
the label $\{i,\lambda_i\}$ for a graph in the five-point amplitude is
specified, the remaining graphs in the triplet are also fixed.

Similar to the  $4\times 4 \times 4$ case,  we also define double-discrepancy functions in 
the spirit of \eqref{DoubleJacobi4x4x4},
\begin{align}
J_{\{i, 1\}, \bullet} &  \equiv \sum_{j=1}^3 J_{\{i, 1\}, j} \,,
\hskip 2 cm 
J_{\{i, 2\}, \bullet}    \equiv \sum_{j=1}^3 J_{\{i, 2\}, j} \,.
\label{DoubleDiscrep5x4A}
\end{align}
Lastly, we also define,
\begin{align}
J_{\{i, 1, 2\}, j} &= 
s(i, 1)_2 J_{\{t(i, 1)_2, 2\}, j}  + s(i, 1)_3 J_{\{t(i, 1)_3, 2\}, j}\,, 
\nonumber \\
J_{\{i, 2, 1\}, j} &= 
s(i, 2)_2 J_{\{t(i, 2)_2,  1\}, j}  + s(i, 2)_3 J_{\{t(i, 2)_3, 1\}, j} \,.
\label{DoubleDiscrep5x4B}
\end{align}
where we did not include terms for $J_{\{t(i, 1)_1, 2\}, j}$ or
$J_{\{t(i, 2)_1,1\}, j}$, because they are already accounted for by
$J_{\{i, 1\}, j}$ and $J_{\{i,2\}, j}$ defined in
\eqn{SingleDiscrep5x4}.  The functions in \eqn{DoubleDiscrep5x4B} can
be interpreted as double-discrepancy functions when the propagators
participating in the two Jacobi relations meet at a vertex.
In total, there are 105  $J$s defined in Eqs.~\eqref{SingleDiscrep5x4}, 
\eqref{DoubleDiscrep5x4A} and \eqref{DoubleDiscrep5x4B}.

As before  $J$s are not independent but satisfy a variety of constraints. 
There are the trivial ones coming from the fact that each Jacobi relation 
has a triplet overcount similar to the ones for the single five-point 
tree amplitude case \eqref{J5SimpleRelations},
\begin{equation}
J_{\{1,1\},j}  =  J_{\{6, 1\},j} \,, \hskip    .5 cm
J_{\{2,1\},j}    =  J_{\{12,1\},j} \,, \hskip .5 cm
J_{\{1,2\},j}    =  J_{\{2,2\},j}  \,, \hskip  .5 cm
J_{\{10,1\},j}   = - J_{\{9,2\},j} \,,
\label{J54SimpleRelations}
\end{equation}
for any value of $j$ corresponding to the three diagrams in the
four-point tree amplitude.  As
for \eqn{J5SimpleRelations}, we can read off all such remaining cases
from \tab{TripletsAndSignsTable}. This gives a total of 60 constraints. 
There are also linear relations such as,
\begin{align}
0 & = -J_{5, \x{j}} -J_{1, \x{j}} + J_{\{1, 2\}, 1} + J_{\{1, 2\}, 2} + J_{\{1, 2\}, 3} - 
       J_{2, \x{j}}\,,  \nonumber\\
0 &= - J_{11, \x{j}} + J_{\{1, 1\}, 1} + J_{\{1, 1\}, 2} 
    + J_{\{1, 1\}, 3} - J_{1, \x{j}} - J_{6, \x{j} } \,,\nonumber \\
0 &= - J_{4, \x{j}} + J_{5, \x{j}}   + J_{\{3, 2\}, 1} 
    + J_{\{3, 2\}, 2} + J_{\{3, 2\}, 3} - J_{3, \x{j}} \, .
\end{align}
They are just special cases of Eq.~\eqref{DoubleDiscrep5x4A} and can also be understood as
corresponding to certain zero eigenvectors of the matrix $\sigma$ defined in Eqs.~\eqref{Jvsn} and \eqref{sigma}.
There are a total of 12 such independent equations. 
Finally, there are generalizations of \eqn{J5Relations} that also involve kinematic variables, for example,
\begin{align}
0 & = \frac{1}{d^{(2)}_{j_1}}
          \biggl(\frac{J_{\{6, 2\}, j_1}}{d^{(1, 1)}_{6}}
               + \frac{J_{\{2, 1\}, j_1}}{d^{(1,2)}_{2}} 
               - \frac{J_{\{1, 2\}, j_1}}{d^{(1,1)}_1} 
               - \frac{J_{\{3, 1\}, j_1}}{d^{(1,2)}_{3}} \biggr)
 \nonumber  \\
 & \hskip 2 cm 
    - \frac{1}{d^{(2)}_{j_2}}
          \biggl(\frac{J_{\{6, 2\}, j_2}}{d^{(1, 1)}_{6}}
               + \frac{J_{\{2, 1\}, j_2}}{d^{(1,2)}_{2}} 
               - \frac{J_{\{1, 2\}, j_2}}{d^{(1,1)}_1} 
               - \frac{J_{\{3, 1\}, j_2}}{d^{(1,2)}_{3}} \biggr)\,,
 \nonumber \\
0 & = \frac{1}{d^{(2)}_{j_1}}
          \biggl(\frac{J_{\{2, 1\},j_1}}{d^{(1,2)}_2} 
               + \frac{J_{\{3, 2\},j_1}}{d^{(1,1)}_{3}}
               - \frac{J_{\{7, 2\},j_1}}{d^{(1,1)}_{7}}   
               - \frac{J_{\{3, 1\},j_1}}{d^{(1,2)}_{3}} \biggr)
 \nonumber  \\
 & \hskip 2 cm 
    - \frac{1}{d^{(2)}_{j_2}}
          \biggl(\frac{J_{\{2, 1\},j_2}}{d^{(1,2)}_2} 
                + \frac{J_{\{3, 2\},j_2}}{d^{(1,1)}_{3}}
                - \frac{J_{\{7, 2\},j_2}}{d^{(1,1)}_{7}}   
                - \frac{J_{\{3, 1\},j_2}}{d^{(1,2)}_{3}} \biggr)\,,
\end{align}
where $j_1, j_2 = 1, 2,3$ and $j_1 \not= j_2$.  
Further similar equations can be obtained from the final three equations in \eqn{J5Relations}
by dividing by $d_{j_1}^{(2)}$, inserting the $j_1$ index into the $J$'s
and subtracting from the result a similar term generated by interchanging $j_1$ and  $j_2$.    
As we shall see, the fact that there is a simple pattern for 
how the constraints are related to the case of the single five-point 
amplitude in the cut will lead to simple relations for the solution.
Altogether there are a total of 10 independent such equations.
In total there are 82 relations between the $J$s leaving 23 independent discrepancy functions.

Constructing and analyzing the $5\times 4$ case of
Eq.~\eqref{JofDelta} reveals that of the remaining 93 $\al$ functions
parametrizing a generalized gauge transformation for such a cut, only
23 are independent and are determined in terms of 23 independent BCJ
discrepancy functions.
The remaining 82 BCJ discrepancy functions are in turn expressed in terms of 23 independent ones (and no $\alpha$ functions).   

After using an ansatz to find an expression with a simple structure,
the extra terms completing a $5\times 4$ cut of the naive double copy to a
cut of a gravity amplitude are
\begin{align}
{\E}^{5 \times 4}_\G 
 = &  \sum_{i=1}^{15}\sum_{j=1}^3 \frac{1}
  {{d_{i}^{(1,1)} d_{i}^{(1,2)} d^{(2)}_{j}}} 
\biggl[
    - \frac{1}{6} {J_{\{i,1\},j} \tJ_{\{i,2\},j} }
 - \left(-\frac{1}{3}\right)\times\frac{1}{6} 
  \Bigl(  {J_{\{i,1\},j} \tJ_{\{i,2\}, \x2}} + 
     {J_{\{i,2\},j} \tJ_{\{i,1\}, \x2}} \Bigr)
\nonumber \\ & \null
 -  a_i {J_{\{i,1\}, j} \tJ_{i, \x2} }
 -  a_i {J_{\{i,2\}, j} \tJ_{i, \x2}}
 +  a^{(1)}_i {J_{\{i, 1, 2\}, j} \tJ_{i ,\x2} }
  +  a^{(2)}_i {J_{\{i, 2, 1\}, j} \tJ_{i, \x2}}
     \biggr]  + \{J \leftrightarrow \tJ\} \,. 
\label{Extra5x4}
\end{align}
The first term is the direct extension of $\E_\G^5$.  The numerical
coefficients $a_i$, $a_i^{(1)}$ and $a_i^{(2)}$ are given in
\tab{Solution5x4Table}.

The values of these coefficients depend critically on the definition
and order of the graphs of the five-point amplitude factor as well as
on the choice of order of propagators for each graph. They moreover
depend on the definitions of $J_{\{i, 1, 2\}, j}$ and $J_{\{i, 2, 1\},
  j}$ in \eqn{DoubleDiscrep5x4B}.  For example, one may choose the 
$a_i$ to be all identical at the expense of modifying these
definitions. It does not appear straightforward, however, to have a
simpler, more systematic form for all $a_i$, $a_i^{(1)}$ and $a_i^{(2)}$
coefficients simultaneously.

%%%%%%%%%%%%%%%  TABLE %%%%%%%%%%%
\begin{table}[tb]
\begin{tabular}{c||c|c|c|c|c|c|c|c|c|c|c|c|c|c|c}
\text{graph}   & 1 & 2  & 3 & 4 & 5 & 6 & 7 & 8 & 9 & 10 & 11 & 12 & 13 & 14 & 15  \cr
   \hline
      \hline
$a_i$ &
$\frac{1}{12}$ & $\frac{1}{6}$  & $\frac{1}{12}$ & $\frac{1}{12}$ & $\frac{1}{12}$ & 
$\frac{1}{12}$ & $\frac{1}{12}$ & $\frac{1}{6} $ & $\frac{1}{12}$ & $\frac{1}{12}$ & 
$\frac{1}{12}$  & $\frac{1}{12}$ & $\frac{1}{12}$ & $\frac{1}{6}$  & $\frac{1}{12}
     \displaystyle{\vphantom{\Big|}}$ \cr
\hline
$a^{(1)}_i$ &  
      $0$      & $0$            & $\frac{1}{12}$ & $\frac{1}{12}$  &     $0$        & 
      $0$      & $\frac{1}{12}$ &       $0$      & $\frac{1}{12}$  &     $0$        & 
      $0$      & $\frac{1}{12}$ & $\frac{1}{12}$ &       $0$       &     $0 
     \displaystyle{\vphantom{\Big|}}$ \cr
\hline      
$a^{(2)}_i$ & 
$\frac{1}{12}$ &      $0$        &     $0$       &        $0$      & $\frac{1}{12}$ & 
$\frac{1}{12}$ &      $0$        &     $0$       &        $0$      & $\frac{1}{12}$ &
$\frac{1}{12}$ &      $0$        &     $0$       &        $0$      & $\frac{1}{12}
     \displaystyle{\vphantom{\Big|}}$ \cr  
\hline
 \end{tabular}
\caption{The coefficients for a particularly simple solution for the $5\times 4$ case.}
\label{Solution5x4Table}
\end{table}
%%%%%%%%%%%%%%%%%%%%%%%%%%%%%%%%%%%%         

As for the $4\times 4\times 4$ case, $\E_\G^{5\times
  4}$ is not local. To extract its corresponding contact term we need
to subtract the contribution of the $4\times 4$- and $5$-contact terms
the cut overlaps with. The discussion in the previous section applies
here as well, so we do not repeat it.

%%%%%%%%

\subsection{One six-point amplitude in cut}

Following the above discussion we also have found a solution for a
single six-point amplitude insertion in a generalized cut.  Our
solution is given in the ancillary file {\tt
  ExtraJ\_6pt.m}~\cite{AttachedFile}.  We follow a similar
organization as for the five-point case discussed in
\sect{sec:5ptcontact}, except that at six points there are 105
diagrams, instead of the 15 at five points.  To apply it in cuts with
a single six-point amplitude, as usual we need to relabel to match the
labels in the cut.  The file lists the Jacobi triplets,
analogous to those of \tab{TripletsAndSignsTable}, as well as the
constraints on $J$'s analogous to those of \eqn{J5Relations}.
Finally, the file contains the formula for the extra terms needed to
correct the naive double-copy contributions in terms of the $J$ and
$\tJ$.  The presented solution is not manifestly crossing symmetric,
but is instead expressed in terms of a set of independent $J$'s
obtained by solving the constraint equations.  Nor are all the
kinematic denominators manifestly organized in terms of diagrams.
Nevertheless, this is adequate for our purpose of simplifying the
analytic structure of \N3 with a single six-point tree amplitude,
compared to directly evaluating the cuts via \eqn{doubleCopyCutKLT}.
It would be an interesting problem to find a more symmetric form that
generalizes to higher points.

As for the earlier cases, one can encounter terms that behave as
$0/0$, when inserted into a cut. These are harmless when the $0$ in
the numerator is manifest, since it corresponds to an absent
contribution.  One extra complication for the six-point case is that
sometimes the $0$ in the numerator is not manifest and requires
cancellation between distinct terms.  When this occurs, the simplest
strategy is to take advantage of the asymmetry in the formula, to
relabel, to avoid these problematic cases. 

\subsection{Formulas for more general cuts}

For the \N4 maximal cuts and beyond, the relative simplicity of the
contact terms can make it advantageous to determine the missing contact
terms by numerical analysis of \eqn{ContactEquation}.
Nevertheless it is important to study the general cases because they
display a pattern which points to the possibility of general simple
solutions for the contact terms at any loop order.  We now generalize the
discussion in the previous sections to the infinite classes of cuts
$4\times\dots\times 4$ and $5\times 4\times\dots\times 4$.

\subsubsection{Multiple four-point tree amplitudes in cuts}

Consider an \N{k} composed of $k$ four-point tree amplitudes.  The analysis for
these cases is very similar to that of the \N2 and \N3 cases with only
four-point tree amplitudes in the cuts.  What
emerges is a simple recursive pattern for generating the extra
corrections terms to the naive double copy \eqref{Extra}.  As in the
$4\times4$  \N2 and in the $4\times4\times4$ \N3 cases, we
label the contributing graphs by the off-shell propagators they
contain.

For \N{k}s we generalize \eqn{Extra4x4x4} by defining the simple, double, triple and so forth 
BCJ discrepancy functions,
\begin{align}
J_{\x1, i_2, i_3, \ldots, i_q}\equiv \sum_{i_1=1}^3 n_{i_1,i_2,i_3,\ldots,i_q} \,,
\nonumber \\
J_{\x1, \x2, i_3, \ldots, i_q}\equiv \sum_{i_1,i_2=1}^3 n_{i_1,i_2,i_3,\ldots,i_q} \,,
\nonumber \\
J_{\x1, \x2, \x3, \ldots, i_q}\equiv \sum_{i_1,i_2,i_3=1}^3 n_{i_1,i_2,i_3,\ldots,i_q} \,,
\end{align}
with similar definitions for the other combinations of indices.

In terms of these quantities, we
can generate the correction to the naive double copy for the case of $q$ four-point tree amplitudes
in the cut by simple substitution rules.  We start with the expression:
\begin{equation}
- \sum_{i_1,i_p, \ldots,i_q =1}^3  \frac{J_{i_1, \ldots, i_p, \ldots ,i_q} \tJ_{i_1, \ldots, i_p, \ldots ,i_q}}
      {d^{(1)}_{i_1} \cdots  d^{(p)}_{i_p} \cdots d^{(q)}_{i_q}}  \,.
\end{equation}
Then one generates new terms by performing the following substitutions repeatedly
until no new terms are generated:
\begin{align}
\sum_{i_p=1}^3  \frac{J_{a_1, \ldots, i_p, \ldots, a_q} \tJ_{b_1, \ldots, i_p, \ldots, b_q}}
      {d^{(p)}_{i_p}} &\rightarrow
- \frac{J_{a_1, \ldots \x{p}, \ldots, a_q} \tJ_{b_1,\ldots, 1, \ldots, b_q} } {d^{(p)}_{1}}
  + \{ J \leftrightarrow \tJ \}
 \,, 
\label{JFourSubs}
\end{align}
where the $a_r$ and $b_r$ are unchanged by the substitution and are either
an $i_r$, 1 or {`\small{$\x{r}$}'}.  We drop all generated terms where
there is not at least one {`\small{$\x{r}$}'} in each $J$ or $\tJ$.
Then we sum over all unique terms generated by repeated substitutions
of \eqn{JFourSubs}.  It is straightforward to see that these
substitutions reproduce the solutions in \eqns{Extra4x4}{Extra4x4x4}
for the $4 \times 4$ and $4\times 4 \time 4 \times 4$ cases.  The
$4\times4\times4\times4$ is given in \app{HigherCutFormulasAppendix}.

%%%%%%%%%%%%%%%%%%%
\subsubsection{One five-point and multiple four-point tree amplitudes in cut}

We now turn to the more intricate case of a single five-point tree
amplitude in the cut along with multiple four-point tree amplitudes.
Again we can give a simple substitution rule for generating such
contributions.

To generate the terms we start from 
\begin{equation}
-\sum_{i=1}^{15} \sum_{j_2,\ldots j_{q} = 1}^3
     \frac{J_{i,j_2, j_3, \ldots, j_q} \tJ_{i,j_2, j_3, \ldots, j_q}}
         {d_{i}^{(1,1)}d_{i}^{(1,2)} d^{(2)}_{j_2} \cdots d^{(q)}_{j_{q}}}\,,
\end{equation}
and perform the following substitutions:
\begin{align}
 J_{i,j_2, j_3, \ldots, j_q} \tJ_{i,j_2, j_3, \ldots, j_q} & \rightarrow
 \frac{1}{6} J_{\{i,1\},j_2, j_3, \ldots, j_q}  \tJ_{\{i,2\},j_2, j_3, \ldots, j_q}  + \{J \leftrightarrow \tJ \}
\,, \nonumber \\
 J_{i,j_2, \ldots, j_p, \ldots, j_q} \tJ_{i,j_2, \ldots, j_p, \ldots, j_q} & \rightarrow
 a_i \sum_{h=1}^2  J_{i,j_2, \ldots, \x{p}, \ldots, j_q}  \tJ_{\{i,h\},j_2, \ldots, j_p, \ldots, j_q} 
     + \{J \leftrightarrow \tJ \}\,, \nonumber \\
 J_{i,j_2, \ldots, j_p, \ldots, j_q} \tJ_{i,j_2, \ldots, j_p, \ldots, j_q} & \rightarrow
 -a_i^{(1)} J_{i,j_2, \ldots, \x{p}, \ldots, j_q}  \tJ_{\{i,1,2\},j_2, \ldots, j_p, \ldots, j_q} 
  + \{J \leftrightarrow \tJ \} \,,
 \nonumber\\
 J_{i,j_2, \ldots, j_p, \ldots, j_q} \tJ_{i,j_2, \ldots, j_p, \ldots, j_q} & \rightarrow
 -a_i^{(2)} J_{i,j_2, \ldots, \x{p}, \ldots, j_q}  \tJ_{\{i,2,1\},j_2, \ldots, j_p, \ldots, j_q} 
   + \{J \leftrightarrow \tJ \}  
\,,
 \nonumber\\
 J_{b,b_2, \ldots, i_p, \ldots, b_q} \tJ_{c,c_2, \ldots, i_p, \ldots, c_q} & \rightarrow -\frac{1}{3} 
 J_{b,b_2, \ldots, \x{p}, \ldots, b_q} \tJ_{c,c_2, \ldots, i_p, \ldots, c_q}  + \{ J \leftrightarrow \tJ \}
\,,
\label{JFiveSubs}
\end{align}
where in the last substitution $b$ and $c$ are one of $i$, $\{i,1\}$,
$\{i,2\}$, $\{i,1,2\}$ or $\{i,2,1\}$ and $b_r$ and $c_r$ are either
$i_r$ or `{\footnotesize$\x{r}$}'.  We drop terms where there is not
at least one such alteration in both $J$ and $\tJ$. The final
substitution rule should be repeatedly applied to all terms until no
new terms are generated. We then sum over the distinct terms generated
this way. As usual, terms should not be double counted. It is
straightforward to see that this generates both $5$ and $5\times 4$
solutions in \eqns{Extra5}{Extra5x4}.  We have also directly confirmed
that this correctly gives the $5\times4\times4$ case given in the
Appendix~\ref{HigherCutFormulasAppendix}.

%%%%%%%%%%%%%%%%%%%%%%%%%%%

\section{Five-loop four-point integrand of $\NeqEight$ Supergravity }
\label{FiveLoopResultsSection}

In this section we present  results for the 
five-loop four-point integrand of $\NeqEight$ supergravity,
obtained using the methods described in the previous sections.
In this case, the two gauge theories used in the construction 
are both $\NeqFour$ super-Yang--Mills theory.

\subsection{$\NeqFour$ super-Yang-Mills starting point}

We start from the five-loop four-point integrand for $\NeqFour$
super-Yang-Mills theory obtained in Ref.~\cite{FiveLoopN4}.  To
make it a bit more useful we rearrange it slightly to remove
the spurious appearance of triangle subdiagrams.
This expresses the five-loop
four-point $\NeqFour$ super-Yang-Mills amplitude in terms of 410
nonvanishing diagrams containing only cubic vertices,
\begin{equation}
{\cal A}_4^{\fiveloop\ \NeqFour} = i g^{12} s t A_4^{\rm tree}
\sum_{{\cal S}_4} \sum_{i = 1}^{410} 
\int \prod_{j = 5}^9 \frac{d^D l_j}{(2 \pi)^D} \frac{1}{S_i} \frac{ c_i n_i}
   { \prod_{m_i = 5}^{20} l_{m_i}^2}  \,.
\label{YMFiveLoopFourPoint}
\end{equation}
The label $i$ runs over the 410
cubic diagrams. Examples of these are shown in \fig{FiveLoopExampleGraphsFigure}.  The other
sum runs over the 24 permutations ${\cal S}_4$ of external leg
labels. As in \eqn{genericGaugeLoopIntegral}, the symmetry factor
$S_i$ for each diagram $i$ removes overcounts, including those arising
from internal automorphism symmetries with external legs fixed.  The
color factor $c_i$ for each graph is obtained by dressing every
three-vertex in the graph with a factor of $\tilde f^{abc}$,
normalized as in \eqn{genericGaugeLoopIntegral}, and the gauge
coupling is $g$.  We denote external momenta by $k_j$ for $j= 1,
\ldots, 4$ and the five independent loop momenta by $l_j$ for $j = 5,
\ldots , 9$.  The remaining $l_j$ for $j=10, \ldots 20$ are linear
combinations of these following the labeling of the diagram.  

The prefactor $A_4^{\rm tree} \equiv A_4^{\rm tree}(1,2,3,4)$ in \eqn{YMFiveLoopFourPoint} is the
color-ordered tree amplitude of $\NeqFour$ super-Yang--Mills
theory, for any states of the theory. The presence of such a universal prefactor is special to the four-point
amplitudes of $\NeqFour$ super-Yang--Mills theory; in general, the
dependence on external states is part of the numerator factors $n_i$. In four dimensions the prefactor is conveniently organized using an on-shell superspace~\cite{Nair}.  The external kinematic invariants
are
\begin{equation}
s = (k_1 + k_2)^2\,, \hskip 1 cm 
t = (k_2 + k_3)^2\,, \hskip 1 cm 
u = (k_1 + k_3)^2\,,
\end{equation}
and the combination $s t A_4^{\rm tree}$ is crossing symmetric.

The diagram, color factors, symmetry factors and kinematic
numerators corresponding to those in \eqn{YMFiveLoopFourPoint} are 
given in the ancillary file {\tt Level0Diagrams.m}~\cite{AttachedFile}.  
Some of the $\NeqFour$ super-Yang-Mills kinematic numerators are rather simple.  For
example, the numerators of the first 15 diagrams are,
\begin{align}
n_1 & =  n_2 = n_3 = n_4 = n_5 = n_6 = n_7 = n_9 = s^4 \,, \nonumber \\
n_{10} & = n_{15} = \frac{1}{2} s^3 (\tau_{3,5} + \tau_{4,15} )\,, \nonumber \\
n_{11} & = n_{13} = \frac{1}{2} s^3  (\tau_{3, 5} +  \tau_{4, 15} + l_{5}^2 + l_{15}^2) \,,  \nonumber\\
n_{12} & = s  \tau_{3, 5}\,,  \nonumber \\
n_{14} &= s^3 s_{3, 5} + s^3 l_5 ^2 - \frac{5}{2} s l_5^2 l_{13}^2 l_{15}^2\,,
\label{NumeratorSamples}
\end{align}
where
\begin{align}
\tau_{i,j} & \equiv 2 k_i \cdot l_j\,,  \hskip 1 cm  (i \le 4, j\ge 5) \nonumber  \\
\tau_{i,j} & \equiv 2 l_i \cdot l_j \,, \hskip 1 cm   (i, j\ge 5) \,.
\end{align}
Two slightly more complicated numerators, are for diagrams 280 and 282,
\begin{align}
n_{280} & = 
 s^4 +  s^3 (\tau_{10, 13} + \tau_{18, 20}) 
 + \frac{1}{2} s^2 (\tau_{10, 13}^2 + \tau_{18, 20}^2) 
 + 2 t (  l_5^2 + l_6^2) ( l_{13}^2 l_{18}^2 + l_{10}^2 l_{20}^2 ) \,, \hskip .4 cm 
\label{NumeratorSamples2} \\
n_{283} & = s^4 + s^3 (\tau_{10, 13} + \tau_{18, 20}) + \frac{1}{2} s^2 (\tau_{10, 13}^2 + \tau_{18, 20}^2) 
  - \Bigl( 2 s + \frac{5}{2} t\Bigr)  (l_5^2 + l_6^2) (l_{13}^2 l_{18}^2 +  l_{10}^2 l_{20}^2)\,,
\nonumber
\end{align}
where these two diagrams are included in \fig{FiveLoopExampleGraphsFigure}.

Some of the remaining kinematic numerators are also relatively simple,
while others are more complicated and contain thousands of terms.  An
important feature of all the numerators is that each term contains at
most three inverse propagators. After factoring out the overall $s t
A_4^\tree$ each numerator term contains four kinematic invariants of
which at least one factor is either $s$ or $t$, leaving at most three
kinematic invariants that can be inverse propagators.  This implies
that the $\NeqFour$ super-Yang--Mills five-loop four-point
 amplitude can be fully constructed from generalized
cuts through the \N3 level~\cite{FiveLoopN4}.

%%%%%%%%%%%%%%%%%%%%%%%%%
\subsection{$\NeqEight$ supergravity naive double copy}

We organize the results for the $\NeqEight$ supergravity five-loop
four-point amplitude into contact term levels starting with the naive
double copy \eqref{NDCLoopIntegral} of the cubic diagrams, which we take as level 0.  Each level $k$ corresponds to the contact diagrams that can be obtained from collapsing $k$ propagators in the cubic diagrams.

At level 0 there are 410 diagrams, in one-to-one correspondence to the
nonvanishing diagrams of the $\NeqFour$ super-Yang--Mills
amplitude \eqref{YMFiveLoopFourPoint},
\begin{equation}
{\cal M}_4^{(0) \fiveloop} = i \Bigl(\frac{\kappa}{2}\Bigr)^{12} s t u M_4^{\rm tree}
  \sum_{{\cal S}_4} \sum_{i = 1}^{410}
\int \prod_{j = 5}^9 \frac{d^D l_j}{(2 \pi)^D} \frac{1}{S_i} \frac{N_i^{(0)}}
       { \prod_{m_i = 5}^{20} l_{m_i}^2}  \,.
\label{GRNaiveFiveLoopFourPoint}
\end{equation}
The $\Neqeight$ supergravity numerators in the naive double copy are
simply squares of the $\NeqFour$ super-Yang-Mills ones,
\begin{equation}
N_i^{(0)} = n_i^2\,,
\label{NumeratorSquare}
\end{equation}
and where we used $[s t A_4^{\rm tree}]^2 = -i s t u M_4^{\rm tree}$ to
re-express the square of the prefactor in \eqn{YMFiveLoopFourPoint} in terms of
four-point supergravity tree amplitude, $M_4^{\rm tree}$.  As for the 
$\NeqFour$ super-Yang--Mills case, this is valid for all states of
the theory.

Since any $\NeqFour$ super-Yang--Mills numerator has at most
three inverse propagators, by squaring them 
in the naive double copy, we obtain no more than six inverse propagators.
This suggests that to construct the supergravity amplitude, the cuts
through level 6 should be sufficient.  Indeed, we have explicitly
confirmed that there is no new information to be found in the \N7s.

\subsection{Contact terms}

%%%%%%%%%%%%%%%  TABLE %%%%%%%%%%%
\begin{table}[tb]
\centering
\begin{tabular}{|c |c|c|c|c|c|c|c|}
\hline
\multirow{2}{*}{level}&\multirow{2}{*}{total $\#$
of diagrams}  &\multicolumn{6}{c|}{number of contact interactions}\\ \cline{3-8} 
  && ~1  ~ &~ 2  ~ & ~ 3 ~ 
  & ~ 4  ~ & ~ 5 ~  &~ 6 ~ \\
\hline 
$ 1 $ & 2,473    &2,473 & 0& 0& 0& 0& 0 \\ \hline
$ 2$ & 7,917   & 1,597 &6,320 &0 &0 &0 &0 \\
\hline
$ 3$ & 15,156  & 940& 6,710 & 7,506& 0& 0& 0\\ \hline
$ 4$ & 19,567  & 434 & 5,232 & 9,510 & 4,391 &0 &0 \\ \hline
$ 5$ & 17,305  & 203& 3,012  & 7,792 & 5,185 & 1,113 &0 \\ \hline
$ 6$ & 10,745  &83 &1,567 & 4,407& 3,694& 896 & 98 \\ \hline
\end{tabular}
\caption{Number of diagrams at each level defined by the number of
  collapsed propagators starting from the 410 top-level diagrams
  (dropping pathological diagrams where a loop has no propagators).
  The columns labeled by the number of contacts $i$ records the number
  of diagrams containing $i$ contact interactions where four or more
  lines meet at a vertex.}
\label{tab:sectors}
\end{table}
%%%%%%%%%%%%%%%%%%%%%%%%%%%%%%%%%        

The next task is to construct the contact term corrections to the
naive double copy. The first level consists of all independent
diagrams generated by collapsing a single propagator in all possible
ways in the 410 level-0 diagrams.  One then removes diagrams that are
identical up to relabelings (the final assembly of the amplitude
accounts for such permutations).  As indicated in \tab{tab:sectors},
at level 1 there are 2,473 independent diagrams, not related by
relabelings.  However, as already noted in \sect{ConstructionSection},
the numerators of these diagrams all vanish because next-to-maximal
cuts of the naive double copy automatically match those of the
supergravity amplitude.

In order to obtain the contact terms for levels 2 and 3 we use the
formulas of
\sects{N2LevelJFormulasSection}{HigherLevelJFormulasSeaction} to
generate expressions for the supergravity cuts.  The contact terms are
then obtained from these. At level 2, the formulas directly give the
contact terms, but beyond this we need to consistently subtract the
previous levels.

The contact diagrams of level  2 are generated by canceling two
propagators in each of the 410 top-level diagrams.  This gives
diagrams with either a single five-point vertex or two four-point
vertices. Examples of such diagrams are given in the first
column of \fig{NkMaxContactsFigure}.
There are 7,917 independent contact diagrams at this level, as
listed in \tab{tab:sectors}.  As can be deduced from
\tab{tab:sectors}, 1,597 of these have a single five-point contact
vertex and 6,320 of these have two four-point contact vertices.  For
the remaining levels, \tab{tab:sectors} gives the number of
independent diagrams at each level, as well as finer information on
the number of diagrams with a given number of contact interactions.
All independent diagrams obtained from
collapsing propagators starting from the 410 top-level diagrams are
included, except for the pathological case where all propagators of a
single loop are canceled.  All other scale-free integrals 
such as diagrams $(4\!: 9)$, $(5\!: 57)$, $(6\!: 983)$ and $(6\!:2669)$ in \fig{NkMaxContactsFigure}
are included.

As noted in \sect{ConstructionSection}, as one proceeds beyond level 3
the contact terms get simpler. However, the cuts themselves become
significantly more complicated.  By level 6 the generalized cuts
can have up to nine-point trees.  These features mean that, at level
4 and beyond it can become efficient to use numerical analysis on the KLT-like
formula for the supergravity cut (\ref{doubleCopyCutKLT}) to determine
the missing contact terms.  Because the contact diagrams have fewer
propagators, the numerator kinematic polynomial is of lower dimension, 
which in turn, implies that it has fewer terms.
For example, at level 2 there are
157,080 potential numerator terms in each diagram prior to imposing
diagram symmetries.  At level level 3 this drops to 17,952 possible terms.
By level 4 this falls to 1,584 potential terms, which is small enough,
especially once diagram symmetries are accounted, 
to make numerical analysis efficient.  By level 6, up to overall
normalization, there are only three possible terms and the contact diagram
numerator are of the form,
\begin{equation}
a_1 s^2 + a_2 s t + a_3 t^2 \,.
\end{equation}
The parameters $a_1, a_2$ and $a_3$ are easily determined 
from three kinematic points that satisfy the cut conditions.

The complete amplitude is given by a sum over diagrams, including the
naive-double-copy ones in \eqn{GRNaiveFiveLoopFourPoint} and contact term diagrams,
\begin{equation}
{\cal M}_4^{\fiveloop} = i \Bigl(\frac{\kappa}{2}\Bigr)^{12} 
 s t u M_4^{\rm tree} \sum_{k=0}^6 \sum_{{\cal S}_4} \sum_{i=1}^{T_k}
 \int \prod_{j = 5}^9 \frac{d^D l_j}{(2 \pi)^D} \frac{1}{S_i} \frac{N_i^{(k)}}
 { \prod_{m_i=5}^{20-k} l_{m_i}^2 } \,,
\label{GRContactFiveLoop}
\end{equation}
where $T_k$ is the total number of diagrams at each level, which can 
be read off from \tab{tab:sectors}.

The results for the diagrams and their numerators at each level are
collected in plain-text {\em Mathematica}-readable ancillary files~\cite{AttachedFile}.  The top-level file {\tt Level0Diagrams.m}
gives the $\NeqEight$ supergravity result via the double copy
\eqref{NumeratorSquare}.  The six other files {\tt Level1Diagrams.m,
  Level3Diagrams.m,\,$\ldots,$\,Level6Diagrams.m} contain the level $1,\ldots, 6$
$\NeqEight$ supergravity contact diagrams, combinatoric factors, and
kinematic numerators. 

Unlike the top-level diagrams, the contact diagrams can have triangle,
bubble and tadpole contributions, as illustrated in
\fig{NkMaxContactsFigure}.  This can be attributed to the poor power counting
of the naive double copy.  Representations with better power counting
without bubbles or triangles should exist, though it would
require further non-trivial work to construct one.

We include diagrams that contain scale-free loop integrals since these
can affect ultraviolet divergences.  If we were to evaluate the
integrals purely in dimensional regularization we could safely ignore
such contributions, since they integrate to zero and in dimensions
$D>4$ there are no infrared singularities to mix with these.  However,
at high loop orders it is much more efficient to extract the
ultraviolet divergences by series expanding in large loop momentum or
equivalently in small external momentum and introducing infrared
regulator, such as a mass for each propagator.  One might think that
scale-free integrals should not be an issue in dimensions
$D>4$, because there are no physical infrared singularities to mix with
the ultraviolet ones.  Unfortunately, this is not correct.  There are
two sources of difficulties.  The first is that the series expansion
of the integrand can generate infrared-singular integrals, even in
higher dimensions where there are no physical infrared singularities.
The second is that in the construction, one can add and subtract scale
free integrals, in such a way that one of the contributions is
manifestly a scale-free integral, such as diagram $(4\! :9)$ of
\fig{NkMaxContactsFigure}, while the contribution that should cancel
it is absorbed into an integral which is not scale free, by
multiplying and dividing by appropriate propagators.  When mixed up
with other terms and momentum conservation is applied, it can be
unobvious that spurious scale-free integrals are mixed in.  While
dimensional regularization would consistently set both contributions
individually to zero, in the presence of a massive infrared regulator
the two contributions can be individually nonzero, but cancel only
after combining them.  If we were to arbitrarily drop integrals that
are manifestly scale free, we would upset this cancellation and obtain
an incorrect result for the ultraviolet divergence.  This phenomenon
is well studied at four loops in $D=11/2$ in Section~IIIC of
Ref.~\cite{SimplifyingBCJ}.  The upshot is that some care is required
to ensure that scale-free integrals that can affect potential
ultraviolet divergences in higher dimensions are properly taken into
account. For example, to ensure that any potential contact term
corresponding to diagram $(4\!\!: 9)$ of \fig{NkMaxContactsFigure} is
properly taken into account we evaluate the corresponding generalized
cut \N4 9 in \fig{NkMaxCutsFigure}.  In both $\NeqFour$
super-Yang--Mills theory and $\NeqEight$ supergravity, all such cuts
vanish, as we directly verified in the latter case using
\eqn{doubleCopyCutKLT}.  The purpose of the contact term is to ensure
that the cut vanishes.

The derived contact terms contained in the {\em Mathematica} files
have some noteworthy properties.  The most striking is that most
vanish.  Specifically, at each level the following number of diagram
numerators vanish:
\begin{align}
&\hbox{Level 2: } \; 6,158  \hbox{ of }  7,917\,, \nonumber \\
&\hbox{Level 3: } \; 11,894 \hbox{ of } 15,156\,, \nonumber \\
&\hbox{Level 4: } \; 14,980 \hbox{ of } 19,567\,, \nonumber \\
&\hbox{Level 5: } \; 13,239 \hbox{ of } 17,305\,, \nonumber \\
&\hbox{Level 6: } \; 7,941  \hbox{ of } 10,745\,.
\end{align}
The precise number of vanishing diagrams depends on the starting point
we used in the naive double copy and also on details of the off-shell
continuation of the contact terms at each level.  The large number of
vanishing contact terms is a consequence of  many dual Jacobi
identities automatically holding.  One reason is that the duality
generally holds automatically around propagators that are part of
one-loop four-point subdiagrams, given these tend not to depend on the
momentum of that loop.  Another reason is that the five-loop ${\cal N}=4$ super-Yang-Mills amplitude was constructed by recycling the corresponding four-loop amplitude on a simple form~\cite{CompleteFourLoopSYM}; it therefore automatically inherits a variety of simplifying properties. 

Another striking property is that the numerators of all level-2
contact diagrams containing two four-point vertices factorize.  This
is a direct consequence of \eqn{Extra4x4}, with $\tJ = J$ since the
two copies are identical.  In fact, these properties were originally
used in Ref.~\cite{GeneralizedDoubleCopy}, as an important clue that a
generalized double-copy construction in terms of BCJ discrepancy
functions should exist.

In order to confirm the correctness of our integrand construction we
performed a number of nontrivial checks.  We computed large numbers of
additional generalized unitarity cuts not used in the construction of
the integrand. This includes cuts generated, not only from releasing
on-shell conditions on the 410 nonvanishing top-level diagrams, but
also those obtained from a larger set of 910 top-level diagrams free of bubble and triangle subdiagrams.  We checked that all such cuts at the \N2 through \N6 levels are correct, without requiring any additional
contributions. Furthermore, we numerically confirmed on a complete set of \N7s---excluding the technically challenging ones containing a ten-point 
tree amplitude---that no further contact terms arise.
 We also numerically confirmed over 300 cuts at the \N8
level without finding any additional contributions.

%%%%%%%%%%%%%%%%
\section{Maximal cut integration check in $D=22/5$}
\label{sec:vacuumExpansion}

In this section we describe a formalism for extracting ultraviolet
divergences and apply it to perform a
nontrivial check on the constructed integrand. We follow the standard
strategy of expanding the integrand at large loop momenta or
equivalently at small external momenta, as used in earlier supergravity
calculations~\cite{SimplifyingBCJ,N4SugraMatter}.  The main
difference is that the integral relations needed to simplify the
results are much more complex, so we employ modern unitarity
compatible integration ideas~\cite{IBPAdvances, Zhang2016kfo,
  Harley2017, Bosma2017} to streamline the computations.

Specifically, we perform checks on the expected finiteness of $\NeqEight$
supergravity in dimensions $D < 24/5$~\cite{BjornssonAndGreen,SevenLoopGravity}.
While there is reason to believe that $\NeqEight$ supergravity may be
finite in $D=24/5$ as well, it is nontrivial to perform the requisite
loop integration, so here we will content ourselves with demonstrating
the expected ultraviolet cancellations in $D=22/5$.  While this result is not a
surprise, it does serve as a nontrivial verification of the integrand.
The integrand we constructed in the previous section does not manifest
the ultraviolet properties term by term.  In fact, some terms contain
up to four extra powers of loop momentum compared to that needed for
manifest finiteness in $D = 22/5$.  For example, if we square either
of the numerators in \eqn{NumeratorSamples2} to obtain the $\NeqEight$
supergravity double-copy numerators (\ref{NumeratorSquare}), we find
terms with up to 6 inverse propagators or 12 powers of loop momentum
in the numerator.  Given that there are 16 propagators and five
independent loop momenta, individual terms do lead to
divergences even in four dimensions.

We could test the cancellation in $D=4$, but there are 
complications for this case: the integrals have subdivergences and
in addition technical difficulties arise with the Baikov representation that we
use.  Since these complications are not relevant for the interesting
case of $D=24/5$, it is much better to perform checks in $D=22/5$
which requires a nontrivial series expansion of the integrand, but is
still far simpler than the $D=24/5$ case.

To carry out this expansion in large loop momenta, we follow
Refs.~\cite{SimplifyingBCJ,N4GravFourLoop,N4SugraMatter} which are
based on Refs.~\cite{Vladimirov}.  Taylor-expanding in small external
momenta (equivalent to large loop momenta) expresses the integrand as
a sum of vacuum integrals. These are Feynman integrals whose
propagator structures are given by graphs without external legs, but
with numerators which can depend on external momenta.  The terms with
six external powers of momenta (not counting the overall $(st A_4^\tree)^2$ factor) are
log divergent in $D=22/5$.  Lorentz invariance can then be used to
perform {\it tensor reduction} which eliminates the appearance of dot
products of loop momenta with external momenta, e.g., using
\begin{equation}
(l_5 \cdot k_1) (l_6 \cdot k_2) \rightarrow 
\frac{1}{D} \, (k_1 \cdot k_2) (l_5 \cdot l_6) \,,
\end{equation}
where $l_i$ is a loop momentum, $k_i$ is an external momentum and $D$
is the spacetime dimension.  This is valid inside the vacuum integrals
and eliminates dot products between loop and external momenta in the
numerators.  Finally, a mass regulator is introduced to deal with
infrared singularities, which are artifacts of the expansion.  See
Ref.~\cite{N4SugraMatter} for further details.

We apply this procedure to each diagram after summing over all 24
permutations of external legs and dividing by the appropriate symmetry
factor that removes double counts and inner automorphisms.  This
leaves us with a large number of vacuum integrals with propagators
 raised to various powers and with different numerators.

To simplify the expression we use IBP relations \cite{IBPRefs}, with
the additional refinement of dropping any ultraviolet finite
integrals, to directly obtain linear relations between ultraviolet
poles of different vacuum integrals~\cite{IntegralRelations}. The
linear relations reduce the ultraviolet divergence of the amplitude to
a linear combination of a small number of master integrals. We will
check whether the coefficients of certain master integrals vanish
individually.  Of course, if the coefficients do not vanish
individually, we would then need to check for additional relations
between master integrals not captured by the reduction procedure,
including those with canceled propagators;
at least for the $D=22/5$ case described here, no further
relations are required.

Given the large number of vacuum integrals generated by the expansion 
procedure, a full IBP reduction would involve solving a large linear
system, which is a nontrivial computational
task. Instead, we will exploit recent advances in IBP reduction on
unitarity cuts~\cite{IBPAdvances, Zhang2016kfo, Harley2017, Bosma2017}
to quickly obtain the coefficients of the two top-level master
integrals---master integrals corresponding to vacuum graphs with only cubic 
vertices---via the maximal cut of vacuum integrals. We will find the
two coefficients to be both zero, which is consistent with
ultraviolet-finiteness of $\NeqEight$ supergravity in $D=22/5$. The zero
coefficients result from non-trivial cancellations between hundreds of
diagrams that contribute to the top-level master integrals after
vacuum expansion and IBP reduction. Therefore this provides a
highly nontrivial check on the five-loop integrand.

In general, for $L$-loop integrals that are dimensionally regularized in $D$
dimensions, there is no ultraviolet subdivergence if none of $D, 2D,
\dots, D(L-1)$ is an even integer.  This implies there are no
subdivergences in $D=22/5$, simplifying our analysis, compared to, for
example, calculations of the five-loop QCD
$\beta$-function~\cite{FiveloopQCDBeta}. Only an overall $1/\eps$ simple
pole in $D=22/5-\eps$ needs to be evaluated.  The high tensor powers
nevertheless make it nontrivial.

%%%%%%%%%%%%%%
\subsection{Warmup: Half-maximal supergravity at two loops in $D=5$}
\label{subsec:twoLoopIBP}

%%%%%%%%%%%%%% FIGURE %%%%%
\begin{figure}
\includegraphics[scale=.5]{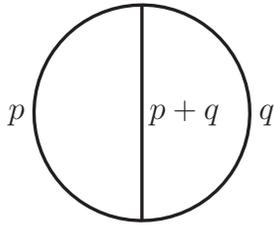}
\caption{The two-loop vacuum diagram corresponding to \eqn{eq:twoLoopIabc}.}
\label{Vacuum2LoopFigure}
\end{figure}
%%%%%%%%%%%%%%%%%%%%%%%%%%%  

In order to explain the machinery that we use at five loops,
we briefly review the treatment in Ref.~\cite{IntegralRelations} of
IBP relations needed to demonstrate the ultraviolet finiteness of
half-maximal supergravity at two loops in $D=5$, and in addition give a more
intuitive treatment by computing cut integrals \cite{CutIntegrals,
  Sogaard2014ila, Harley2017, Bosma2017} following the method of
Ref.~\cite{Bosma2017}.

As explained in Ref.~\cite{IntegralRelations}, after vacuum expansion and
tensor reduction, the potential ultraviolet divergence for this case
is given by,
\begin{equation}
I_{1,1,3} + 2 I_{1,2,2} \,, \label{eq:N4Sugra2loopDiv}
\end{equation}
where we omit an overall constant factor and 
\begin{equation}
I_{A,B,C} = \int d^D l_1\, d^D l_2 \, \frac{1}{(l_1^2-m^2)^A (l_2^2-m^2)^B \left[ (l_1+l_2)^2-m^2 \right]^C}\,, 
 \label{eq:twoLoopIabc}
\end{equation}
with $ D = 5 - 2\epsilon$.  This corresponds to
\fig{Vacuum2LoopFigure} with $m$ being a uniform mass to regulate the
infrared divergences. $I_{A,B,C}$ is invariant under ${\cal S}_3$
permutations of $(A,B,C)$, a fact which we will use without further
mention.  The task we are interested in here is to show that the
combination of integrals in \eqn{eq:N4Sugra2loopDiv} is ultraviolet
finite.  Explicit calculation gives~\cite{IntegralRelations}
\begin{equation}
I_{1,1,3}\big|_{\text{UV div.}} = - \frac {\pi}{192 \epsilon} \, , \quad I_{1,2,2}\big|_{\text{UV div.}} 
= \frac {\pi}{96 \epsilon} \, ,
\end{equation}
which shows that the divergence in \eqn{eq:N4Sugra2loopDiv} cancels.

However, explicit evaluation becomes overwhelmingly more challenging at five loops, 
and it is generally easier to find relations between integrals rather than to 
evaluate them explicitly. An example of a useful IBP identity is
\begin{align}
0 &= \int d^D l_1 \int d^D l_2 \left( l_1^\mu \frac{\partial}{\partial l_1^\mu} 
- l_2^\mu \frac{\partial}{\partial l_2^\mu} \right) 
\frac{1}{(l_1^2-m^2)^A (l_2^2-m^2)^B \left[ (l_1+l_2)^2-m^2 \right]^C} \nn \\
&=  (-2A+2B) I_{A,B,C} - 2C\, I_{A-1,B,C+1} + 2 C I_{A,B-1,C+1} \nn \\
 &\hskip 2 cm \null  + m^2 \left( -2A \, I_{A+1,B,C} + 2B \, I_{A,B+1,C} \right) \, . \label{eq:twoLoopIBPABC}
\end{align}
With $A+B+C=5,\, A,B,C >0$ we have a leading ultraviolet divergence.
While the second line of \eqn{eq:twoLoopIBPABC} is logarithmically
ultraviolet divergent, the terms proportional to $m^2$ are ultraviolet
convergent by power counting.  (Only the overall power counting is
needed because dimensional regularization does not yield one-loop
subdivergences near $D=5$.) Therefore, we need only keep 
terms without an explicit factor of $m^2$ to
obtain linear relations between ultraviolet poles of different vacuum
integrals.  The same relations can be obtained by setting 
$m^2=0$ from the beginning. Furthermore, since this IBP relation has no
explicit $D$ dependence in the coefficient of the integrals on the
right-hand side, we can set $D=5$ instead of $D=5-2\epsilon$. In
summary, explicit appearances of $\epsilon$ and $m^2$ may be discarded
at the start of the calculation, leaving only implicit dependence
in the integrals.  We employ this vast simplification at
five loops.

Following the above logic and setting $A=1,\, B=C=2$, Eq.~\eqref{eq:twoLoopIBPABC} becomes,
\begin{align}
0 &= 2 I_{1,2,2} - 4 I_{0,2,3} + 4 I_{1,1,3} + \text{ ultraviolet finite}\nn \\
&= 2 I_{1,2,2} + 4 I_{1,1,3} + \text{ ultraviolet finite} \,, \label{eq:twoLoopI22andI113}
\end{align}
where in the second line we dropped $I_{0,2,3}$, because the integral
factorizes into two one-loop integrals and is therefore ultraviolet
finite in dimensional regularization near five dimensions. This completes
the IBP-based proof that Eq.~\eqref{eq:N4Sugra2loopDiv} is
ultraviolet finite in dimensional regularization.

By five loops, the number of potential integrals and IBP relations
explodes, so it becomes important to find further simplifications.  A
more direct derivation of the ultraviolet finiteness of
\eqn{eq:N4Sugra2loopDiv} comes from the study of ``cut integrals'',
i.e.\ Feynman integrals computed on unitarity cuts.  In carrying this
out some care is required, because integration contours need to be
chosen carefully to preserve integral relations such as IBP
identities. The Baikov representation \cite{BaikovRep} of Feynman
integrals, which uses inverse propagators as integration variables, is
the natural representation to use for cut integrals in arbitrary dimensions.  For the two-loop
vacuum integral Eq.~\eqref{eq:twoLoopIabc} with $m^2=0$ (as justified
in the discussions above), the Baikov representation can be derived
using the following change of variables,
\begin{equation}
z_1 = l_1^2, \quad z_2 = l_2^2, \quad z_3 = (l_1+l_2)^2 \, .
\end{equation}
Using polar coordinates one can show that 
\begin{equation}
I_{A,B,C} \propto \int \frac{dz_1}{z_1^A} \int \frac{dz_2}{z_2^B}
 \int \frac{dz_3}{z_3^C} \left[ P(z_i) \right]^{(D-3)/2}, \label{eq:twoLoopBaikov}
\end{equation}
where we omitted a constant of proportionality which depends on only the dimension $D$.
The \emph{Baikov polynomial} $P(z_i)=P(z_1,z_2,z_3)$ is defined as
\begin{equation}
P(z_i) = \det (2 l_i \cdot l_j) = 4 \left[ l_1^2 \, l_2^2 - (l_1 \cdot l_2)^2 \right] 
= 2z_1 z_2 + 2 z_2 z_3 + 2 z_3 z_1 - z_1^2-z_2^2-z_3^2 \, . \label{eq:defBaikovPoly}
\end{equation}
The integration boundary in Eq.~\eqref{eq:twoLoopBaikov} is $P(z_i)=0$ because with real $l_1$ and $l_2$, the triangle inequality implies that $P(z_i) \geq 0$.
As discussed following Eq.~\eqref{eq:twoLoopI22andI113}, if any of
the three propagators of the integral is canceled, the integral
factorizes into two one-loop integrals and therefore becomes
ultraviolet finite. This leads us to compute
Eq.~\eqref{eq:twoLoopBaikov} on the maximal cut $z_1 = z_2
=z_3$=0. An obvious prescription for imposing the maximal cut is turning
each propagator into a Dirac delta function,
\begin{equation}
\int \frac {dz_i}{z_i} \rightarrow \int dz_i \, \delta(z_i) \,. \label{eq:cutPrescription1}
\end{equation}
However, such a prescription breaks down because the denominators $z_i$ are raised to general integer powers. A consistent prescription is to turn each $dz_i$ integral into a contour integral around $z_i=0$ \cite{Sogaard2014ila},
\begin{equation}
\int \frac {dz_i}{z_i^A} \rightarrow \frac {1}{2\pi i} \oint \frac {dz_i}{z_i^A} = \frac{dz_i}{(A-1)!} \left( \frac{\partial}{\partial z_i} \right)^{A-1} \Bigg|_{z_i = 0} \, , \label{eq:cutPrescription2}
\end{equation}
which matches the naive prescription Eq.~\eqref{eq:cutPrescription1} when $A=1$. Using the contour integral prescription, we compute Eq.~\eqref{eq:twoLoopBaikov} on the maximal cut in $D=5$,
\begin{align}
I_{A,B,C}\big|_{\rm cut} &\propto \left (\frac 1 {2\pi i} \right)^3 \oint \frac {dz_1}{z_1^A} \oint \frac {dz_2}{z_2^B} \oint \frac {dz_3}{z_3^C} \, P(z_i) \nn \\
&= \text{ coefficient of } z_1^{A-1} z_2^{B-1} z_3^{C-1} \text{ in} \nn \\
&\quad (2z_1 z_2 + 2 z_2 z_3 + 2 z_3 z_1 - z_1^2-z_2^2-z_3^2 ) \,,
\end{align}
which directly gives
\begin{equation}
I_{1,1,3}\big|_{\rm cut} \propto -1, \quad I_{1,2,2}\big|_{\rm cut} \propto 2 \,,
\end{equation}
reproducing the IBP relation Eq.~\eqref{eq:twoLoopI22andI113}, up to
ultraviolet finite terms that are dropped because we imposed the
maximal cut along with $m^2=0$ and $\epsilon=0$.

The above calculation can be straightforwardly generalized to $D=7$,
which is the ultraviolet critical dimension of $\NeqEight$ supergravity at two
loops, by changing the power of the Baikov polynomial to
$(D-3)/2=2$. We reproduce the relation between ultraviolet poles in
$7$ dimensions~\cite{BDDPR},
\begin{equation}
\frac{I_{3,1,3}\big|_{\rm div}} { I_{2,2,3}\big|_{\rm div}} = \frac 3 2 \,,
\end{equation}
without a full evaluation of the two integrals.  By five loops this approach 
becomes enormously beneficial.

%%%%%%%%%%%%%%%%%%%%%%%%%%%%%%%%%
\subsection{Ultraviolet cancellation in $\NeqEight$ super gravity at five loops in $D=22/5$}

For our check of ultraviolet properties in $D=22/5$ we will impose
maximal cuts on the vacuum diagrams, similar to the two-loop example
above.  While this is justified at two loops because the daughter
integrals are all ultraviolet finite in dimensional regularization, in the five-loop case there is no such
argument.  Nevertheless, we can simply ignore the daughter vacuum
diagrams and ask whether the coefficient of the parent master
integrals vanish.  The case of $D=22/5$ should be especially
straightforward because it is very likely that an integrand
representation exists which is term-by-term finite in this dimension,
even if nontrivial to construct.

\begin{figure}
  \centering
  \includegraphics[width=0.5\textwidth]{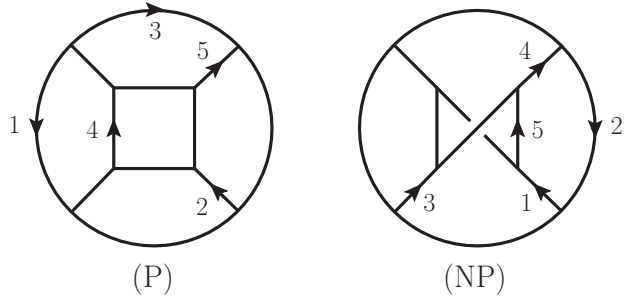}
  \caption{The five-loop planar ``cube'' and nonplanar ``crossed cube'' 
      topologies are the top-level vacuum integrals divergent in $D=22/5$.}
  \label{fig:Vacuum5Loop}
\end{figure}

As usual we organize the integration by parts identities according to
the topology of the vacuum integral.  By topology we mean the set of
propagators, but not the powers which the propagators are raised
to. Therefore every vacuum topology can be defined by a vacuum diagram
in which no two propagators have the same momentum.  At five loops there are four
top-level vacuum diagrams without repeated propagators,
among which only the ``cube'' and the ``crossed cube'' (see 
Fig.~\ref{fig:Vacuum5Loop}) turn out to have non-trivial top-level
master integrals that cannot be reduced to integrals with fewer 
propagators.

%%%%%%%%%%% FIGURE %%%%%%%%%%%%%%
%%%%%%%%%%%%%%%%%%%%%%%%%

Here we will discuss the nonplanar crossed cube in some detail.
The planar cube topology can be treated similarly. The first 12
integration variables 
are defined to be the inverse propagators,
\begin{equation}
z_i = l_i^2, \quad 1 \leq i \leq 12 \,,
\end{equation}
where $l_1,l_2,l_3,l_4$ and $l_5$ are the five independent loop momenta labeled
in the second diagram of Fig.\ \ref{fig:Vacuum5Loop}, while $l_6,
l_7,\dots,l_{12}$ are the momenta of the remaining propagators, each
being a linear combination of the five independent momenta. There are
three irreducible numerators which cannot be written as linear
combinations of inverse propagators,
\begin{equation}
z_{13} = l_1 \cdot l_3, \quad z_{14} = (l_4-l_5)\cdot (l_2-l_1-l_3), \quad z_{15} = -l_3 \cdot l_4 \, ;
\end{equation}
these are the last three integration variables.
We consider the maximal cut which does not allow any propagator to be
canceled. With the 12 propagators $1/z_i$ raised to the powers $p_i$
and with the three irreducible numerators $z_{13}, z_{14}, z_{15}$
raised to the powers $y_1, y_2, y_3$, respectively, the Baikov
representation of the vacuum integral is (again omitting an overall
constant factor)
\begin{equation}
I_{p_i,y_i}=\left( \prod_{i=1}^{12} \int \frac{dz_i}{z_i^{p_i}} \right)\, 
dz_{13} dz_{14}\, dz_{15}\, z_{13}^{y_1} z_{14}^{y_2} z_{15}^{y_3} \left[ P(z_i) \right]^{(D-6)/2} \, .
 \label{eq:baikov5Loop}
\end{equation}
In Eq.~\eqref{eq:baikov5Loop}, $P(z_i)$ is the Baikov polynomial which has uniform degree 5. It is defined as a determinant in a way similar to Eq.~\eqref{eq:defBaikovPoly},
\begin{equation}
P(z_i) = \det (2 l_i \cdot l_j) \, .
\end{equation}
The full expression of the Baikov polynomial, in terms of $z_1, z_2,
\dots, z_{15}$, is needed in the calculation prior to differentiating,
 but omitted here as
it can be easily reproduced.  On the maximal cut, we set $z_i=0, \, 1
\leq i \leq 12$, leaving
\begin{equation}
P(z_i)\big|_{\rm cut} = 64 z_{13} z_{14} z_{15} (z_{13}-z_{14}) (z_{13}+z_{14}+z_{15}) \, . \label{eq:baikovPolyCrossedCubeMaxCut}
\end{equation}

The vacuum expansion of the amplitude at five loops will produce a
linear combination of a large number of vacuum integrals with
different $p_i$ and $y_i$ indices in Eq.~\eqref{eq:baikov5Loop}. For
each of these integrals, imposing the maximal cut using the contour
prescription Eq.~\eqref{eq:cutPrescription2}, we obtain
\begin{equation}
I_{p_i,y_i} \big|_{\rm cut} = \prod_{i=1}^{12} \frac{dz_i}{(p_i-1)!} \left( \frac{\partial}{\partial z_i} \right)^{p_i-1}\bigg|_{z_i=0} \left[  z_{13}^{y_1} z_{14}^{y_2} z_{15}^{y_3} \left[ P(z_i) \right]^{(D-6)/2} \right] \,.
\label{eq:baikov5LoopCut}
\end{equation}
Since all the inverse propagator variables are set to zero after
taking derivatives, Eq.~\eqref{eq:baikov5LoopCut} will become a linear
combination of different integrals (with different $y_i$ exponents and
$\Delta_d$ parameters below) of the following form,
\begin{equation}
\int dz_{13} \int dz_{14} \int dz_{15}\, z_{13}^{y_1} z_{14}^{y_2} z_{15}^{y_3} P(z_i)\big|_{\rm cut}^{(D-2\Delta_D-6)/2}\,, 
\label{eq:baikov5LoopCut2}
\end{equation}
where the power of the Baikov polynomial has decreased by some integer $\Delta_D$ compared with the expression Eq.~\eqref{eq:baikov5Loop}. The value of $\Delta_D$ is in fact correlated with the $y_1,y_2,y_3$,
since the logarithmic power counting of the integrals are preserved by the maximal cut. This turns Eq.~\eqref{eq:baikov5LoopCut2} into
\begin{equation}
\int dz_{13} \int dz_{14} \int dz_{15}\, z_{13}^{y_1} z_{14}^{y_2} z_{15}^{y_3} P(z_i)\big|_{\rm cut}^{-(3+y_1+y_2+y_3)/5} \, . \label{eq:baikov5LoopCut2a}
\end{equation}
The above integral has logarithmic power counting since $P(z_i)\big|_{\rm cut}$ has uniform degree 5 in the $z_i$ variables, and each $z_i$ variable has mass dimension 2.

The top-level master integral for the crossed cube
topology, $V^{(\NP)}$, is defined as the integral with a unit numerator and with no
propagator denominator raised to more than its first power.  We use integration-by-parts identities to reduce all integrals Eq.~\eqref{eq:baikov5LoopCut2a} to the master integral $V^{(\NP)}$ with $y_1=y_2=y_3=0$. The integration-by-parts identities on the maximal cut are given by
\begin{equation}
0 = \int dz_{13} \int dz_{14} \int dz_{15}\, \frac{\partial}{\partial z_j} 
\Bigg[ z_{13}^{y_1} z_{14}^{y_2} z_{15}^{y_3} P(z_i)\big|_{\rm cut}^{(D-2\Delta_D-6)/2} \Bigg] \,,
\end{equation}
where the value of $j$ may be $13$, $14$, or $15$, and the values of
$y_1,y_2,y_3, \Delta_D$ are any integers. We find it convenient to adopt the strategy suggested in
Ref.~\cite{Zhang2016kfo} to first use dimension-shifting identities
and then generate IBP identities in the same spacetime dimension by
solving syzygy equations \cite{Larsen2015ped}. We omit the technical
details and refer the reader to the literature.  As an alternative to
IBP reduction, in Appendix~\ref{sec:IntegrationAppendix} we directly
integrate Eq.~\eqref{eq:baikov5LoopCut2a} with appropriate
integration limits.

We carried out the same IBP procedure for the planar
cube topology to reduce all integrals on the maximal cut to the master
integral $V^{(\P)}$ which again has a unit numerator and no propagator raised to more than its first power.

%%%%%%%%%%%%%%%  TABLE %%%%%%%%%
\begin{table}[tb]
\begin{tabular}{c|c||c|c}
\text{level} & \text{ diagram number } & $V^{(\P)}$ & $V^{(\NP)}$ \cr \hline \hline
0 & 280 & $ \frac{9792}{55}$ & $0$ \cr 
0 & 283 & $0$ & $-\frac{1908}{5}$ \cr
0 &  285 & $-\frac{648}{11}$ & $0$ \cr 
0 &  335 & $\frac{28734}{175}$ & $0$ \cr
0 &  404 & $ \frac{604752003}{123200}$ & $0$ \cr 
0 & 410 & $0$ &$\frac{127594657}{9600}$\cr
2 & 448 & $ -\frac{196356}{385} $  & $0$ \cr
2 & 141 & $ -\frac{819501}{19250}$ & $0$ \cr
2 & 102 & $0$ & $-96$ \cr
2 & 617 & $0$ & $ -\frac{168}{5}$ \cr
3 & 196 & $-\frac{15}{7}$ & $0$ \cr
3 & 186 & $-\frac{75039}{3080}$ & $0$ \cr
3 & 91 & $0$ &$-\frac{75039}{3080} $ \cr 
3 & 68 & $0$ & $\frac{338664}{1925}$ \cr
4 & 42 & $\frac{4453833}{3080}$ & $0$ \cr
4 & 46 & $\frac{32842137}{6160}$ & $0$ \cr
4 & 9 & $0$ &  $-\frac{219}{11}$ \cr
 \end{tabular}
\caption{Ultraviolet divergences in $D=22/5$ from a sample of the diagrams
  given in \figs{FiveLoopExampleGraphsFigure}{NkMaxContactsFigure}.
  The coefficients in the last two columns corresponds to the
  contributions from the two vacuum diagrams in \fig{fig:Vacuum5Loop}.
}
\label{VacuumDiagramsTable}
\end{table}
%%%%%%%%%%%%%%%%%%%%%%%%%%%%%%%%

%%%%%%%%%%%%%%%  TABLE %%%%%%%%%
\begin{table}[tb]
\begin{tabular}{c||c|c}
\text{ level }  & $V^{(\P)}$ & $V^{(\NP)}$  \cr
 \hline
 \hline
  0 &  $\frac{2439779211}{154000}$ & $\frac{2911616507}{7392000}$ \cr
  2 & $\frac{374402283}{308000}$ & $\frac{8846490651}{224000}$ \cr
  3 & $\frac{3535277}{800}$  & $\frac{791440021}{35200}$ \cr
  4 &  $-\frac{18900121}{880}$  & $-\frac{1152620531}{18480}$ \cr
\hline 
 sum & 0 & 0 \cr
\hline
 \end{tabular}
\caption{Ultraviolet divergence in $D=22/5$ after summing up individual
  contributions for two vacuum diagrams.  For each contact level we
  give the coefficients of the planar and nonplanar top-level master
  integrals. The columns sum to zero confirming the expected
  cancellation in $D=22/5$.  Level 0 contributions are from the naive
  double copy, levels 2, 3, 4 are contact term corrections, and levels
  5, 6 do not contribute to the top-level vacuum integrals.}
\label{VacuumResultsTable}
\end{table}
%%%%%%%%%%%%%%%%%%%%%%%%%%%%%%%%

After performing vacuum-diagram expansion and IBP reduction, summing over
permutations, and dividing by appropriate symmetry factors, there are
a total of 152 diagrams that give 
nonvanishing contributions to the coefficient of the planar vacuum
integral $V^{(\P)}$; they originate from contact levels 0, 2, 3 and 4. Similarly, 366 diagrams in total give
nonvanishing contributions to the nonplanar vacuum integral
$V^{(\NP)}$.  Of the diagrams listed in
\fig{FiveLoopExampleGraphsFigure}, six give nontrivial
contributions. Similarly of the contact diagrams in
\fig{NkMaxContactsFigure}, eleven give nontrivial contributions.
These seventeen contributions are collected in \tab{VacuumDiagramsTable}, as examples of the numbers that appear.  Since we are keeping only the two parent master vacuum diagrams, we obtain
contribution only from levels $k \le 4$.  Beyond this there are too
few propagators to contribute.

To find a cancellation we must sum over all the $152+366$ contributions.  In
\tab{VacuumResultsTable} we give the results for each cut level, as
well as the sum over all levels.  It is noteworthy that while the
coefficients coming from individual contributions involve ratios of
large numbers, they completely cancel in the sum over contributions.
The nontriviality of this cancellation strongly suggests that the
terms contributing to this potential ultraviolet divergence are correct.

\section{Conclusions and outlook}
\label{ConclusionSection}

In this paper we described in some detail a generalized double-copy
construction, previously outlined in
Ref.~\cite{GeneralizedDoubleCopy}, for obtaining gravity loop
amplitudes from corresponding gauge-theory loop amplitudes.  It
bypasses the task of finding forms of gauge-theory amplitudes that
satisfy color-kinematics duality, which has proven difficult in
particular situations, but retains our ability to obtain multiloop
gravity integrands in a useful form directly from gauge theory.  We
applied this new method to construct the five-loop four-point
amplitude of $\NeqEight$ supergravity.  At present, only methods that
rely on the double-copy principle are capable of obtaining
supergravity loop integrands at such high loop orders.

Our construction starts with a slightly reorganized version of the
five-loop four-point $\NeqFour$ super-Yang--Mills integrand given in
Ref.~\cite{FiveLoopN4}. By taking a naive double copy of this
integrand, even though it does not manifest the duality between color
and kinematics, we obtain an expression whose maximal and
next-to-maximal cuts automatically match those of the corresponding
supergravity amplitude.  Using the double copy and generalized gauge
symmetry, as outlined in Ref.~\cite{GeneralizedDoubleCopy} and fleshed
out here, we derive generic corrections that are bilinear in the
gauge-theory discrepancy functions that account for the lack of
manifest duality. These correction terms are generic in the sense that
they give explicit formulas that apply to large numbers of different
cuts that we encounter at five loops; more generally, they apply to any loop order and generic double-copy theory. For the case where a generalized cut involve at most one
five-point tree and an arbitrary number of four-point trees, we found
a simple universal pattern for the correction terms. At five loops,
out of all the N$^{k}$MCs that require corrections, slightly more than
half of them are of this type. For the remaining cuts, which typically
have corrections with simpler analytic structure, we used a mixture of
analytical and numerical methods.

To ensure the reliability of the five-loop integrand we carried out a
number of checks. In particular, we verified a large number of
unitarity cuts that are redundant compared to the ones used in the
construction of the integrand.  As a further nontrivial check we
confirmed that in the large loop-momentum limit the ultraviolet
divergences in $D=22/5$ cancel for the top-level master integrals.
While this cancellation is completely expected, the individual
diagrams can be superficially divergent in $D\ge4$, and thus it provides
a nontrivial confirmation not only for our integration techniques but
also for our integrand.

There are a number of open problems. The most obvious application of
the results presented here would be to integrate the expression for
large loop momenta in the next-higher spacetime dimension where an ultraviolet
divergence is possible; that is, in $D=24/5$ dimensions. Knowing the
ultraviolet behavior in $D=24/5$ is of critical importance. Arguments have
suggested that $\NeqEight$ supergravity should diverge in this
dimension at five loops~\cite{BjornssonAndGreen}, and at the same time
we know that similar arguments for $\NeqFour$ supergravity at three
loops and $\NeqFive$ supergravity at four loops imply divergences in
$D=4$ where none exist~\cite{N4GravThreeLoops,N5GravFourLoops}.  In addition,
in the case of half-maximal supergravity in $D=5$ analogous cancellations
have been explicitly linked to the double-copy
structure~\cite{HalfMaxD5}.

The $D=24/5$ integration requires analyzing contributions that are
four momentum powers suppressed compared to the superficial divergence
of the integrand, giving an enormous proliferation of contributing
terms compared to the $D=22/5$ integration. The sheer number of
contributions is a computational challenge, but also the greater
number of relations needed for the various vacuum diagrams encountered
is technically demanding.  In this paper we presented new efficient
methods based on modern developments~\cite{IBPAdvances, Zhang2016kfo,
  Harley2017, Bosma2017} that are suitable for carrying out these
integrations, and we tested them for the simpler case of
$D=22/5$. Further refinements would be important for streamlining
this.

The complications encountered in extracting the ultraviolet behavior
are not surprising given that the representation of the five-loop four-point
amplitude we constructed has a far worse diagram-by-diagram power
counting than ideal. As mentioned, individual terms are ultraviolet
divergent even in four dimensions where there should be no divergences
in the full amplitude~\cite{SevenLoopGravity,BjornssonAndGreen}.  This
poor behavior is inherited from our starting point: the representation of
the $\NeqFour$
super-Yang-Mills integrand~\cite{FiveLoopN4}.  An obvious approach to
this problem would be to find a representation of the $\NeqFour$
super-Yang-Mills amplitude whose naive double copy would be manifestly
ultraviolet finite in $D<24/5$. Then, as the contact term corrections
are added to the integrand, some care would be needed to ensure that
they do not increase the power counting.  If this could be done it
would enormously simplify the loop integration, especially in
$D=24/5$.

A further important open issue is to find explicit formulas for the
contact-term corrections, such that they are manifestly local without
requiring nontrivial cancellations. For the contact terms with two
canceled propagators, our derived formulas have this property.  Beyond
this, the simple patterns of corrections to the cuts with zero or
one five-point tree amplitudes and the rest three- or four-point
amplitudes hints that it may be possible to find such formulas.

Although we focused here on $\NeqEight$ supergravity and $\NeqFour$
super-Yang--Mills, the construction generalizes in the obvious way to
different theories which obey BCJ duality at tree level, and thus to
all gravitational and
non-gravitational~\cite{NLSMBCJ,abelianZ,NLSMaction} double-copy
theories obtained from them.  In particular, we expect similar ideas
to hold for all double-copy theories whose single copies include
fields in the fundamental representation of the gauge
group~\cite{FundMatter, DoubleCopyTheoriesFund}.  A specific
application of our generalized double-copy procedure would be to
construct the five-loop four-point integrand of $\NeqFive$
supergravity, which is important for studying ultraviolet properties
of supergravity theories.  While $\NeqFour$ supergravity does diverge
at four loops, this appears tied to a U(1) anomaly~\cite{Anomaly,
  N4GravFourLoop}.  Such anomalies do not occur in ${\cal N} \ge 5$
supergravity, so it would be important to test whether $\NeqFive$
supergravity diverges at five loops, given that the four-loop
four-point amplitude of this theory is ultraviolet
finite~\cite{N5GravFourLoops}.

Another interesting direction is that our results suggest that it may
be possible to convert any gauge-theory classical solution to a
gravitational one without needing special generalized
gauges.  In particular, it will be interesting to see if the ideas
presented in this paper are helpful for the problem of gravitational
radiation, which has recently been shown to have a double-copy
structure~\cite{RadiationSolutions}.

We expect that the ideas presented in this paper will be useful, not
only for investigating the ultraviolet behavior of perturbative
quantum gravity, but also for understanding general physical properties
of gravity theories.  We look forward to exploring this in the coming
years.

\vskip .15 cm 

\subsection*{Acknowledgments}
We thank Jacob Bourjaily, Lance Dixon, Alex Edison, Michael Enciso,
Enrico Hermann, David Kosower, Julio Parra-Martinez, Chia-Hsien Shen,
Jaroslav Trnka and Yang Zhang for many useful and interesting
discussions.  This work is supported by the Department of Energy under
Award Numbers DE-SC0009937 and DE-SC0013699.  We acknowledge the
hospitality of KITP at UC Santa Barbara in the program ``Scattering
Amplitudes and Beyond'', during various stages of this work.  While at
KITP this work was also supported by US NSF under Grant
No.~PHY11-25915.  J.\,J.\,M.\,C. is supported by the European Research
Council under ERC-STG-639729, {\it preQFT: Strategic Predictions for
  Quantum Field Theories}.  The research of H.\,J. is supported in
part by the Swedish Research Council under grant 621-2014-5722, the
Knut and Alice Wallenberg Foundation under grant KAW 2013.0235, and
the Ragnar S\"{o}derberg Foundation under grant S1/16.

%%%%%%%%%%%%%%%%%%%%%%%%%%%%%

\appendix

\section{Some explicit higher-cut formulas}
\label{HigherCutFormulasAppendix}

In this appendix we give the explicit result of applying the
substitution formulas~\eqref{JFourSubs} and~\eqref{JFiveSubs}
to the ${4\times 4\times 4\times 4}$ and $5\times 4 \times 4$ cases.

\subsection{Four four-point tree amplitudes in the cut}

To obtain the ${4\times 4\times 4\times 4}$ case, we start with the expression
\begin{equation}
  \sum_{i_1,i_2,i_3,i_4 = 1}^3  \frac{J_{i_1, i_2, i_3, i_4} \tJ_{i_1, i_2, i_3,i_4}}
      { d_{i_1}^{(1)}d_{i_2}^{(2)}d_{i_3}^{(3)}d_{i_4}^{(4)}}  \,,
\end{equation}
and then apply the substitution in \eqn{JFourSubs} repeatedly until no
further terms are found to generate the terms needed to correct the
cut of the naive double copy.  This gives,
\begin{equation}
{\E}^{4\times 4\times 4\times 4}_\G = T_1 + T_2 + T_3\,,
\end{equation}
where 
\be
T_1&=&
-  \sum_{i, j=1}^3\left(\frac{J_{\x1, 1, i, j}\tJ_{1, \x2, i, j} }
  {d^{(1)}_1 d^{(2)}_1 d^{(3)}_{i} d^{(4)}_{j} }
+\frac{J_{\x1, i, 1, j}\tJ_{1, i, \x3, j} }
  {d^{(1)}_1 d^{(2)}_{i} d^{(3)}_{1} d^{(4)}_{j}}
+\frac{J_{\x1, i,  j, 1}\tJ_{1, i,j, \x4} }
  {d^{(1)}_1 d^{(2)}_{i} d^{(3)}_{j} d^{(4)}_{1}}+\frac{J_{i, \x2,  1, j}\tJ_{i, 1, \x3, j} }
  {d^{(1)}_{i} d^{(2)}_{1} d^{(3)}_{1} d^{(4)}_{j}}\right.\\
&&\hspace{4cm}\left. \null +
\frac{J_{i, \x2,  j, 1}\tJ_{i, 1, j, \x4} }
  {d^{(1)}_{i} d^{(2)}_{1} d^{(3)}_{j} d^{(4)}_{1}}
+\frac{J_{i, j,  \x3, 1}\tJ_{i, j, 1, \x4} }
  {d^{(1)}_{i} d^{(2)}_{j} d^{(3)}_{1} d^{(4)}_{1}}\right) + \{J\leftrightarrow\tJ\} \,, \cr
%%%%
T_2&=& \sum_{i=1}^3 \left(\frac{J_{1, 1, \x3, i} \tJ_{\x1, \x2, 1, i}+J_{1, \x2, 1, i} \tJ_{\x1, 1, \x3, i} +J_{\x1, 1, 1, i} \tJ_{1, \x2, \x3, i} }
  {d^{(1)}_1 d^{(2)}_1 d^{(3)}_1 d^{(4)}_{i}}\right.\\
%%%%%%%%%%%%%%%
&&\hspace{3cm} \null +
\frac{J_{1, 1, i, \x4} \tJ_{\x1, \x2, i, 1}+J_{1, \x2, i, 1} \tJ_{\x1, 1, i, \x4}+J_{\x1, 1, i, 1} \tJ_{1, \x2, i, \x4}  }{d^{(1)}_1 d^{(2)}_1 d^{(3)}_{i} d^{(4)}_1}\cr
%%%%%%%%%%%%%%%
&&\hspace{3cm}\null +
\frac{J_{1, i, 1, \x4} \tJ_{\x1, i, \x3, 1} +J_{1, i, \x3, 1} \tJ_{\x1, i, 1, \x4}+J_{\x1, i, 1, 1} \tJ_{1, i, \x3, \x4} }{d^{(1)}_1 d^{(2)}_{i} d^{(3)}_{1} d^{(4)}_1}\cr
%%%%%%%%%%%%%%%
&&\hspace{3cm}\null +\left.
\frac{J_{i, 1, 1, \x4} \tJ_{i,  \x2, \x3, 1} +J_{i, 1, \x3, 1} \tJ_{i,  \x2, 1, \x4}+J_{i, \x2, 1, 1} \tJ_{i, 1, \x3, \x4}}{d^{(1)}_{i} d^{(2)}_{1} d^{(3)}_{1} d^{(4)}_1}\right)
+ \{J\leftrightarrow\tJ\}
 \,, \cr
T_3&=&-\frac{1}{d^{(1)}_{1} d^{(2)}_{1} d^{(3)}_{1} d^{(4)}_1}\left(J_{\x1, 1, 1, 1}\tJ_{1, \x2, \x3, \x4}+J_{1, \x2, 1, 1}\tJ_{\x1, 1, \x3, \x4}+J_{1, 1, \x3, 1}\tJ_{\x1, \x2, 1, \x4}+J_{1, 1, 1, \x4}\tJ_{\x1, \x2, \x3, 1} \right.\\
&&\hspace{4cm}\null +\left. J_{1, 1, \x3, \x4}\tJ_{\x1, \x2, 1, 1}+J_{1, \x2, 1, \x4}\tJ_{\x1, 1, \x3, 1}+J_{1, \x2, \x3, 1}\tJ_{\x1, 1, 1, \x4}\right)
+ \{J\leftrightarrow\tJ\} \,.
\nonumber
\ee

By solving the generalized gauge transformations in term of the BCJ
discrepancy functions, we have explicitly confirmed that this indeed
is a solution for the extra contributions correcting the naive
double copy.  In fact, this pattern appears to continue for any number
additional four-point tree amplitudes in the cut.

%%%%%%%%%%%%%%%%%%%%%%%%%%
\subsection{One five-point and two four-point amplitudes in the cut} 

To obtain the ${5\times 4\times 4}$ case, we start with the expression
\begin{equation}
-\sum_{i=1}^{15} \sum_{j_2,j_3 = 1}^3
     \frac{J_{i,j_2, j_3} \tJ_{i,j_2, j_3}}
         {d_{i}^{(1,1)}d_{i}^{(1,2)} d^{(2)}_{j_2} d^{(3)}_{j_{3}}}\,.
\end{equation}
Applying the substitutions in \eqn{JFiveSubs} generates the terms
\begin{equation}
{\cal C}^{5 \times 4\times 4}_\G =
\sum_{i=1}^{15} \sum_{j_2,j_3 = 1}^3
     \frac{n_{i,j_2, j_3} \tn_{i,j_2, j_3}}
         {d_{i}^{(1,1)}d_{i}^{(1,2)} d^{(2)}_{j_2} d^{(3)}_{j_{3}}}
-  \sum_{i=1}^{6} T_i \,,
\end{equation}
where 
\begin{align}
% single $\times$ single:
T_1 &=
\sum_{i,j_2,j_3} \frac{1}{\mathcal{D}_{ij_2j_3}} \Big[\frac{1}{6} 
J_{\{i,1\},j_2, j_3} \tJ_{\{i,2\},j_2, j_3}
         +\left(-\frac{1}{3}\right)^2 
        J_{i,\x2, j_3} \tJ_{i,j_2 , \x3} \nonumber \\ 
& \hskip 3 cm  \null 
  +\sum_{h=1}^2 \left(
 a_i J_{\{i,h\},j_2, j_3} \tJ_{i,\x2, j_3}    +
         \, a_i J_{\{i,h\},j_2, j_3} \tJ_{i,j_2, \x3} \right)\Big]
+ \{J\leftrightarrow \tJ \} \,,\nonumber \\         
%
% single $\times$ double:
T_2 &=
\left(-\frac{1}{3}\right)\times\frac{1}{6}
\sum_{{h_1\ne h_2}\atop{h_1, h_2 = 1}}^2\sum_{i,j_2,j_3}\frac{1}{\mathcal{D}_{ij_2j_3}}\left(
 J_{\{i,h_1\},j_2, j_3} \tJ_{\{i,h_2\},\x2, j_3} 
  + J_{\{i,h_1\},j_2, j_3} \tJ_{\{i,h_2\},j_2, \x3}\right) +  \{J\leftrightarrow \tJ \} \,, \nonumber \\
T_3 &=\left(-\frac{1}{3}\right)\times\sum_{h=1}^2\sum_{i,j_2,j_3}\frac{1}{\mathcal{D}_{ij_2j_3}} \left(
\, a_i J_{\{i,h\},j_2, j_3} \tJ_{i,\x2, \x3}  +
a_i J_{i,\x2, j_3} \tJ_{\{i, h\},j_2, \x3}   +
\, a_i J_{i,j_2, \x3} \tJ_{\{i, h\},j_2, \x3}  \right) \nonumber \\
 & \hskip 4 cm \null  + \{J\leftrightarrow \tJ \} \,, \nonumber\\
T_4&=(-1)\sum_{{h_1\not = h_2}\atop {h_1,h_2=1}}^2 \sum_{i,j_2,j_3} \frac{1}{\mathcal{D}_{ij_2j_3}} \left(
a^{(h_1)}_i     J_{i,\x2, j_3} \tJ_{\{i, h_1, h_2\},j_2, j_3} +
 a^{(h_1)}_i     J_{i,j_2, \x3} \tJ_{\{i, h_1, h_2\},j_2, j_3} \right) + \{J\leftrightarrow \tJ \} 
\,, \nonumber \\
% single $\times$ triple:
T_{5}&=\hspace{-0.2cm}
\sum_{{h_1 \not = h_2} \atop {h_1,h_2=1}}^2 \sum_{i,j_2,j_3}\frac{1}{\mathcal{D}_{ij_2j_3}}
\Bigg[\left(-\frac{1}{3}\right)^2\times\frac{1}{6} J_{\{i, h_1\},j_2, j_3} \tJ_{\{i, h_2\},\x2, \x3} 
+\left(-\frac{1}{3}\right)\times (-1)a^{(h_1)}_i J_{i,j_2, \x3} \tJ_{\{i, h_1, h_2\},\x2, j_3}
 \Bigg] \nonumber \\
 & \hskip 4 cm \null + \{J\leftrightarrow \tJ \} \,, \nonumber \\
% double $\times$ double:
T_{6}&=\hspace{-0.2cm}
\sum_{{h_1 \not = h_2}\atop{h_1,h_2=1}}^2 
 \sum_{i,j_2,j_3} \frac{1}{\mathcal{D}_{ij_2j_3}}\Bigg[ \left(-\frac{1}{3}\right)^2\times\frac{1}{6}
J_{\{i, h_1\},j_2, \x3} \tJ_{\{i, h_2\},\x2, j_3} +\left(-\frac{1}{3}\right)\times (-1) a^{(h_1)}_i J_{i,\x2, \x3} \tJ_{\{i, h_1, h_2\},j_2, j_3} \Bigg]
 \nonumber \\
& \null \hskip 4cm + \{J\leftrightarrow \tJ \} \,, 
\end{align}
where we use the shorthand notation,
\be
 \sum_{i,j_2,j_3}&\equiv& \sum_{i=1}^{15}\sum_{j_2, j_3 =1}^3\,,
\hskip 2cm 
\mathcal{D}_{ij_2j_3}\equiv 
d_{i}^{(1,1)}d_{i}^{(1,2)} d^{(2)}_{j_2}d^{(3)}_{j_3}\,.
\ee 
The $a_i$, $a^{(1)}_i$ and $a^{(2)}_i$ coefficients are the same as for the $5\times 4$ cut, given 
in \tab{Solution5x4Table}.

%%%%%%%%%%%%%%%%%%%%%%%%%%%%%%%%%%
\section{Direct evaluation of five-loop cut vacuum integrals}
\label{sec:IntegrationAppendix}

In this appendix we present an alternative direct integration of the
five-loop cut vacuum integrals.  We work out the crossed cube in
\fig{fig:Vacuum5Loop} in detail.  The task here is to evaluate
Eq.~\eqref{eq:baikov5LoopCut2}.

We set the integration region in Eq.~\eqref{eq:baikov5LoopCut2} to be
\begin{equation}
z_{13} >0, \quad z_{14}>0, \quad z_{15} > 0 \,. \label{eq:zBoundary}
\end{equation}
In the original uncut integral in the Baikov representation, the
boundary of the integration region is defined by the points at which
the Baikov polynomial vanishes, so there is no boundary term in
integration-by-parts identities \cite{Frellesvig2017aai, Bosma2017,
  Harley2017}. In the cut integrals, the boundary of the region
Eq.~\eqref{eq:zBoundary} is
\begin{equation}
\{(z_{13},z_{14},z_{15}) \, | z_{13}=0 \text{ or } z_{14}=0 \text{ or } z_{15}=0 \}\,,
\end{equation}
on which the Baikov polynomial $P(z_i)|_{\rm cut}$ in
Eq.~\eqref{eq:baikov5LoopCut} evaluates to zero, as is the case for
the uncut integral. This means the cut integrals will inherit IBP
identities of the uncut integrals, which is crucial for the
consistency of this approach and for demonstrating ultraviolet
cancellations. In Eq.~\eqref{eq:baikov5LoopCut2}, we make a change of
variables
\begin{equation}
z_{13}=z \alpha, \quad z_{14} = z \beta, \quad z_{15} = z (1- \alpha - \beta) \,,
\end{equation}
and factor out the overall integral independent of $y_1,y_2,y_3$ 
(with the $\epsilon$ dependence reinstated for the purpose of illustration),
\begin{equation}
\int_0^\infty \frac{dz}{z^{1+5\epsilon}}\,,
\end{equation}
whose ultraviolet divergence is $-1/(5\epsilon)$. This leaves us with an integral
\begin{align}
&\quad \int_0^1 d\alpha \int_0^{1-\alpha} d\beta \, \alpha^{y_1} \beta^{y_2} (1-\alpha-\beta)^{y_3} \left[ 64 \, \alpha \beta (1-\alpha-\beta)(1-\beta) \right]^{-(3+y_1+y_2+y_3)/5} \nn \\
&= (64)^{-(3+y_1+y_2+y_3)/5} \frac{\Gamma \left( \big(1 +2y_1-3y_2 +2y_3\big)/5 \right)} {\Gamma\left( \big(3+y_1+y_2+y_3 \big) / 5 \right) \Gamma\left( \big(4+3y_1-2y_2+3y_3 \big) / 5 \right)} \nn \\
&\quad \times \Gamma \left( \big(2+4y_1-y_2-y_3\big)/5 \right)\Gamma \left( \big(2 - y_1 +4y_2 -y_3\big)/5 \right) \Gamma \left( \big(2 -y_1 -y_2 +4y_3\big)/5 \right), \label{eq:crossedCubeGammaFuncs}
\end{align}
which evaluates to non-singular values with the values of $(y_1,y_2,y_3)$ appearing in our calculation.

The top-level master integral for the crossed cube
topology, $V^{(\NP)}$, is defined as the integral with a unit numerator and with no
propagator denominator raised to more than its first power. The
coefficient of the top-level master integral is obtained by dividing
Eq.~\eqref{eq:crossedCubeGammaFuncs} by its value at $y_1=y_2=y_3=0$.
This method gives exactly the same results for the coefficients of the 
top-level crossed cube integral as the IBP method outlined in \sect{sec:vacuumExpansion}. 
As before, adding up all contributions to the coefficients of crossed-cube gives
a vanishing result, providing a nontrivial check on the integrand.

For the planar cube topology, the Baikov polynomial no longer
factorizes into linear polynomials as in
Eq.~\eqref{eq:baikovPolyCrossedCubeMaxCut}, so direct integration to
obtain a closed form expression is more difficult. This is not a
problem for the IBP reduction method in \sect{sec:vacuumExpansion} which
is sufficient for our purposes.  In any case, this direct approach 
gives a powerful alternative for dealing with five-loop vacuum integrals.

%%%%%%%%%%%%%%%%%%%%


\begin{thebibliography}{99}

%+% 3 refs
\bibitem{GeneralizedUnitarity}
Z.~Bern, L.~J.~Dixon, D.~C.~Dunbar and D.~A.~Kosower,
%``One loop n point gauge theory amplitudes, unitarity and collinear limits,''
Nucl.\ Phys.\ B {\bf 425}, 217 (1994)
%doi:10.1016/0550-3213(94)90179-1
[hep-ph/9403226];\\
%%CITATION = doi:10.1016/0550-3213(94)90179-1;%%
%
Z.~Bern, L.~J.~Dixon, D.~C.~Dunbar and D.~A.~Kosower,
%``Fusing gauge theory tree amplitudes into loop amplitudes,''
Nucl.\ Phys.\ B {\bf 435}, 59 (1995)
%doi:10.1016/0550-3213(94)00488-Z
[hep-ph/9409265].
%%CITATION = doi:10.1016/0550-3213(94)00488-Z;%%

%+% 5 refs
\bibitem{MaximalCutMethod}
Z.~Bern, J.~J.~M.~Carrasco, H.~Johansson and D.~A.~Kosower,
%``Maximally supersymmetric planar Yang-Mills amplitudes at five loops,''
Phys.\ Rev.\ D {\bf 76}, 125020 (2007)
%doi:10.1103/PhysRevD.76.125020
[arXiv:0705.1864 [hep-th]].
%%CITATION = doi:10.1103/PhysRevD.76.125020;%%

%+% 10 refs
\bibitem{BCJ}
Z.~Bern, J.~J.~M.~Carrasco and H.~Johansson,
%``New Relations for Gauge-Theory Amplitudes,''
Phys.\ Rev.\ D {\bf 78}, 085011 (2008)
%doi:10.1103/PhysRevD.78.085011
[arXiv:0805.3993 [hep-ph]].
%%CITATION = doi:10.1103/PhysRevD.78.085011;%%

%+% 10 refs
\bibitem{BCJLoop}
Z.~Bern, J.~J.~M.~Carrasco and H.~Johansson,
%``Perturbative Quantum Gravity as a Double Copy of Gauge Theory,''
Phys.\ Rev.\ Lett.\  {\bf 105}, 061602 (2010)
%doi:10.1103/PhysRevLett.105.061602
[arXiv:1004.0476 [hep-th]].
%%CITATION = doi:10.1103/PhysRevLett.105.061602;%%

%+% 1 ref
\bibitem{SmirnovBook}
V.~A.~Smirnov,
  {\it Analytic tools for Feynman integrals,}
  Springer Tracts Mod.\ Phys.\  {\bf 250}, 1 (2012).
%  doi:10.1007/978-3-642-34886-0
  %%CITATION = doi:10.1007/978-3-642-34886-0;%%

%+% 2 refs
\bibitem{IBPRefs}
K.G.~Chetyrkin and F.V.~Tkachov,
%``Integration by parts: the algorithm to calculate beta functions 
%in 4 loops,''                                                                   
Nucl.\ Phys.\ B {\bf 192}, 159 (1981);\\
%%CITATION = NUPHA,B192,159;%%
%
K.~G.~Chetyrkin and F.~V.~Tkachov,
%``Integration by Parts: The Algorithm to Calculate beta Functions in 4 Loops,''
Nucl.\ Phys.\ B {\bf 192}, 159 (1981);\\
%doi:10.1016/0550-3213(81)90199-1;
%%CITATION = doi:10.1016/0550-3213(81)90199-1;%%
%
S.~Laporta,
%``High precision calculation of multiloop Feynman integrals by difference equations,''
Int.\ J.\ Mod.\ Phys.\ A {\bf 15}, 5087 (2000)
%doi:10.1016/S0217-751X(00)00215-7, 10.1142/S0217751X00002157
[hep-ph/0102033];\\
%%CITATION = doi:10.1016/S0217-751X(00)00215-7, 10.1142/S0217751X00002157;%%
%
S.~Laporta and E.~Remiddi,
%``The Analytical value of the electron (g-2) at order alpha**3 in QED,''
Phys.\ Lett.\ B {\bf 379}, 283 (1996)
%doi:10.1016/0370-2693(96)00439-X
[hep-ph/9602417];\\
%%CITATION = doi:10.1016/0370-2693(96)00439-X;%%
%
C.~Anastasiou and A.~Lazopoulos,
%``Automatic integral reduction for higher order perturbative calculations,''
JHEP {\bf 0407}, 046 (2004)
%doi:10.1088/1126-6708/2004/07/046
[hep-ph/0404258];\\
%%CITATION = doi:10.1088/1126-6708/2004/07/046;%%
%
A.~V.~Smirnov,
%``FIRE5: a C++ implementation of Feynman Integral REduction,''
Comput.\ Phys.\ Commun.\  {\bf 189}, 182 (2015)
%doi:10.1016/j.cpc.2014.11.024
[arXiv:1408.2372 [hep-ph]];\\
%%CITATION = doi:10.1016/j.cpc.2014.11.024;%%
%
A.~von Manteuffel and C.~Studerus,
%``Reduze 2 - Distributed Feynman Integral Reduction,''
arXiv:1201.4330 [hep-ph];\\
%%CITATION = ARXIV:1201.4330;%%
R.~N.~Lee,
%``Presenting LiteRed: a tool for the Loop InTEgrals REDuction,''
arXiv:1212.2685 [hep-ph];\\
%%CITATION = ARXIV:1212.2685;%%
%
B.~Ruijl, T.~Ueda and J.~A.~M.~Vermaseren,
%``Forcer, a FORM program for the parametric reduction of four-loop massless propagator diagrams,'' 
arXiv:1704.06650 [hep-ph];\\
%%CITATION = ARXIV:1704.06650;%%
%
P.~Maierhoefer, J.~Usovitsch and P.~Uwer,
%``Kira - A Feynman Integral Reduction Program,''
arXiv:1705.05610 [hep-ph].
%%CITATION = ARXIV:1705.05610;%%

%+% 5 refs
\bibitem{IBPAdvances}
% bibitem{Gluza:2010ws}
J.~Gluza, K.~Kajda and D.~A.~Kosower,
%``Towards a Basis for Planar Two-Loop Integrals,''  
Phys.\ Rev.\ D {\bf 83}, 045012 (2011)
%doi:10.1103/PhysRevD.83.045012
[arXiv:1009.0472 [hep-th]];\\
%%CITATION = doi:10.1103/PhysRevD.83.045012;%%
%
%bibitem{Schabinger:2011dz}
R.~M.~Schabinger,
%``A New Algorithm For The Generation Of Unitarity-Compatible Integration By Parts Relations,''
JHEP {\bf 1201}, 077 (2012)
%doi:10.1007/JHEP01(2012)077
[arXiv:1111.4220 [hep-ph]];\\
%%CITATION = doi:10.1007/JHEP01(2012)077;%%
%
%bibitem{Sogaard:2014jla}
M.~S{\o}gaard and Y.~Zhang,
%``Elliptic Functions and Maximal Unitarity,''                                
Phys.\ Rev.\ D {\bf 91}, no. 8, 081701 (2015)
%doi:10.1103/PhysRevD.91.081701
[arXiv:1412.5577 [hep-th]];\\
 %%CITATION = doi:10.1103/PhysRevD.91.081701;%%
%
%{Georgoudis:2015hca}                                                      
A.~Georgoudis and Y.~Zhang,
  %``Two-loop Integral Reduction from Elliptic and Hyperelliptic Curves,''      
  JHEP {\bf 1512}, 086 (2015)
%  doi:10.1007/JHEP12(2015)086
  [arXiv:1507.06310 [hep-th]];\\
  %%CITATION = doi:10.1007/JHEP12(2015)086;%%
%
%bibitem{Ita:2015tya}
H.~Ita,
%``Two-loop Integrand Decomposition into Master Integrals and Surface Terms,''
Phys.\ Rev.\ D {\bf 94}, no. 11, 116015 (2016),
%doi:10.1103/PhysRevD.94.116015
[arXiv:1510.05626 [hep-th]];\\
%%CITATION = doi:10.1103/PhysRevD.94.116015;%%
%
%\bibitem{Georgoudis:2016wff} 
  A.~Georgoudis, K.~J.~Larsen and Y.~Zhang,
  %``Azurite: An algebraic geometry based package for finding bases of loop integrals,''
  Comput.\ Phys.\ Commun.\  {\bf 221}, 203 (2017)
  %doi:10.1016/j.cpc.2017.08.013
  [arXiv:1612.04252 [hep-th]];\\
  %%CITATION = doi:10.1016/j.cpc.2017.08.013;%%
%
H.~Ita,
%``Towards a Numerical Unitarity Approach for Two-loop Amplitudes in QCD,''
PoS LL {\bf 2016}, 080 (2016)
[arXiv:1607.00705 [hep-ph]];\\
%%CITATION = ARXIV:1607.00705;%%
%
%\bibitem{Abreu:2017xsl} 
  S.~Abreu, F.~Febres Cordero, H.~Ita, M.~Jaquier, B.~Page and M.~Zeng,
  %``Two-Loop Four-Gluon Amplitudes with the Numerical Unitarity Method,''
  Phys.\ Rev.\ Lett.\  {\bf 119}, no. 14, 142001 (2017)
  %doi:10.1103/PhysRevLett.119.142001
  [arXiv:1703.05273 [hep-ph]].
  %%CITATION = doi:10.1103/PhysRevLett.119.142001;%%

%+% 2 refs
\bibitem{Larsen2015ped}
K.~J.~Larsen and Y.~Zhang,
%``Integration-by-parts reductions from unitarity cuts and algebraic geometry,''
Phys.\ Rev.\ D {\bf 93}, no. 4, 041701 (2016),
%doi:10.1103/PhysRevD.93.041701
[arXiv:1511.01071 [hep-th]].
%%CITATION = doi:10.1103/PhysRevD.93.041701;%%

%+% 6 refs
\bibitem{Zhang2016kfo}
Y.~Zhang,
%``Lecture Notes on Multi-loop Integral Reduction and Applied Algebraic Geometry,''
arXiv:1612.02249 [hep-th].
%%CITATION = ARXIV:1612.02249;%%

%+% 2 refs
\bibitem{ThreeFourloopN8}
Z.~Bern, J.~J.~Carrasco, L.~J.~Dixon, H.~Johansson, D.~A.~Kosower and R.~Roiban,
%``Three-Loop Superfiniteness of N=8 Supergravity,''
Phys.\ Rev.\ Lett.\  {\bf 98}, 161303 (2007)
%doi:10.1103/PhysRevLett.98.161303
[hep-th/0702112];\\
%%CITATION = doi:10.1103/PhysRevLett.98.161303;%%
%
Z.~Bern, J.~J.~M.~Carrasco, L.~J.~Dixon, H.~Johansson and R.~Roiban,
%``Manifest Ultraviolet Behavior for the Three-Loop Four-Point Amplitude of N=8 Supergravity,''
Phys.\ Rev.\ D {\bf 78}, 105019 (2008)
%doi:10.1103/PhysRevD.78.105019
[arXiv:0808.4112 [hep-th]];\\
%%CITATION = doi:10.1103/PhysRevD.78.105019;%%
%
Z.~Bern, J.~J.~Carrasco, L.~J.~Dixon, H.~Johansson and R.~Roiban,
%``The Ultraviolet Behavior of N=8 Supergravity at Four Loops,''
Phys.\ Rev.\ Lett.\  {\bf 103}, 081301 (2009)
%doi:10.1103/PhysRevLett.103.081301
[arXiv:0905.2326 [hep-th]].
%%CITATION = doi:10.1103/PhysRevLett.103.081301;%%

%+% 9 refs
\bibitem{SimplifyingBCJ}
Z.~Bern, J.~J.~M.~Carrasco, L.~J.~Dixon, H.~Johansson and R.~Roiban,
%``Simplifying Multiloop Integrands and Ultraviolet Divergences of Gauge Theory and Gravity Amplitudes,''
Phys.\ Rev.\ D {\bf 85}, 105014 (2012)
%doi:10.1103/PhysRevD.85.105014
[arXiv:1201.5366 [hep-th]].
%%CITATION = doi:10.1103/PhysRevD.85.105014;%%

%+% 4 refs
\bibitem{N4GravFourLoop}
Z.~Bern, S.~Davies, T.~Dennen, A.~V.~Smirnov and V.~A.~Smirnov,
%``Ultraviolet Properties of N=4 Supergravity at Four Loops,''
Phys.\ Rev.\ Lett.\  {\bf 111}, no. 23, 231302 (2013)
% doi:10.1103/PhysRevLett.111.231302
[arXiv:1309.2498 [hep-th]].
%%CITATION = doi:10.1103/PhysRevLett.111.231302;%%

%+% 5 refs
\bibitem{N4GravThreeLoops}
Z.~Bern, S.~Davies, T.~Dennen and Y.~t.~Huang,
%``Absence of Three-Loop Four-Point Divergences in N=4 Supergravity,''
Phys.\ Rev.\ Lett.\  {\bf 108}, 201301 (2012)
%doi:10.1103/PhysRevLett.108.201301
[arXiv:1202.3423 [hep-th]].
 %%CITATION = doi:10.1103/PhysRevLett.108.201301;%%

%+% 5 refs
\bibitem{N4SugraMatter}
Z.~Bern, S.~Davies and T.~Dennen,
  %``The Ultraviolet Structure of Half-Maximal Supergravity with Matter Multiplets at Two and Three Loops,''
  Phys.\ Rev.\ D {\bf 88}, 065007 (2013)
%  doi:10.1103/PhysRevD.88.065007
  [arXiv:1305.4876 [hep-th]].
  %%CITATION = doi:10.1103/PhysRevD.88.065007;%%

%+% 7 refs
\bibitem{N5GravFourLoops}
Z.~Bern, S.~Davies and T.~Dennen,
%``Enhanced ultraviolet cancellations in $\mathcal N=5$ supergravity at four loops,''
Phys.\ Rev.\ D {\bf 90}, no. 10, 105011 (2014)
%doi:10.1103/PhysRevD.90.105011
[arXiv:1409.3089 [hep-th]].
%%CITATION = doi:10.1103/PhysRevD.90.105011;%%

%+% 11 refs
\bibitem{GeneralizedDoubleCopy}
Z.~Bern, J.~J.~Carrasco, W.~M.~Chen, H.~Johansson and R.~Roiban,
%``Gravity Amplitudes as Generalized Double Copies of Gauge-Theory Amplitudes,''
Phys.\ Rev.\ Lett.\  {\bf 118}, no. 18, 181602 (2017)
%doi:10.1103/PhysRevLett.118.181602
[arXiv:1701.02519 [hep-th]].
%%CITATION = doi:10.1103/PhysRevLett.118.181602;%%

%+% 1 ref
\bibitem{NeqEightSugra}
E.~Cremmer and B.~Julia,
%``The N=8 Supergravity Theory. 1. The Lagrangian,''
Phys.\ Lett.\  {\bf 80B}, 48 (1978);\\
%  doi:10.1016/0370-2693(78)90303-9;
 %%CITATION = doi:10.1016/0370-2693(78)90303-9;%%
%
% E.~Cremmer and B.~Julia,
  %``The SO(8) Supergravity,''
Nucl.\ Phys.\ B {\bf 159}, 141 (1979).
%  doi:10.1016/0550-3213(79)90331-6
%%CITATION = doi:10.1016/0550-3213(79)90331-6;%%

%+% 1 ref
\bibitem{OtherUnitarity}
J.~L.~Bourjaily, E.~Herrmann and J.~Trnka,
%``Prescriptive Unitarity,''
JHEP {\bf 1706}, 059 (2017)
%doi:10.1007/JHEP06(2017)059
[arXiv:1704.05460 [hep-th]];\\
%%CITATION = doi:10.1007/JHEP06(2017)059;%%
%
N.~Arkani-Hamed, J.~L.~Bourjaily, F.~Cachazo, S.~Caron-Huot and J.~Trnka,
%``The All-Loop Integrand For Scattering Amplitudes in Planar N=4 SYM,''
JHEP {\bf 1101}, 041 (2011)
%doi:10.1007/JHEP01(2011)041
[arXiv:1008.2958 [hep-th]].
%%CITATION = doi:10.1007/JHEP01(2011)041;%%

%+% 1 ref
\bibitem{Bootstrap}
 L.~J.~Dixon and M.~von Hippel,
  %``Bootstrapping an NMHV amplitude through three loops,''
  JHEP {\bf 1410}, 065 (2014)
%  doi:10.1007/JHEP10(2014)065
  [arXiv:1408.1505 [hep-th]];\\
  %%CITATION = doi:10.1007/JHEP10(2014)065;%%
%
 L.~J.~Dixon, M.~von Hippel and A.~J.~McLeod,
  %``The four-loop six-gluon NMHV ratio function,''
  JHEP {\bf 1601}, 053 (2016)
%  doi:10.1007/JHEP01(2016)053
  [arXiv:1509.08127 [hep-th]];\\
  %%CITATION = doi:10.1007/JHEP01(2016)053;%%
%
 S.~Caron-Huot, L.~J.~Dixon, A.~McLeod and M.~von Hippel,
  %``Bootstrapping a Five-Loop Amplitude Using Steinmann Relations,''
  Phys.\ Rev.\ Lett.\  {\bf 117}, no. 24, 241601 (2016)
%  doi:10.1103/PhysRevLett.117.241601
  [arXiv:1609.00669 [hep-th]].
  %%CITATION = doi:10.1103/PhysRevLett.117.241601;%%

%+% 2 refs
\bibitem{DoubleCopyTheories}
J.~J.~M.~Carrasco, M.~Chiodaroli, M.~Gunaydin and R.~Roiban,
%``One-loop four-point amplitudes in pure and matter-coupled $N \le 4$ supergravity,''
JHEP {\bf 1303}, 056 (2013)
% doi:10.1007/JHEP03(2013)056
[arXiv:1212.1146 [hep-th]];\\
  %%CITATION = doi:10.1007/JHEP03(2013)056;%%
  %
M.~Chiodaroli, M.~Gunaydin, H.~Johansson and R.~Roiban,
%``Scattering amplitudes in $ \mathcal{N}=2 $ Maxwell-Einstein and Yang-Mills/Einstein supergravity,''
JHEP {\bf 1501}, 081 (2015)
%doi:10.1007/JHEP01(2015)081
[arXiv:1408.0764 [hep-th]].
  %%CITATION = doi:10.1007/JHEP01(2015)081;%%

%+% 3 refs
\bibitem{FundMatter}
H.~Johansson and A.~Ochirov,
  %``Pure Gravities via Color-Kinematics Duality for Fundamental Matter,''
  JHEP {\bf 1511}, 046 (2015)
%  doi:10.1007/JHEP11(2015)046
  [arXiv:1407.4772 [hep-th]];\\
  %%CITATION = doi:10.1007/JHEP11(2015)046;%%
%
H.~Johansson and A.~Ochirov,
%``Color-Kinematics Duality for QCD Amplitudes,''
JHEP {\bf 1601}, 170 (2016)
%doi:10.1007/JHEP01(2016)170
[arXiv:1507.00332 [hep-ph]].
%%CITATION = doi:10.1007/JHEP01(2016)170;%%

%+% 3 refs
\bibitem{DoubleCopyTheoriesFund}
  M.~Chiodaroli, M.~Gunaydin, H.~Johansson and R.~Roiban,
  %``Spontaneously Broken Yang-Mills-Einstein Supergravities as Double Copies,''
  JHEP {\bf 1706}, 064 (2017)
  %doi:10.1007/JHEP06(2017)064
  [arXiv:1511.01740 [hep-th]];\\
  %%CITATION = doi:10.1007/JHEP06(2017)064;%%
%
M.~Chiodaroli, M.~Gunaydin H.~Johansson and R.~Roiban,
%``Complete construction of magical, symmetric and homogeneous N=2 supergravities as double copies of gauge theories,''
Phys.\ Rev.\ Lett.\  {\bf 117}, no. 1, 011603 (2016)
%doi:10.1103/PhysRevLett.117.011603
[arXiv:1512.09130 [hep-th]];\\
%%CITATION = doi:10.1103/PhysRevLett.117.011603;%
%
A.~Anastasiou, L.~Borsten, M.~J.~Duff, M.~J.~Hughes, A.~Marrani, S.~Nagy and M.~Zoccali,
  %``Twin supergravities from Yang-Mills theory squared,''
  Phys.\ Rev.\ D {\bf 96}, no. 2, 026013 (2017)
  %doi:10.1103/PhysRevD.96.026013
  [arXiv:1610.07192 [hep-th]];\\
  %%CITATION = doi:10.1103/PhysRevD.96.026013;%%
%
  A.~Anastasiou, L.~Borsten, M.~J.~Duff, A.~Marrani, S.~Nagy and M.~Zoccali,
  %``Are all supergravity theories Yang-Mills squared?,''
  arXiv:1707.03234 [hep-th].
  %%CITATION = ARXIV:1707.03234;%%
  %2 citations counted in INSPIRE as of 17 Aug 2017

%+% 1 ref
\bibitem{Conformal}
  H.~Johansson and J.~Nohle,
  %``Conformal Gravity from Gauge Theory,''
  arXiv:1707.02965 [hep-th].
  %%CITATION = ARXIV:1707.02965;%%

%+% 4 refs
\bibitem{Square}
Z.~Bern, T.~Dennen, Y.~t.~Huang and M.~Kiermaier,
%``Gravity as the Square of Gauge Theory,''
Phys.\ Rev.\ D {\bf 82}, 065003 (2010)
%doi:10.1103/PhysRevD.82.065003
[arXiv:1004.0693 [hep-th]].
%%CITATION = doi:10.1103/PhysRevD.82.065003;%%

%+% 2 refs
\bibitem{KiermaierTalk}
 M.~Kiermaier\footnote{\url{http://www.strings.ph.qmul.ac.uk/~theory/Amplitudes2010/Talks/MK2010.pdf}},
 Amplitudes 2010, Queen Mary, University of London.

%+% 3 refs
\bibitem{BjerrumMomKernel}
N.~E.~J.~Bjerrum-Bohr, P.~H.~Damgaard, T.~Sondergaard and P.~Vanhove,
%``The Momentum Kernel of Gauge and Gravity Theories,''
JHEP {\bf 1101}, 001 (2011)
%  doi:10.1007/JHEP01(2011)001
[arXiv:1010.3933 [hep-th]].
  %%CITATION = doi:10.1007/JHEP01(2011)001;%%

%+% 1 ref
\bibitem{BCJTreeProof}
C.~R.~Mafra, O.~Schlotterer and S.~Stieberger,
%``Explicit BCJ Numerators from Pure Spinors,''
JHEP {\bf 1107}, 092 (2011)
%  doi:10.1007/JHEP07(2011)092
[arXiv:1104.5224 [hep-th]];\\
  %%CITATION = doi:10.1007/JHEP07(2011)092;%%
%
Y.~J.~Du and C.~H.~Fu,
  %``Explicit BCJ numerators of nonlinear sigma model,''
JHEP {\bf 1609}, 174 (2016)
%  doi:10.1007/JHEP09(2016)174
[arXiv:1606.05846 [hep-th]];\\
%%CITATION = doi:10.1007/JHEP09(2016)174;%%
%
N.~E.~J.~Bjerrum-Bohr, J.~L.~Bourjaily, P.~H.~Damgaard and B.~Feng,
  %``Manifesting Color-Kinematics Duality in the Scattering Equation Formalism,''
  JHEP {\bf 1609}, 094 (2016)
  %doi:10.1007/JHEP09(2016)094
  [arXiv:1608.00006 [hep-th]];\\
  %%CITATION = doi:10.1007/JHEP09(2016)094;%%
%
Y.~J.~Du and F.~Teng,
  %``BCJ numerators from reduced Pfaffian,''
  JHEP {\bf 1704}, 033 (2017)
  %doi:10.1007/JHEP04(2017)033
  [arXiv:1703.05717 [hep-th]];\\
  %%CITATION = doi:10.1007/JHEP04(2017)033;%%
  %2 citations counted in INSPIRE as of 17 Aug 2017
%
  Y.~J.~Du, B.~Feng and F.~Teng,
  %``Expansion of All Multitrace Tree Level EYM Amplitudes,''
  arXiv:1708.04514 [hep-th].
  %%CITATION = ARXIV:1708.04514;%%

%+% 1 ref
\bibitem{BCJRelationProof}
N.~E.~J.~Bjerrum-Bohr, P.~H.~Damgaard and P.~Vanhove,
  %``Minimal Basis for Gauge Theory Amplitudes,''
  Phys.\ Rev.\ Lett.\  {\bf 103}, 161602 (2009)
  %doi:10.1103/PhysRevLett.103.161602
  [arXiv:0907.1425 [hep-th]];\\
  %%CITATION = doi:10.1103/PhysRevLett.103.161602;%%
  %207 citations counted in INSPIRE as of 17 Aug 2017
  %
  S.~Stieberger,
  %``Open & Closed vs. Pure Open String Disk Amplitudes,''
  arXiv:0907.2211 [hep-th];\\
  %%CITATION = ARXIV:0907.2211;%%
  %224 citations counted in INSPIRE as of 17 Aug 2017
  %
  Y.~X.~Chen, Y.~J.~Du and B.~Feng,
  %``A Proof of the Explicit Minimal-basis Expansion of Tree Amplitudes in Gauge Field Theory,''
  JHEP {\bf 1102}, 112 (2011)
  %doi:10.1007/JHEP02(2011)112
  [arXiv:1101.0009 [hep-th]];\\
  %%CITATION = doi:10.1007/JHEP02(2011)112;%%
  %
  L.~de la Cruz, A.~Kniss and S.~Weinzierl,
  %``Proof of the fundamental BCJ relations for QCD amplitudes,''
  JHEP {\bf 1509}, 197 (2015)
  % doi:10.1007/JHEP09(2015)197
  [arXiv:1508.01432 [hep-th]].
  %%CITATION = doi:10.1007/JHEP09(2015)197;%%
  %16 citations counted in INSPIRE as of 17 Aug 2017

%+% 1 ref
\bibitem{SelfDualYM}
R.~Monteiro and D.~O'Connell,
  %``The Kinematic Algebra From the Self-Dual Sector,''
  JHEP {\bf 1107}, 007 (2011)
  %doi:10.1007/JHEP07(2011)007
  [arXiv:1105.2565 [hep-th]].
  %%CITATION = doi:10.1007/JHEP07(2011)007;%%

%+% 3 refs
\bibitem{NLSMaction}
 C.~Cheung and C.~H.~Shen,
  %``Symmetry for Flavor-Kinematics Duality from an Action,''
  Phys.\ Rev.\ Lett.\  {\bf 118}, no. 12, 121601 (2017)
  %doi:10.1103/PhysRevLett.118.121601
  [arXiv:1612.00868 [hep-th]].
  %%CITATION = doi:10.1103/PhysRevLett.118.121601;%%

%+% 1 ref
\bibitem{OtherExamples}
Z.~Bern, C.~Boucher-Veronneau and H.~Johansson,
  %``N >= 4 Supergravity Amplitudes from Gauge Theory at One Loop,''
  Phys.\ Rev.\ D {\bf 84}, 105035 (2011)
%  doi:10.1103/PhysRevD.84.105035
  [arXiv:1107.1935 [hep-th]];\\
  %%CITATION = doi:10.1103/PhysRevD.84.105035;%%
%
 C.~Boucher-Veronneau and L.~J.~Dixon,
  %``N >- 4 Supergravity Amplitudes from Gauge Theory at Two Loops,''
  JHEP {\bf 1112}, 046 (2011)
% doi:10.1007/JHEP12(2011)046
  [arXiv:1110.1132 [hep-th]];\\
  %%CITATION = doi:10.1007/JHEP12(2011)046;%%
%
J.~J.~Carrasco and H.~Johansson,
  %``Five-Point Amplitudes in N=4 Super-Yang-Mills Theory and N=8 Supergravity,''
  Phys.\ Rev.\ D {\bf 85}, 025006 (2012)
%  doi:10.1103/PhysRevD.85.025006
  [arXiv:1106.4711 [hep-th]];\\
  %%CITATION = doi:10.1103/PhysRevD.85.025006;%%
%
Z.~Bern, S.~Davies, T.~Dennen, Y.~t.~Huang and J.~Nohle,
  %``Color-Kinematics Duality for Pure Yang-Mills and Gravity at One and Two Loops,''
  Phys.\ Rev.\ D {\bf 92}, no. 4, 045041 (2015)
%  doi:10.1103/PhysRevD.92.045041
  [arXiv:1303.6605 [hep-th]];\\
  %%CITATION = doi:10.1103/PhysRevD.92.045041;%%
%
 S.~He, R.~Monteiro and O.~Schlotterer,
  %``String-inspired BCJ numerators for one-loop MHV amplitudes,''
  JHEP {\bf 1601}, 171 (2016)
%  doi:10.1007/JHEP01(2016)171
  [arXiv:1507.06288 [hep-th]];\\
  %%CITATION = doi:10.1007/JHEP01(2016)171;%%
    %
 E.~Herrmann and J.~Trnka,
  %``Gravity On-shell Diagrams,''
  JHEP {\bf 1611}, 136 (2016)
%  doi:10.1007/JHEP11(2016)136
  [arXiv:1604.03479 [hep-th]].
  %%CITATION = doi:10.1007/JHEP11(2016)136;%%

%+% 1 ref
\bibitem{TwoLoopSQCD}
  H.~Johansson, G.~K{\"a}lin and G.~Mogull,
  %``Two-loop supersymmetric QCD and half-maximal supergravity amplitudes,''
  JHEP {\bf 1709}, 019 (2017)
  %doi:10.1007/JHEP09(2017)019
  [arXiv:1706.09381 [hep-th]].
  %%CITATION = doi:10.1007/JHEP09(2017)019;%%

%+% 1 ref
\bibitem{ClassicalSolutions}
R.~Monteiro, D.~O'Connell and C.~D.~White,
%``Black holes and the double copy,''
JHEP {\bf 1412}, 056 (2014)
%doi:10.1007/JHEP12(2014)056
[arXiv:1410.0239 [hep-th]];\\
 %%CITATION = doi:10.1007/JHEP12(2014)056;%%
%
A.~Luna, R.~Monteiro, D.~O'Connell and C.~D.~White,
%``The classical double copy for Taub-€"NUT spacetime,''
Phys.\ Lett.\ B {\bf 750}, 272 (2015)
%doi:10.1016/j.physletb.2015.09.021
[arXiv:1507.01869 [hep-th]];\\
%%CITATION = doi:10.1016/j.physletb.2015.09.021;%%
%
G.~Cardoso, S.~Nagy and S.~Nampuri,
%``Multi-centered $ \mathcal{N}=2 $ BPS black holes: a double copy description,''
JHEP {\bf 1704}, 037 (2017)
%  doi:10.1007/JHEP04(2017)037
[arXiv:1611.04409 [hep-th]];\\
 %%CITATION = doi:10.1007/JHEP04(2017)037;%%
%
T.~Adamo, E.~Casali, L.~Mason and S.~Nekovar,
%``Scattering on plane waves and the double copy,''
arXiv:1706.08925 [hep-th].
%%CITATION = ARXIV:1706.08925;%%

%+% 2 refs
\bibitem{RadiationSolutions}
A.~Luna, R.~Monteiro, I.~Nicholson, D.~O'Connell and C.~D.~White,
%``The double copy: Bremsstrahlung and accelerating black holes,''
JHEP {\bf 1606}, 023 (2016)
%doi:10.1007/JHEP06(2016)023
[arXiv:1603.05737 [hep-th]];\\
%%CITATION = doi:10.1007/JHEP06(2016)023;%%
%
W.~D.~Goldberger and A.~K.~Ridgway,
%``Radiation and the classical double copy for color charges,''
Phys.\ Rev.\ D {\bf 95}, no. 12, 125010 (2017)
% doi:10.1103/PhysRevD.95.125010
[arXiv:1611.03493 [hep-th]];\\
%%CITATION = doi:10.1103/PhysRevD.95.125010;%%
%
A.~Luna, R.~Monteiro, I.~Nicholson, A.~Ochirov, D.~O'Connell, N.~Westerberg and C.~D.~White,
  %``Perturbative spacetimes from Yang-Mills theory,''
  JHEP {\bf 1704}, 069 (2017)
  %doi:10.1007/JHEP04(2017)069
  [arXiv:1611.07508 [hep-th]];\\
  %%CITATION = doi:10.1007/JHEP04(2017)069;%%
%
%\bibitem{Goldberger:2017frp} 
  W.~D.~Goldberger, S.~G.~Prabhu and J.~O.~Thompson,
  %``Classical gluon and graviton radiation from the bi-adjoint scalar double copy,''
  Phys.\ Rev.\ D {\bf 96}, no. 6, 065009 (2017)
  %doi:10.1103/PhysRevD.96.065009
  [arXiv:1705.09263 [hep-th]].
  %%CITATION = doi:10.1103/PhysRevD.96.065009;%%

%+% 1 ref
\bibitem{Donoghue}
 N.~E.~J.~Bjerrum-Bohr, J.~F.~Donoghue, B.~R.~Holstein, L.~Planté and P.~Vanhove,
  %``Bending of Light in Quantum Gravity,''
  Phys.\ Rev.\ Lett.\  {\bf 114}, no. 6, 061301 (2015)
%  doi:10.1103/PhysRevLett.114.061301
  [arXiv:1410.7590 [hep-th]];\\
  %%CITATION = doi:10.1103/PhysRevLett.114.061301;%%
%
  N.~E.~J.~Bjerrum-Bohr, J.~F.~Donoghue, B.~R.~Holstein, L.~Plante and P.~Vanhove,
  %``Light-like Scattering in Quantum Gravity,''
  JHEP {\bf 1611}, 117 (2016)
%  doi:10.1007/JHEP11(2016)117
  [arXiv:1609.07477 [hep-th]];\\
  %%CITATION = doi:10.1007/JHEP11(2016)117;%%
%
%\bibitem{Bjerrum-Bohr:2017dxw} 
  N.~E.~J.~Bjerrum-Bohr, B.~R.~Holstein, J.~F.~Donoghue, L.~Planté and P.~Vanhove,
  %``Illuminating Light Bending,''
  PoS CORFU {\bf 2016}, 077 (2017)
  [arXiv:1704.01624 [gr-qc]].
  %%CITATION = ARXIV:1704.01624;%%

%+% 1 ref
\bibitem{SugraSyms}
L.~Borsten, M.~J.~Duff, L.~J.~Hughes and S.~Nagy,
%``Magic Square from Yang-Mills Squared,''
Phys.\ Rev.\ Lett.\  {\bf 112}, no. 13, 131601 (2014)
% doi:10.1103/PhysRevLett.112.131601
[arXiv:1301.4176 [hep-th]];\\
%%CITATION = doi:10.1103/PhysRevLett.112.131601;%%
%
A.~Anastasiou, L.~Borsten, M.~J.~Duff, L.~J.~Hughes and S.~Nagy,
%``A magic pyramid of supergravities,''
JHEP {\bf 1404}, 178 (2014)
%doi:10.1007/JHEP04(2014)178
[arXiv:1312.6523 [hep-th]];\\
%%CITATION = doi:10.1007/JHEP04(2014)178;%%
%
A.~Anastasiou, L.~Borsten, M.~J.~Duff, L.~J.~Hughes and S.~Nagy,
%``Yang-Mills origin of gravitational symmetries,''
Phys.\ Rev.\ Lett.\  {\bf 113}, no. 23, 231606 (2014)
%doi:10.1103/PhysRevLett.113.231606
[arXiv:1408.4434 [hep-th]].
%%CITATION = doi:10.1103/PhysRevLett.113.231606;%%

%+% 2 refs
\bibitem{MultiLoopFormFactor}
  R.~H.~Boels, B.~A.~Kniehl, O.~V.~Tarasov and G.~Yang,
  %``Color-kinematic Duality for Form Factors,''
  JHEP {\bf 1302}, 063 (2013)
  %doi:10.1007/JHEP02(2013)063
  [arXiv:1211.7028 [hep-th]];\\
  %%CITATION = doi:10.1007/JHEP02(2013)063;%%
  %71 citations counted in INSPIRE as of 17 Aug 2017
  %
G.~Yang,
%``Color-kinematics duality and Sudakov form factor at five loops for N=4 supersymmetric Yang-Mills theory,''
Phys.\ Rev.\ Lett.\  {\bf 117}, no. 27, 271602 (2016)
% doi:10.1103/PhysRevLett.117.271602
[arXiv:1610.02394 [hep-th]];\\
%%CITATION = doi:10.1103/PhysRevLett.117.271602;%%
%
  R.~H.~Boels, T.~Huber and G.~Yang,
  %``The four-loop non-planar cusp anomalous dimension in N = 4 SYM,''
  arXiv:1705.03444 [hep-th].
  %%CITATION = ARXIV:1705.03444;%%

%+% 1 ref
\bibitem{BLGBCJ}
T.~Bargheer, S.~He and T.~McLoughlin,
%``New Relations for Three-Dimensional Supersymmetric Scattering Amplitudes,''
Phys.\ Rev.\ Lett.\  {\bf 108}, 231601 (2012)
%doi:10.1103/PhysRevLett.108.231601
[arXiv:1203.0562 [hep-th]];\\
%%CITATION = doi:10.1103/PhysRevLett.108.231601;%%
%
Y.~t.~Huang and H.~Johansson,
%``Equivalent D=3 Supergravity Amplitudes from Double Copies of Three-Algebra and Two-Algebra Gauge Theories,''
Phys.\ Rev.\ Lett.\  {\bf 110}, 171601 (2013)
%doi:10.1103/PhysRevLett.110.171601
[arXiv:1210.2255 [hep-th]];\\
%%CITATION = doi:10.1103/PhysRevLett.110.171601;%%
%
Y.~t.~Huang, H.~Johansson and S.~Lee,
%``On Three-Algebra and Bi-Fundamental Matter Amplitudes and Integrability of Supergravity,''
JHEP {\bf 1311}, 050 (2013)
%doi:10.1007/JHEP11(2013)050
[arXiv:1307.2222 [hep-th]].
%%CITATION = doi:10.1007/JHEP11(2013)050;%%
%

%+% 2 refs
\bibitem{NLSMBCJ}
G.~Chen and Y.~J.~Du,
%``Amplitude Relations in Non-linear Sigma Model,''
JHEP {\bf 1401}, 061 (2014)
%doi:10.1007/JHEP01(2014)061
[arXiv:1311.1133 [hep-th]];\\
%%CITATION = doi:10.1007/JHEP01(2014)061;%%
%
F.~Cachazo, S.~He and E.~Y.~Yuan,
%``Scattering Equations and Matrices: From Einstein To Yang-Mills, DBI and NLSM,''
JHEP {\bf 1507}, 149 (2015)
% doi:10.1007/JHEP07(2015)149
[arXiv:1412.3479 [hep-th]];\\
%%CITATION = doi:10.1007/JHEP07(2015)149;%%
%
F.~Cachazo, P.~Cha and S.~Mizera,
%``Extensions of Theories from Soft Limits,''
JHEP {\bf 1606}, 170 (2016)
% doi:10.1007/JHEP06(2016)170
[arXiv:1604.03893 [hep-th]];\\
  %%CITATION = doi:10.1007/JHEP06(2016)170;%%
%
C.~R.~Mafra and O.~Schlotterer,
%``Non-abelian $Z$-theory: Berends-Giele recursion for the $\alpha'$-expansion of disk integrals,''
JHEP {\bf 1701}, 031 (2017)
% doi:10.1007/JHEP01(2017)031
[arXiv:1609.07078 [hep-th]];\\
%%CITATION = doi:10.1007/JHEP01(2017)031;%%
%
%\bibitem{Carrasco:2016ygv} 
  J.~J.~M.~Carrasco, C.~R.~Mafra and O.~Schlotterer,
  %``Semi-abelian Z-theory: NLSM$+?^{3}$ from the open string,''
  JHEP {\bf 1708}, 135 (2017)
  %doi:10.1007/JHEP08(2017)135
  [arXiv:1612.06446 [hep-th]];\\
  %%CITATION = doi:10.1007/JHEP08(2017)135;%%
%
C.~Cheung, C.~H.~Shen and C.~Wen,
%``Unifying Relations for Scattering Amplitudes,''
arXiv:1705.03025 [hep-th].
%%CITATION = ARXIV:1705.03025;%%

%+% 3 refs
\bibitem{abelianZ}
  J.~J.~M.~Carrasco, C.~R.~Mafra and O.~Schlotterer,
  %``Abelian Z-theory: NLSM amplitudes and $\alpha$'-corrections from the open string,''
  JHEP {\bf 1706}, 093 (2017)
%  doi:10.1007/JHEP06(2017)093
  [arXiv:1608.02569 [hep-th]].
  %%CITATION = doi:10.1007/JHEP06(2017)093;%%

%+% 1 ref
\bibitem{DiskandHeteroticStringBCJ}
  J.~Broedel, O.~Schlotterer and S.~Stieberger,
  %``Polylogarithms, Multiple Zeta Values and Superstring Amplitudes,''
  Fortsch.\ Phys.\  {\bf 61}, 812 (2013)
  %doi:10.1002/prop.201300019
  [arXiv:1304.7267 [hep-th]];\\
  %%CITATION = doi:10.1002/prop.201300019;%%
  %
  S.~Stieberger and T.~R.~Taylor,
  %``Closed String Amplitudes as Single-Valued Open String Amplitudes,''
  Nucl.\ Phys.\ B {\bf 881}, 269 (2014)
  %doi:10.1016/j.nuclphysb.2014.02.005
  [arXiv:1401.1218 [hep-th]];\\
  %%CITATION = doi:10.1016/j.nuclphysb.2014.02.005;%%
  %
  Y.~t.~Huang, O.~Schlotterer and C.~Wen,
  %``Universality in string interactions,''
  JHEP {\bf 1609}, 155 (2016)
  %doi:10.1007/JHEP09(2016)155
  [arXiv:1602.01674 [hep-th]].
  %%CITATION = doi:10.1007/JHEP09(2016)155;%%

%+% 2 refs
\bibitem{Review}
J.~J.~M.~Carrasco and H.~Johansson,
%``Generic multiloop methods and application to N=4 super-Yang-Mills,''
J.\ Phys.\ A {\bf 44}, 454004 (2011)
%doi:10.1088/1751-8113/44/45/454004
[arXiv:1103.3298 [hep-th]];\\
%%CITATION = doi:10.1088/1751-8113/44/45/454004;%%
%
J.~J.~M.~Carrasco,
%``Gauge and Gravity Amplitude Relations,''
%doi:10.1142/9789814678766\_0011
arXiv:1506.00974 [hep-th];\\
%%CITATION = doi:10.1142/9789814678766_0011;%%
%
M.~Chiodaroli,
%``Simplifying amplitudes in Maxwell-Einstein and Yang-Mills-Einstein supergravities,''
arXiv:1607.04129 [hep-th];\\
%%CITATION = ARXIV:1607.04129;%%
%
C.~Cheung,
  %``TASI Lectures on Scattering Amplitudes,''
  arXiv:1708.03872 [hep-ph].
  %%CITATION = ARXIV:1708.03872;%%

%+% 2 refs
\bibitem{BCJDifficulty}
Z.~Bern, S.~Davies and J.~Nohle,
%``Double-Copy Constructions and Unitarity Cuts,''
Phys.\ Rev.\ D {\bf 93}, no. 10, 105015 (2016)
%doi:10.1103/PhysRevD.93.105015
[arXiv:1510.03448 [hep-th]];\\
%%CITATION = doi:10.1103/PhysRevD.93.105015;%%
%
G.~Mogull and D.~O'Connell,
%``Overcoming Obstacles to Colour-Kinematics Duality at Two Loops,''
JHEP {\bf 1512}, 135 (2015)
%doi:10.1007/JHEP12(2015)135
[arXiv:1511.06652 [hep-th]].
%%CITATION = doi:10.1007/JHEP12(2015)135;%%

%+% 9 refs
\bibitem{FiveLoopN4}
Z.~Bern, J.~J.~M.~Carrasco, H.~Johansson and R.~Roiban,
%``The Five-Loop Four-Point Amplitude of N=4 super-Yang-Mills Theory,''
Phys.\ Rev.\ Lett.\  {\bf 109}, 241602 (2012)
%doi:10.1103/PhysRevLett.109.241602
[arXiv:1207.6666 [hep-th]].
%%CITATION = doi:10.1103/PhysRevLett.109.241602;%%

%+% 6 refs
\bibitem{AttachedFile}
See the ancillary files of this manuscript.

%+% 6 refs
\bibitem{Harley2017}
  M.~Harley, F.~Moriello and R.~M.~Schabinger,
  %``Baikov-Lee Representations Of Cut Feynman Integrals,''                     
  JHEP {\bf 1706}, 049 (2017)
%  doi:10.1007/JHEP06(2017)049
  [arXiv:1705.03478 [hep-ph]].
  %%CITATION = doi:10.1007/JHEP06(2017)049;%%

%+% 7 refs
\bibitem{Bosma2017}
  J.~Bosma, M.~S{\o}gaard and Y.~Zhang,
  %``Maximal Cuts in Arbitrary Dimension,''
  JHEP {\bf 1708}, 051 (2017)
  %doi:10.1007/JHEP08(2017)051
  [arXiv:1704.04255 [hep-th]].
  %%CITATION = doi:10.1007/JHEP08(2017)051;%%

%+% 4 refs
\bibitem{BjornssonAndGreen}
J.~Bj\"{o}rnsson and M.~B.~Green,
%``5 loops in 24/5 dimensions,''                                                
JHEP {\bf 1008}, 132 (2010) [arXiv:1004.2692 [hep-th]];\\
%%CITATION = JHEPA,1008,132;%%
%
J.~Bj\"ornsson,
%``Multi-loop amplitudes in maximally supersymmetric pure spinor                
% field theory,''                                                               
JHEP {\bf 1101}, 002 (2011)
[arXiv:1009.5906 [hep-th]].
%%CITATION = ARXIV:1009.5906;%%

%+% 3 refs
\bibitem{SevenLoopGravity}
M.~B.~Green, J.~G.~Russo and P.~Vanhove,
%``String-theory dualities and supergravity divergences,''                      
JHEP {\bf 1006}, 075 (2010) [arXiv:1002.3805 [hep-th]];\\
%
G.~Bossard, P.~S.~Howe and K.~S.~Stelle,
%``On duality symmetries of supergravity invariants,''                          
JHEP {\bf 1101}, 020 (2011) [arXiv:1009.0743 [hep-th]];\\
%%CITATION = ARXIV:1009.0743;%%
%
N.~Beisert, H.~Elvang, D.~Z.~Freedman, M.~Kiermaier, A.~Morales
and S.~Stieberger,
%``E7(7) constraints on counterterms in N=8 supergravity,''                     
Phys.\ Lett.\ B {\bf 694}, 265 (2010)
[arXiv:1009.1643 [hep-th]];\\
%%CITATION = ARXIV:1009.1643;%%
%
G.~Bossard, P.~S.~Howe, K.~S.~Stelle and P.~Vanhove,
%``The vanishing volume of D=4 superspace,''                                    
Class.\ Quant.\ Grav.\  {\bf 28}, 215005 (2011)
[arXiv:1105.6087 [hep-th]].
%%CITATION = ARXIV:1105.6087;%%

%+% 2 refs
\bibitem{KLT}
H.~Kawai, D.~C.~Lewellen and S.~H.~H.~Tye,
%``A Relation Between Tree Amplitudes of Closed and Open Strings,''             
Nucl.\ Phys.\ B {\bf 269}, 1 (1986).
%doi:10.1016/0550-3213(86)90362-7;                                            
%%CITATION = doi:10.1016/0550-3213(86)90362-7;%%

%+% 1 ref
\bibitem{BGK}
  F.~A.~Berends, W.~T.~Giele and H.~Kuijf,
  %``On relations between multi - gluon and multigraviton scattering,''
  Phys.\ Lett.\ B {\bf 211}, 91 (1988).
%  doi:10.1016/0370-2693(88)90813-1
  %%CITATION = doi:10.1016/0370-2693(88)90813-1;%%

%+% 3 refs
\bibitem{OneloopN8}
Z.~Bern, L.~J.~Dixon, M.~Perelstein and J.~S.~Rozowsky,
%``Multileg one loop gravity amplitudes from gauge theory,''
Nucl.\ Phys.\ B {\bf 546}, 423 (1999)
%doi:10.1016/S0550-3213(99)00029-2
[hep-th/9811140].
%%CITATION = doi:10.1016/S0550-3213(99)00029-2;%%

%+% 2 refs
\bibitem{BDDPR}
 Z.~Bern, L.~J.~Dixon, D.~C.~Dunbar, M.~Perelstein and J.~S.~Rozowsky,
  %``On the relationship between Yang-Mills theory and gravity and its implication for ultraviolet divergences,''
  Nucl.\ Phys.\ B {\bf 530}, 401 (1998)
%  doi:10.1016/S0550-3213(98)00420-9
  [hep-th/9802162];\\
  %%CITATION = doi:10.1016/S0550-3213(98)00420-9;%%
%
 Z.~Bern, L.~J.~Dixon, M.~Perelstein and J.~S.~Rozowsky,
  %``One loop n point helicity amplitudes in (selfdual) gravity,''
  Phys.\ Lett.\ B {\bf 444}, 273 (1998)
%  doi:10.1016/S0370-2693(98)01397-5
  [hep-th/9809160].
  %%CITATION = doi:10.1016/S0370-2693(98)01397-5;%%

%+% 1 ref
\bibitem{DDM}
V.~Del Duca, L.~J.~Dixon and F.~Maltoni,
%``New color decompositions for gauge amplitudes at tree and loop level,''
Nucl.\ Phys.\ B {\bf 571}, 51 (2000)
%doi:10.1016/S0550-3213(99)00809-3
[hep-ph/9910563].
%%CITATION = doi:10.1016/S0550-3213(99)00809-3;%%

%+% 1 ref
\bibitem{Nonplanar5PtN4}
Z.~Bern, E.~Herrmann, S.~Litsey, J.~Stankowicz and J.~Trnka,
%``Evidence for a Nonplanar Amplituhedron,''
JHEP {\bf 1606}, 098 (2016)
%doi:10.1007/JHEP06(2016)098
[arXiv:1512.08591 [hep-th]].
%%CITATION = doi:10.1007/JHEP06(2016)098;%%

%+% 1 ref
\bibitem{SuperSums}
H.~Elvang, D.~Z.~Freedman and M.~Kiermaier,
%``Recursion Relations, Generating Functions, and Unitarity Sums in N=4 SYM Theory,''
JHEP {\bf 0904}, 009 (2009)
%doi:10.1088/1126-6708/2009/04/009
[arXiv:0808.1720 [hep-th]];\\
  %%CITATION = doi:10.1088/1126-6708/2009/04/009;%%
%
Z.~Bern, J.~J.~M.~Carrasco, H.~Ita, H.~Johansson and R.~Roiban,
%``On the Structure of Supersymmetric Sums in Multi-Loop Unitarity Cuts,''
Phys.\ Rev.\ D {\bf 80}, 065029 (2009)
%doi:10.1103/PhysRevD.80.065029
[arXiv:0903.5348 [hep-th]].
%%CITATION = doi:10.1103/PhysRevD.80.065029;%%

%+% 2 refs
\bibitem{DDimensions}
Z.~Bern, J.~J.~Carrasco, T.~Dennen, Y.~t.~Huang and H.~Ita,
%``Generalized Unitarity and Six-Dimensional Helicity,''
Phys.\ Rev.\ D {\bf 83}, 085022 (2011)
%  doi:10.1103/PhysRevD.83.085022
[arXiv:1010.0494 [hep-th]].
  %%CITATION = doi:10.1103/PhysRevD.83.085022;%%

%+% 1 ref
\bibitem{BCJGaugeSym}
  R.~H.~Boels and R.~Medina,
  %``Graviton and gluon scattering from first principles,''
  Phys.\ Rev.\ Lett.\  {\bf 118}, no. 6, 061602 (2017)
  %doi:10.1103/PhysRevLett.118.061602
  [arXiv:1607.08246 [hep-th]];\\
  %%CITATION = doi:10.1103/PhysRevLett.118.061602;%%
%
R.~W.~Brown and S.~G.~Naculich,
%``BCJ relations from a new symmetry of gauge-theory amplitudes,'' 
JHEP {\bf 1610}, 130 (2016)
%doi:10.1007/JHEP10(2016)130
[arXiv:1608.04387 [hep-th]];\\
%%CITATION = doi:10.1007/JHEP10(2016)130;%%
%
N.~Arkani-Hamed, L.~Rodina and J.~Trnka,
%``Locality and Unitarity from Singularities and Gauge Invariance,''
arXiv:1612.02797 [hep-th].
%%CITATION = ARXIV:1612.02797;%%

%+% 1 ref
\bibitem{CGJR}
M.~Chiodaroli, M.~Gunaydin, H.~Johansson and R.~Roiban,
% ``Explicit Formulae for Yang-Mills-Einstein Amplitudes from the Double Copy,''
  JHEP {\bf 1707}, 002 (2017)
%  doi:10.1007/JHEP07(2017)002
  [arXiv:1703.00421 [hep-th]].
  %%CITATION = doi:10.1007/JHEP07(2017)002;%%

%+% 1 ref
\bibitem{KK}
R.~Kleiss and H.~Kuijf,
%``Multi - Gluon Cross-sections and Five Jet Production at Hadron Colliders,''
Nucl.\ Phys.\ B {\bf 312}, 616 (1989).
% doi:10.1016/0550-3213(89)90574-9
%%CITATION = doi:10.1016/0550-3213(89)90574-9;%%

%+% 2 refs
\bibitem{HenryConstraints}
S.~H.~Henry Tye and Y.~Zhang,
%``Dual Identities inside the Gluon and the Graviton Scattering Amplitudes,''
JHEP {\bf 1006}, 071 (2010)
Erratum: [JHEP {\bf 1104}, 114 (2011)]
%doi:10.1007/JHEP06(2010)071, 10.1007/JHEP04(2011)114
[arXiv:1003.1732 [hep-th]].
 %%CITATION = doi:10.1007/JHEP06(2010)071, 10.1007/JHEP04(2011)114;%%

%+% 1 ref
\bibitem{PierreConstraints}
N.~E.~J.~Bjerrum-Bohr, P.~H.~Damgaard, T.~Sondergaard and P.~Vanhove,
%``Monodromy and Jacobi-like Relations for Color-Ordered Amplitudes,''
JHEP {\bf 1006}, 003 (2010)
%doi:10.1007/JHEP06(2010)003
[arXiv:1003.2403 [hep-th]].
%%CITATION = doi:10.1007/JHEP06(2010)003;%%

%+% 1 ref
\bibitem{Nair}
V.~P.~Nair,
%``A Current Algebra for Some Gauge Theory Amplitudes,''
Phys.\ Lett.\ B {\bf 214}, 215 (1988).
%doi:10.1016/0370-2693(88)91471-2
%%CITATION = doi:10.1016/0370-2693(88)91471-2;%%


\bibitem{CompleteFourLoopSYM} 
  Z.~Bern, J.~J.~M.~Carrasco, L.~J.~Dixon, H.~Johansson and R.~Roiban,
  %``The Complete Four-Loop Four-Point Amplitude in N=4 Super-Yang-Mills Theory,''
  Phys.\ Rev.\ D {\bf 82}, 125040 (2010)
  %doi:10.1103/PhysRevD.82.125040
  [arXiv:1008.3327 [hep-th]].
  %%CITATION = doi:10.1103/PhysRevD.82.125040;%%

%+% 1 ref
\bibitem{Vladimirov}
A.~A.~Vladimirov,
%``Method For Computing Renormalization Group Functions In Dimensional
%Renormalization Scheme,''
Theor.\ Math.\ Phys.\  {\bf 43}, 417 (1980)
[Teor.\ Mat.\ Fiz.\  {\bf 43}, 210 (1980)];\\
%%CITATION = TMFZA,43,210;%%
%
N.~Marcus and A.~Sagnotti,
%``A Simple Method For Calculating Counterterms,''
Nuovo Cim.\ A {\bf 87}, 1 (1985).
%%CITATION = NUCIA,A87,1;%%

%+% 4 refs
\bibitem{IntegralRelations}
Z.~Bern, M.~Enciso, J.~Parra-Martinez and M.~Zeng,
%``Manifesting enhanced cancellations in supergravity: 
% integrands versus integrals,''
JHEP {\bf 1705}, 137 (2017)
%doi:10.1007/JHEP05(2017)137
[arXiv:1703.08927 [hep-th]].
%%CITATION = doi:10.1007/JHEP05(2017)137;%%

%+% 1 ref
\bibitem{FiveloopQCDBeta}
P.~A.~Baikov, K.~G.~Chetyrkin and J.~H.~K\"uhn,
 %``Five-Loop Running of the QCD coupling constant,''
Phys.\ Rev.\ Lett.\  {\bf 118}, no. 8, 082002 (2017)
%  doi:10.1103/PhysRevLett.118.082002
[arXiv:1606.08659 [hep-ph]];\\
%%CITATION = doi:10.1103/PhysRevLett.118.082002;%%
%
F.~Herzog, B.~Ruijl, T.~Ueda, J.~A.~M.~Vermaseren and A.~Vogt,
%``The five-loop beta function of Yang-Mills theory with fermions,''
JHEP {\bf 1702}, 090 (2017)
%  doi:10.1007/JHEP02(2017)090
[arXiv:1701.01404 [hep-ph]];\\
%%CITATION = doi:10.1007/JHEP02(2017)090;%%
%
T.~Luthe, A.~Maier, P.~Marquard and Y.~Schroder,
%``Complete renormalization of QCD at five loops,''
JHEP {\bf 1703}, 020 (2017)
%  doi:10.1007/JHEP03(2017)020
[arXiv:1701.07068 [hep-ph]].
%%CITATION = doi:10.1007/JHEP03(2017)020;%%

%+% 1 ref
\bibitem{CutIntegrals}
% Kosower:2011ty
D.~A.~Kosower and K.~J.~Larsen,
%``Maximal Unitarity at Two Loops,''                                          
Phys.\ Rev.\ D {\bf 85}, 045017 (2012)
%doi:10.1103/PhysRevD.85.045017
[arXiv:1108.1180 [hep-th]];\\
%%CITATION = doi:10.1103/PhysRevD.85.045017;%%
%
%CaronHuot:2012ab
S.~Caron-Huot and K.~J.~Larsen,
%``Uniqueness of two-loop master contours,''                                  
JHEP {\bf 1210}, 026 (2012)
%doi:10.1007/JHEP10(2012)026
[arXiv:1205.0801 [hep-ph]];\\
%%CITATION = doi:10.1007/JHEP10(2012)026;%%
%
% Sogaard:2013yga
M.~S{\o}gaard,
  %``Global Residues and Two-Loop Hepta-Cuts,''                                 
  JHEP {\bf 1309}, 116 (2013)
%  doi:10.1007/JHEP09(2013)116
  [arXiv:1306.1496 [hep-th]];\\
  %%CITATION = doi:10.1007/JHEP09(2013)116;%%
%
% Johansson:2013sda
H.~Johansson, D.~A.~Kosower and K.~J.~Larsen,
%``Maximal Unitarity for the Four-Mass Double Box,''                          
Phys.\ Rev.\ D {\bf 89}, no. 12, 125010 (2014)
%doi:10.1103/PhysRevD.89.125010
[arXiv:1308.4632 [hep-th]];\\
%%CITATION = doi:10.1103/PhysRevD.89.125010;%%
%
% Sogaard:2013fpa
M.~S{\o}gaard and Y.~Zhang,
%``Multivariate Residues and Maximal Unitarity,''                             
JHEP {\bf 1312}, 008 (2013)
% doi:10.1007/JHEP12(2013)008
[arXiv:1310.6006 [hep-th]];\\
%%CITATION = doi:10.1007/JHEP12(2013)008;%%
%
% {Abreu:2017ptx}
S.~Abreu, R.~Britto, C.~Duhr and E.~Gardi,
%``Cuts from residues: the one-loop case,''                                   
JHEP {\bf 1706}, 114 (2017)
% doi:10.1007/JHEP06(2017)114
[arXiv:1702.03163 [hep-th]].
%%CITATION = doi:10.1007/JHEP06(2017)114;%%

%+% 2 refs
\bibitem{Sogaard2014ila}
M.~S{\o}gaard and Y.~Zhang,
%``Unitarity Cuts of Integrals with Doubled Propagators,''                    
JHEP {\bf 1407}, 112 (2014)
%doi:10.1007/JHEP07(2014)112
[arXiv:1403.2463 [hep-th]].
%%CITATION = doi:10.1007/JHEP07(2014)112;%%

%+% 1 ref
\bibitem{BaikovRep}
P.~A.~Baikov,
%``Explicit solutions of the three loop vacuum integral recurrence relations,''
Phys.\ Lett.\ B {\bf 385}, 404 (1996)
%doi:10.1016/0370-2693(96)00835-0
[hep-ph/9603267];\\
%%CITATION = doi:10.1016/0370-2693(96)00835-0;%%
%
  P.~A.~Baikov,
  %``Explicit solutions of the multiloop integral recurrence relations and its application,''
  Nucl.\ Instrum.\ Meth.\ A {\bf 389}, 347 (1997)
%  doi:10.1016/S0168-9002(97)00126-5
  [hep-ph/9611449];\\
  %%CITATION = doi:10.1016/S0168-9002(97)00126-5;%%
%
  R.~E.~Cutkosky,
  %``Singularities and discontinuities of Feynman amplitudes,''                 
  J.\ Math.\ Phys.\  {\bf 1}, 429 (1960);\\
% doi:10.1063/1.1703676
  %%CITATION = doi:10.1063/1.1703676;%%
%
  A.~G.~Grozin,
  %``Integration by parts: An Introduction,''                                   
  Int.\ J.\ Mod.\ Phys.\ A {\bf 26}, 2807 (2011)
%  doi:10.1142/S0217751X11053687
  [arXiv:1104.3993 [hep-ph]].
  %%CITATION = doi:10.1142/S0217751X11053687;%%

%+% 1 ref
\bibitem{HalfMaxD5}
Z.~Bern, S.~Davies, T.~Dennen and Y.-t.~Huang,
%``Ultraviolet Cancellations in Half-Maximal Supergravity 
% as a Consequence of the Double-Copy Structure,''
Phys.\ Rev.\ D {\bf 86}, 105014 (2012)
%doi:10.1103/PhysRevD.86.105014  
[arXiv:1209.2472 [hep-th]].
%%CITATION = doi:10.1103/PhysRevD.86.105014;%%

%+% 1 ref
\bibitem{Anomaly}
N.~Marcus,
  %``Composite Anomalies in Supergravity,''
  Phys.\ Lett.\  {\bf 157B}, 383 (1985);\\
%  doi:10.1016/0370-2693(85)90385-5
  %%CITATION = doi:10.1016/0370-2693(85)90385-5;%%
%
J.~J.~M.~Carrasco, R.~Kallosh, R.~Roiban and A.~A.~Tseytlin,
%``On the U(1) duality anomaly and the S-matrix of N=4 supergravity,'' 
  JHEP {\bf 1307}, 029 (2013)
  %doi:10.1007/JHEP07(2013)029   
  [arXiv:1303.6219 [hep-th]];\\
  %%CITATION = doi:10.1007/JHEP07(2013)029;%%
%
 R.~Kallosh,
  %``Cancellation of Conformal and Chiral Anomalies in $\mathcal{N} \geq 5$ supergravities,''
  Phys.\ Rev.\ D {\bf 95}, no. 4, 041701 (2017)
%  doi:10.1103/PhysRevD.95.041701
  [arXiv:1612.08978 [hep-th]];\\
  %%CITATION = doi:10.1103/PhysRevD.95.041701;%%
%
 D.~Z.~Freedman, R.~Kallosh, D.~Murli, A.~Van Proeyen and Y.~Yamada,
  %``Absence of U(1) Anomalous Superamplitudes in $\mathcal{N}\geq 5$ Supergravities,''
  JHEP {\bf 1705}, 067 (2017)
%  doi:10.1007/JHEP05(2017)067
  [arXiv:1703.03879 [hep-th]];\\
  %%CITATION = doi:10.1007/JHEP05(2017)067;%%
%
%\bibitem{Bern:2017tuc} 
  Z.~Bern, A.~Edison, D.~Kosower and J.~Parra-Martinez,
  %``Curvature-squared multiplets, evanescent effects, and the U(1) anomaly in $N=4$ supergravity,''
  Phys.\ Rev.\ D {\bf 96}, no. 6, 066004 (2017)
  %doi:10.1103/PhysRevD.96.066004
  [arXiv:1706.01486 [hep-th]].
  %%CITATION = doi:10.1103/PhysRevD.96.066004;%%

%+% 1 ref
\bibitem{Frellesvig2017aai}
H.~Frellesvig and C.~G.~Papadopoulos,
%``Cuts of Feynman Integrals in Baikov representation,''                      
JHEP {\bf 1704}, 083 (2017)
%doi:10.1007/JHEP04(2017)083
[arXiv:1701.07356 [hep-ph]].

\end{thebibliography}
\end{document}